\documentclass[journal,12pt,draftclsnofoot,onecolumn,english]{IEEEtran}

\usepackage{epsfig}
\usepackage{times}
\usepackage{float}
\usepackage{afterpage}
\usepackage{amsmath}
\usepackage{amstext,cite}
\usepackage{amssymb,bm}
\usepackage{latexsym}
\usepackage{color}
\usepackage{graphicx}
\usepackage{amsmath}
\usepackage{amsthm}
\usepackage{graphicx}
\usepackage[center]{caption}
\usepackage{pstricks}
\usepackage{caption}
\usepackage{subcaption}
\usepackage{booktabs}
\usepackage{multicol}
\usepackage{lipsum}
\usepackage[T1]{fontenc}
\usepackage{hyperref}
 \usepackage{aecompl}

\allowdisplaybreaks

\long\def\comment#1{}

\newcommand{\dv}{{\mathbf d}}

\newcommand{\Zm}{{\mathbf Z}}

\newcommand{\Dc}{{\mathcal D}}

\newcommand{\Fc}{{\mathcal F}}
\newcommand{\Gc}{{\mathcal G}}

\newcommand{\Kc}{{\mathcal K}}

\newcommand{\Rc}{{\mathcal R}}
\newcommand{\Sc}{{\mathcal S}}

\newcommand{\Uc}{{\mathcal U}}
\newcommand{\Wc}{{\mathcal W}}
\newcommand{\Vc}{{\mathcal V}}

\newcommand{\hsf}{{\mathsf h}}

\newcommand{\ksf}{{\mathsf k}}
\newcommand{\lsf}{{\mathsf l}}

\newcommand{\Bsf}{{\mathsf B}}

\newcommand{\Hsf}{{\mathsf H}}

\newcommand{\Ksf}{{\mathsf K}}
\newcommand{\Lsf}{{\mathsf L}}
\newcommand{\Msf}{{\mathsf M}}
\newcommand{\Nsf}{{\mathsf N}}

\newcommand{\Rsf}{{\mathsf R}}

\renewcommand{\arg}{{\hbox{arg}}}

\newtheorem{thm}{Theorem}
\newtheorem{cor}{Corollary}

\newtheorem{rem}{Remark}
\newtheorem{example}{Example}
\newtheorem{defn}{\protect\definitionname}

\providecommand{\definitionname}{Definition}

\newcommand{\Kbc}{\mathsf{K}_\mathrm{mbs}}
\newcommand{\Kdd}{\mathsf{K}_\mathrm{sbs}}
\newcommand{\Rbc}{\Rsf_{\mathrm{mbs}}}
\newcommand{\Rdd}{\Rsf_{\mathrm{sbs}}}

\begin{document}

\title{On the Fundamental Limits of Fog-RAN Cache-aided Networks with Downlink and Sidelink Communications}

\author{
Kai~Wan,~\IEEEmembership{Member,~IEEE,} 
Daniela Tuninetti,~\IEEEmembership{Senior~Member,~IEEE,}
Mingyue~Ji,~\IEEEmembership{Member,~IEEE,}
and~Giuseppe Caire,~\IEEEmembership{Fellow,~IEEE}
\thanks{
K.~Wan and G.~Caire are with the 
EECS Dept., Technische Universit\"at Berlin, 10587 Berlin, Germany
(e-mail:  kai.wan@tu-berlin.de; caire@tu-berlin.de). The work of K.~Wan and G.~Caire was partially funded by the ERC Advanced Grant N. 789190, CARENET.}
\thanks{
D.~Tuninetti is with the ECE Dept., University of Illinois at Chicago, Chicago, IL 60607, USA (e-mail: danielat@uic.edu). The work of   D.~Tuninetti was supported by   NSF 1527059.}
\thanks{
M.~Ji is with the ECE Dept., University of Utah, Salt Lake City, UT 84112, USA (e-mail: mingyue.ji@utah.edu). The work of M.~Ji was supported by NSF 1817154  and 1824558.}
\thanks{
A short version of this paper is presented at IEEE International Symposium on Information  Theory (ISIT) 2019. 
}
}
\maketitle

\begin{abstract}
Maddah-Ali and Niesen (MAN) in 2014 showed that coded caching in single bottleneck-link broadcast networks allows serving an arbitrarily large number of cache-equipped users with a total link load (bits per unit time) that does not scale with the number of users. 
Since then, the general topic of coded caching has generated enormous interest both from the information theoretic and (network) coding theoretic viewpoint, and 
from the viewpoint of applications.  Building on the MAN work, this paper considers a particular network topology  referred to as
cache-aided Fog Radio Access Network (Fog-RAN), that includes a Macro-cell Base Station (MBS) co-located with the content server, 
several cache-equipped Small-cell Base Stations (SBSs), and many users without caches. Some users are served directly by the MBS broadcast downlink, 
while other users are served by the SBSs. The SBSs can also exchange data via rounds of direct communication via a side channel, referred to as ``sidelink''.  
For this novel Fog-RAN model, the fundamental tradeoff among
(a) the amount of cache memory at the SBSs, 
(b) the load on the downlink (from MBS to directly served users and SBSs), and  (c) the aggregate load on the sidelink
is studied, under the standard worst-case demand scenario. Several existing results are recovered as special cases of 
this network model and byproduct results of independent interest are given. 
Finally, the role of topology-aware versus topology-agnostic caching is discussed. 
\end{abstract}

\section{Introduction}
\label{sec:intro}

In content distribution networks, traffic can be smoothed out  by placing content 
in local caches ``closer'' to the end users during off-peak hours ({\it placement phase}),
with the hope that the pre-fetched content will be requested during peak hours, in which case the load 
from the server to the users ({\it delivery phase}) will be reduced.
Coded caching was originally considered in~\cite{dvbt2fundamental} by Maddah-Ali and Niesen (MAN) for a 
single bottleneck-link broadcast network model, where a server 
communicates to $\Ksf$ users 
through a shared noiseless channel of finite capacity;
the server contains a library of $\Nsf$ content files; each user  has a cache able to store the equivalent size of $\Msf$ files.
In the original MAN scheme \cite{dvbt2fundamental}, each of the $\Nsf$ files is partitioned into a number of segments (subfiles), 
and each subfile is stored into a number of user caches that depends on how many times the library can be replicated across 
all the aggregate cache memory.  
After this symmetric uncoded cache placement phase, MAN generates coded multicast messages by bit-wise XOR multiple requested subfiles, 
in such a way that each XOR-ed transmission (multicast message) is simultaneously useful to many users; 
these coded multicast message delivery drastically reduces the download time, or network traffic load, 
and harness the so-called global caching gain, or multicasting gain (which is proportional to the total amount of memory in the system), 
compared to traditional caching strategies such as multiple unicast. 
The MAN scheme is known to be exactly optimal for single bottleneck-link broadcast networks under the constraint of uncoded cache placement~\cite{ontheoptimality,exactrateuncoded}, and optimal to within a factor of~$2$ otherwise~\cite{yas2}.\footnote{%
In terms of terminology, known caching schemes can be divided into three classes: 
those with {\it uncoded cache placement} (where bits from the library are simply copied within the caches), 
those with {\it intra-file coded placement} (where coding only occurs within the bits of the same file), and
those with {\it inter-file coded placement} (where coding occurs across the bits of all files).
}

In~\cite{d2dcaching}, Ji {\it et al.} extended the MAN model in~\cite{dvbt2fundamental} to the case where 
the cache-aided users communicate among each other in the delivery phase (and no communication from the server is allowed). 
The Device-to-Device (D2D) caching scheme in~\cite{d2dcaching} is optimal to within a factor of~$8$, and asymptotically achieves the same load as that of the MAN scheme when the number of users goes to infinity. This last fact is somewhat counterintuitive since, in D2D caching networks, each node can only encode 
based on the content of its own storage (instead of the entire library), which in principle limits the amount of multicasting opportunities. 

As argued in~\cite{bastug2014caching5g,poularakis2016explotin5g,tondon2016cloudedge} and references therein, to bring the caching idea into the reality of next generation cellular networks, hybrid architectures are of great interest. In such models, the operator/Internet provider caches popular content files at the 
Fog Radio Access Networks (Fog-RAN), which is then delivered in two hops. 
In the first hop, the Macro-cell Base Station (MBS) transmits packets to Small-cell Base Stations (SBSs);  
in the second hop, the SBSs communicate  
 with their locally connected users. 
The works in~\cite{cacheaidedinter,interferencemanagement,interferenceXu2017,degreesHachem2018,sengupta2017fog,zhang2017multi,tandon2016cloud}  modeled the second hop as a Gaussian interference channel; the fundamental limits of such a network were investigated in the high SNR regime, by either using the Degrees of Freedom (DoF) metric~\cite{cacheaidedinter,interferencemanagement,interferenceXu2017,degreesHachem2018}, or the  Normalized Delivery Time (NDT) metric~\cite{sengupta2017fog,zhang2017multi,tandon2016cloud,karasik2019howmuch} that captures the worst-case coding latency. 
In~\cite{karasik2019howmuch}, a similar Fog-RAN over a Gaussian interference channel is studied, with the twist that D2D communication 
among the cache-less users is allowed. 

The proposed Fog-RAN model is not meant to capture the interference-limited aspect of the last hop of wireless cellular systems. 
Instead, we focus on {\it the interplay between the cellular downlink (from the MBS to some users and the SBSs) and other co-existing 
``side links'' (among SBSs and from SBSs to their own local users), over which part of the cellular traffic may be offloaded.}
In particular, we refer to the ensemble of communication links between SBSs as {\em sidelink}. This can be implemented via wires (e.g., 
the SBSs may be connected to the same Ethernet cable, as in segment of an enterprise WiFi network) or 
via radio, on a separate band with respect to the cellular downlink. Furthermore, we consider the local access between each SBS and its connected users
as very high rate, such that these links are never the system bottleneck.

\subsection{Brief Description of the Proposed Fog-RAN Model}  
\label{sub:Brief Description}

\begin{figure}
\centerline{\includegraphics[scale=0.3]{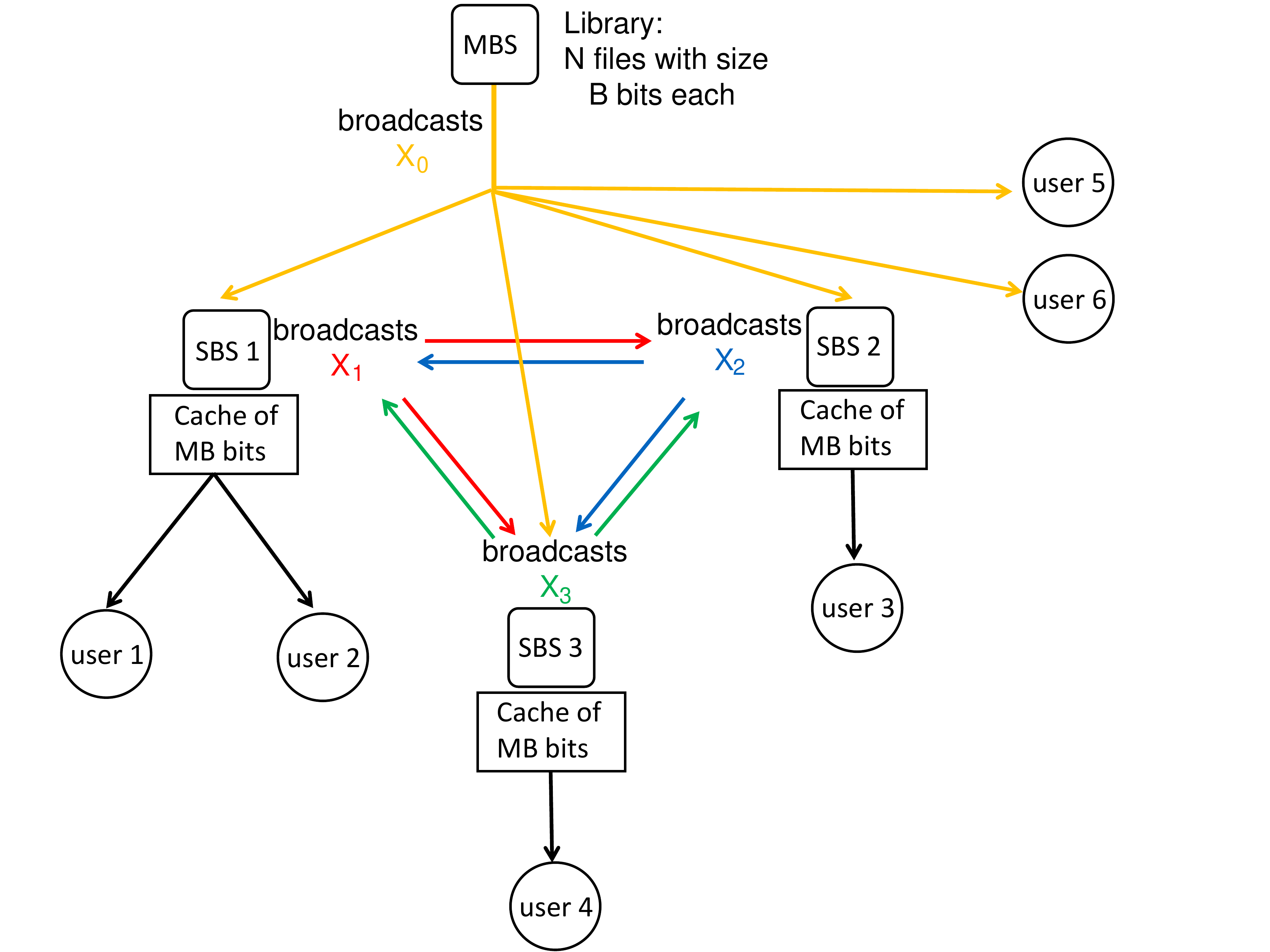}}
\caption{\small The considered Fog-RAN architecture with one MBS and $\Hsf=3$ SBSs. There are $\Kdd=4$ users connected to some SBS, and $\Kbc=2$ users directly connected to the MBS.}
\label{fig: system_model}
\end{figure}

This paper considers the Fog-RAN network model as illustrated in Fig.~\ref{fig: system_model}. 
An MBS with a library of $\Nsf$ files, is connected to $\Hsf$ SBSs and to $\Kbc$ users via standard cellular downlink, modeled here as 
a noiseless broadcast channel of unit capacity of one file per unit time. 
Each SBS has a local memory to cache the equivalent size of $\Msf$ files.  SBSs can communicate between each others via another broadcast channel
referred to as {\em sidelink}, orthogonal to the cellular downlink.  There are also $\Kdd$ users, each of which is connected to one of the SBSs. The local ``access network'' between each SBS and its connected users
has very high capacity, such that this is never the system bottleneck, and there is no interference between such local access networks and
either the cellular downlink and the Fog-RAN sidelink.  
Users have no local cache memory.  The two populations of $\Kbc$ and $\Kdd$ users reflect the practical scenario where the SBSs 
offer high throughput connectivity over a limited coverage area (hot-spots),  and therefore there are some users outside the reach of the SBSs, 
which must be served directly by the MBS. 

There are two phases in such a network: a cache placement phase, done without knowledge of the later users' demands, and a delivery phase. 
In the proposed Fog-RAN model, the delivery phase contains three steps. 
(1) In the first step, the MBS broadcasts packets to the SBSs and the $\Kbc$ directly served users via the downlink.
(2) In the second step, the SBSs exchange coded packets among each other via the sidelink. Such coded packets are
functions of their cached content and of the received packets from the MBS. 
(3) In the third step, based on the sidelink exchange and on the received packets from the MBS, each SBS delivers the demanded files to its connected users via the local access network.  The goal is to find the tradeoff among the memory size at the SBSs, 
the MBS load (i.e., number of broadcasted bits by the MBS in the first step), and the load of the SBS sidelink communication (i.e., number of total exchanged 
bits in the second step), for the worst-case among all possible demands.

It is reasonable and practical to consider the MBS downlink load and SBSs sidelink load separately, because the transmissions over these two subnetworks typically takes place  in orthogonal channels (as said before) with possibly very different operational costs (e.g., the MBS uses prime cellular frequencies, 
while the SBS sidelink may use some mmWave band or an Ethernet cable). 
 
For certain configuration of our network model parameters, one can recover special models studied in the literature.  
When $\Kdd=0$ (i.e., there are no users directly connected to the SBSs) our model reduces to the MAN single bottleneck-link coded 
caching problem with {\it cache-less users} \cite{cless2018Em}.
When $\Kdd=\Hsf$, and each of the $\Kdd$ users is connected to a distinct SBS, and there is no sidelink communication among the SBSs, 
we have the single bottleneck-link coded caching problem with {\it heterogeneous cache sizes}~\cite{wang2015hetero,bidokhti2018noisycaching,ibrahim2017hetero,asadi2018unequal,daniel2017hetero,liao2017smallcell,cao2018twousers,ibrahim2018coded}, where some users have no cache. 
When $\Kbc=0$ (i.e., there are no users directly connected to the MBS) and there is no sidelink communication among the SBSs, 
we obtain the single bottleneck-link coded caching problem with {\it shared caches, or with users making multiple requests}~\cite{parrinello2018sharedcache,multishared2018Karat,multiJi2014,Sengupta2017multirequest,xu2018codedsharedcache}. 
When $\Kdd=\Hsf$, $\Kbc=0$ and the MBS load is equal to $0$, our model reduces to the D2D coded caching problem~\cite{d2dcaching}.
Next we shall summarize the main results for these various coded caching models subsumed by our model.
 
\subsection{Related Works}  
\label{sub:related works}

\paragraph{Systems with Cache-less Users or with Heterogeneous Cache sizes}
Since the users directly served by the MBS in our model do not have caches (while the SBSs do), 
our Fog-RAN model is related to the single bottleneck-link caching problem with cache-less users.

%
For such heterogeneous system with cache-less and cache-aided users, 
the scheme of \cite{cless2018Em} uses MAN uncoded cache placement to fill the caches of the cache-aided users. This 
scheme is optimal under the constraint of uncoded cache placement if $\Nsf\geq \Ksf$.
For the single bottleneck-link caching problem with heterogeneous cache sizes, 
caching schemes based on uncoded placement and various delivery schemes using linear coding and random coding 
were proposed in~\cite{wang2015hetero,bidokhti2018noisycaching,ibrahim2017hetero,asadi2018unequal,daniel2017hetero}.
The achievable scheme in~\cite{wang2015hetero} was proved to be generally order optimal within a factor of $12$ if $\Nsf\geq \Ksf$. 
Also for $\Nsf\geq \Ksf$, the caching scheme in~\cite{ibrahim2017hetero} is optimal under the constraint of uncoded cache placement if the aggregate cache memory 
is not larger than the size of library. 

Other schemes have considered coded cache placement. For example, 
intra-file MDS (Maximum Distance Separable) pre-coded placement phase and a linear-programming based coded delivery 
was proposed in~\cite{liao2017smallcell}, where during the delivery phase the server directly transmits some packets of each file. 
The MDS pre-coding ensures that the packets delivered from the server are different from the cached contents. 
For the case of heterogeneous cache sizes and only two users, the authors in~\cite{cao2018twousers} proposed an optimal 
scheme using inter-file coded cache placement for the memory size pair $(\Msf_1,\Msf_2)=(\Nsf-1,0)$, where $\Msf_1$ and $\Msf_2$ represent 
the memory sizes of the two users, respectively. 
In~\cite{ibrahim2018coded}, the authors proposed an MDS  inter-file coded cache placement scheme for the case of heterogeneous 
cache sizes and three users 
that strictly outperforms existing schemes based on uncoded cache placement.  It was also extended to small memory size regime (total cache sizes of all users are less than $\Nsf$).
Intuitively, coded placement benefits systems with heterogeneous cache sizes because the transmissions intended to serve the users with small 
cache size are useful in decoding the coded cache content of the users with large cache size. 

\paragraph{Systems with Shared Caches or with Multiple Requests}
Since the throughput from each SBS to its connected users can be regarded as infinite in our model, 
our model is related to the single bottleneck-link caching problem with shared caches, that is, group of users with possibly different demands 
are connected to the same cache.

In~\cite{parrinello2018sharedcache}, the shared cache model with more files than users and uncoded placement is considered. 
The authors used MAN uncoded placement together with a multi-round delivery phase to achieve the optimal worst-case load under the constraint 
that the placement is both uncoded and topology-agnostic, i.e., the placement cannot be a function of the number of users connected to each cache 
(referred to in the following as {\em occupancy number}). 
When there exist files demanded by several users, the authors in~\cite{multishared2018Karat} proposed to use the multi-round method in~\cite{parrinello2018sharedcache} and in each round to use the single bottleneck-link caching scheme in~\cite{exactrateuncoded} 
to leverage the multicast opportunities, which improves on the performance of the caching scheme in~\cite{parrinello2018sharedcache}. 

If each cache is shared by the same number of users, the single bottleneck-link caching problem with shared caches is equivalent to the single bottleneck-link caching problem with multiple requests~\cite{multiJi2014}, where each user equipped with a cache requests $\Lsf$ files. 
It was proved in~\cite{Sengupta2017multirequest} that dividing this problem into $\Lsf$ independent MAN systems can achieve a load which 
is order optimal within a factor of $11$.

Moving to coded placement strategies, it is worthwhile to remark that 
a caching scheme based on an inter-file coded cache placement was proposed in~\cite{xu2018codedsharedcache} for $\Hsf\Msf\in [1, \Nsf]$, 
to improve the caching scheme in~\cite{parrinello2018sharedcache} for small memory size regime but without any optimality guarantees.

\subsection{Contributions}  
\label{sub:contributions}
In this paper, we study the fundamental tradeoffs for the novel Fog-RAN architecture formally described in Section~\ref{sec:model}. 
The topology of our network is defined by the number $\Hsf$ of SBSs, by the number of users $\Kbc$ directly served  by the MBS downlink, 
and by the number of users connected to each SBS (occupancy number). 
It is reasonable to assume that the cache placement may depend on $\Hsf$,  since the number of SBSs within a macro-cell 
is known by the system designer and does not change over time (unless on a very slowly time scale). 
In contrast, the knowledge of $\Kbc$ and of the SBS occupancy numbers may or may not be available at the placement phase, depending on 
the network topology dynamics which in turn depend on the specific application.
Hence, we distinguish  the proposed cache-aided Fog-RAN systems into two classes:
{\em topology-agnostic Fog-RAN systems}, for which the placement phase may depend on $\Hsf$  but cannot exploit
the knowledge of  $\Kbc$ and the SBS occupancy numbers, and  {\it topology-aware Fog-RAN systems}, for which the placement phase can exploit the full topology information.

In addition, we also distinguish the achievable caching schemes into three classes depending on the use of $\Kbc$  and  the SBS occupancy numbers in the placement phase:  {\em topology-agnostic caching schemes} which do not leverage $\Kbc$  and  the SBS occupancy numbers, {\em topology-partially-agnostic caching schemes}
which   only exploit    the knowledge of  $\Kbc$, and {\em topology-aware caching schemes} which exploit $\Kbc$  and  the SBS occupancy numbers.
One should notice that for topology-agnostic Fog-RAN systems,  only topology-agnostic caching schemes can  be used while for topology-aware Fog-RAN systems, any class of caching schemes can be used.

Our main contributions are as follows.
\begin{itemize}

\item {\em Topology-aware Systems.}

\paragraph{{\bf Converse}}
We derive a converse bound based on  Han's inequality~\cite[Theorem 17.6.1]{hanineq}\footnote{Han's inequality was   used  in~\cite{Sengupta2017multirequest} for the single bottleneck-link caching problem with multiple requests,   when each user demands the same number of files. It will be explained later in Section~\ref{sub:main converse} that  the proof in~\cite{Sengupta2017multirequest} based on Han's inequality can be not used in our problem  because of the asymmetry in the problem. Our strategy based on   Han's inequality is thus different to the one in~\cite{Sengupta2017multirequest}. }, which is tighter than the one based on a straightforward use of the cut-set bound idea in~\cite{dvbt2fundamental}. In simple terms, our technique allows one to remove the ``floor operator'' in the cut-set bound. In addition, with the goal of bounding the sidelink load, 
we develop a novel lower bound of the D2D caching problem \cite{d2dcaching}. 
 
 \paragraph{{\bf By-product~1}}
The proposed bounding approach can also improve state-of-the-art  cut-set converse bounds for other cache-aided networks  by removing the ``floor operator'' in those converse bounds, e.g., the converse bound for   single bottleneck-link caching system in~\cite{dvbt2fundamental}, the converse bound for single bottleneck-link caching system wih multi-requests in~\cite{multiJi2014}, the converse bound for single bottleneck-link caching system with heterogeneous cache sizes in~\cite{wang2015hetero}, etc.
In addition, with our new lower bound for D2D caching networks eliminates the constraint on symmetric load transmitted 
by each D2D user which limits the applicability of the converse bound for D2D caching with multi-request in~\cite{Sengupta2017multirequest}.

 \paragraph{{\bf  Topology-partially-agnostic Caching Scheme}}
Observing that there are users  directly served by the MBS downlink, for which the MBS must broadcast the whole demanded files, 
we propose an inter-file coded cache placement with the MAN symmetric file subpacketization, such that the SBSs can leverage the broadcasted 
files by the MBS to its directly served users in order to ``decode'' the coded content in their cache.

The delivery phase consists of two steps. 
In the first step, we let the $\Kbc$ users directly served by the MBS downlink to recover their desired file, 
and let the SBSs decode the coded content in their cache. Then the MBS broadcasts packets based on the single bottleneck-link delivery with shared caches in~\cite{parrinello2018sharedcache}. In the second step, instead of directly extending the idea of~\cite{parrinello2018sharedcache} to the D2D scenario, we propose a novel group-based coded D2D delivery with shared caches which outperforms the direct extension of~\cite{parrinello2018sharedcache}. 
 \paragraph{{\bf By-product~2}}
The proposed inter-file coded cache placement scheme can be used for the single bottleneck-link caching model with heterogeneous cache sizes if there are cache-less users. 
The resulting performance is strictly better than that of the scheme with uncoded cache 
placement in~\cite{cless2018Em,bidokhti2017cacheassignment}. 
Different from existing coded cache placement schemes for the standard MAN single bottleneck-link problem (which can only lead to a small gain in a small memory size regime compared to the uncoded cache placement as shown in~\cite{smallbufferusers,codedcachingviaintef,ibrahim2018coded,amiri2016improved,vilar2018coded}), the proposed inter-file coded cache placement for this heterogeneous model provides a significant gain in all memory size regimes compared to the uncoded cache placement.  Compared to the inter-file coded cache placements in~\cite{cao2018twousers,ibrahim2018coded},  
the proposed placement in this paper  can work for any number of users and for any memory sizes (please also refer to Remark~\ref{rem:extension to heterogeneous}).\footnote{The results in this paper are independent from the very recent paper~\cite{ibrahim2018coded}, which was posted on arXiv two days before we posted the arXiv version of this paper~\cite{franarxiv}, which was then submitted to ISIT 2019.}


 \paragraph{{\bf  Optimality}}
We prove the following (exact or order) optimality results by comparing the combination of our achievable schemes to our converse bound: 
\begin{itemize}
\item $\Nsf\leq \Kbc$. The proposed scheme is exactly optimal.
\item $\Kbc<\Nsf\leq \Kbc+\Lsf_{\max}$, where $\Lsf_{\max}$ is the maximum occupancy number over all the SBSs.  
The proposed scheme is order optimal to within a constant factor of $2$.
\item $\Nsf> \Kbc+\Lsf_{\max}$. The proposed scheme is order optimal to within a factor of $22$ if the SBS have balanced occupancy numbers (i.e., 
each SBS serves the same number $\Lsf_{\max}$ of users). If the occupancy numbers are unbalanced, 
we can show the order optimality within a factor of  $2g$, where $g:=\min\left\{\Hsf,\frac{\Nsf-\Kbc}{\Msf}\right\}$. 
\end{itemize}

 \paragraph{{\bf   Topology-aware Caching Scheme}}
The above  achievable scheme is topology-partially-agnostic (in the sense defined at the beginning of this section). We also consider a topology-aware
achievable scheme that exploit the detailed knowledge of the SBS occupancy numbers in the placement phase. 
Our topology-aware cache placement combines the proposed inter-file coded cache placement with a novel asymmetric file subpacketization 
method that leverages the network structure. In this case, we can prove  {\em exactly optimality} for some memory size regime. 

 \paragraph{{\bf  By-product~3}}
For the single bottleneck-link caching problem with shared caches, our novel topology-aware  scheme can be shown to   reduce the load compared to the scheme in~\cite{parrinello2018sharedcache} whose placement is designed without leveraging the detailed network topology,  as well as the subpacketization level.

\item {\em  Topology-agnostic Systems.}

 \paragraph{{\bf   Topology-agnostic Caching Scheme}}
The fact that our new inter-file coded cache placement with asymmetric file subpacketization 
provide a gain when $\Kbc$ and the SBS occupancy numbers are known at the placement phase, suggests a possible application of such approach
also in the   random topologies with fixed $\Hsf$ and
$\Kbc$ and  the SBS occupancy numbers are random variables with known statistics. 
For example, in a mobile user scenario it is reasonable to assume that each hot-spot has an occupancy number that fluctuates statistically 
around some typical value, and that the histogram of the occupancy numbers can be learned over time, since it remains stable across time, 
possibly following some cyclostationary statistics (e.g., imagine a hot-spot located at a subway station, whose occupancy number distribution changes
over time according to the rush hours pattern).
Hence, we extend the proposed cache placement and delivery schemes to the topology-agnostic systems where, 
during the placement phase, only the distributions of $\Kbc$ and  the occupancy numbers is known a priori, such that the new 
asymmetric placement  can be designed based on such statistical knowledge. 
 
\end{itemize}

\subsection{Paper Organization}
The rest of this paper is organized as follows.
Section~\ref{sec:model} defines the problem.
Section~\ref{sec:main converse} presents our main results  on the converse bound  for topology-aware systems.
Sections~\ref{sec:main symmetric} and~\ref{sec:improvement} describe the proposed achievable schemes 
 with symmetric and asymmetric file subpacketizations for  topology-aware systems, respectively.
Section~\ref{sub:decentralized} extends the proposed ideas to topology-agnostic systems. Finally,
Section~\ref{sec:conclusion} concludes the paper and some proofs are in the Appendices.

\subsection{Notation Convention}
Calligraphic symbols denote sets, 
bold symbols denote vectors, 
and sans-serif symbols denote system parameters.
We use $|\cdot|$ to represent the cardinality of a set or the length of a vector;
$[a:b]:=\left\{ a,a+1,\ldots,b\right\}$ and $[n] := [1,2,\ldots,n]$; 
$\oplus$ represents bit-wise XOR; $\mathbb{E}[\cdot]$ represents the expectation value of a random variable; 
$[a]^+:=\max\{a,0\}$; 
we let $\binom{x}{y}=0$ if $x<0$ or $y<0$ or $x<y$;
the number of $k$-permutations of $n, n\geq k,$ is indicated as $P(n,k):=n\cdot(n-1)\cdots(n-k+1)$.

\section{System Model of A  $(\Kbc,\Kdd,\Hsf,\Nsf,\Msf)$ Topology-Aware Fog-RAN System} 
\label{sec:model}

A  $(\Kbc,\Kdd,\Hsf,\Nsf,\Msf)$ topology-aware Fog-RAN system is defined as follows.

An MBS has access to $\Nsf$ files. 
The set of files is denoted by $\{F_1, \cdots, F_\Nsf\}$. 
Each file is composed of $\Bsf$ i.i.d. bits.
The MBS is connected to $\Hsf$ SBSs and to $\Kbc$ users through a single bottleneck-link of finite capacity. 
The~$h$-th SBS, $h\in [\Hsf],$ serves $\Lsf_h$ users through a local broadcast link with infinite throughput (i.e., 
$\Lsf_h$ is the occupancy number of SBS $h$).
The total number of users in the system is $\Ksf:=\Kdd+\Kbc$, of which $\Kbc$ are directly served by the MBS and $\Kdd:=\sum_{h \in [\Hsf]}\Lsf_h$ are connected to SBSs.
The set of user directly connected to the MBS as $\Uc_0:=\big[\Kbc \big]$, while
the set of users connected to the~$h$-th SBS is denoted as $\Uc_h:=\big[1+\sum_{i \in [0:h-1]}\Lsf_i, \sum_{i \in [0:h]}\Lsf_i\big]$ for $h\in [\Hsf]$ and with $\Lsf_0=\Kbc$. 
Without loss of generality, we let $\Lsf_1\geq \dots \geq \Lsf_{\Hsf}$.
We shall denote the SBS serving user $k\in[\Kdd]$ as $\hsf_{k}$,
i.e., $\hsf_{k}=h$ if and only if $k\in\Uc_h$.
To simplify the notation, 
let 
\begin{align}
\Lsf_{\Sc}:=\sum_{h\in \Sc}\Lsf_h, \ \forall \Sc \subseteq [\Hsf], 
\label{eq:Lsf_Sc def}
\end{align} 
be the total occupancy number of  the  SBSs in $\Sc$.

The system operates in two phases.

{\it Placement Phase.}
During the placement phase, the~$h$-th SBS, $h\in [\Hsf],$ stores information about the $\Nsf$ files in its local cache of size $\Msf\Bsf$ bits, where $\Msf \in[0,\Nsf]$. The placement is done with knowledge of  $\Hsf$, $\Kbc$, and the occupancy number of each SBS, while the placement is  without knowledge of users' future demands.
The content in the cache of SBS $h\in[\Hsf]$ is denoted by $Z_{h}$.  We let $\Zm:=(Z_{1},\ldots,Z_{\Hsf})$.
In other words, for placement functions $\phi_h, h\in[\Hsf]$,  we have 
\begin{align}
Z_{h} = \phi_h(F_1, \cdots, F_\Nsf) : H(Z_{h}) \leq \Bsf \Msf, \ \forall h\in[\Hsf].
\label{eq: placement functions def}
\end{align}

{\it Delivery Phase.}
Once the placement phase is completed, each user demands one file.
The file demanded by user $k\in[\Ksf]$ is denoted by $d_{k}\in[\Nsf]$. 
We denote the demand vector as $\dv:=(d_1,\ldots,d_\Ksf)$. 
The delivery phase contains the following steps:
\begin{enumerate}

\item Broadcast from MBS via cellular downlink.
Given demand vector $\dv$, the MBS broadcasts a message $X_{0}$ of $\Bsf \, \Rbc(\dv)$ bits to each SBS $h\in [\Hsf]$ and each user $k\in \Uc_0$.
In other words, for encoding function $\psi_0$, we have
\begin{align}
X_{0} = \psi_0(\dv,F_1, \cdots, F_\Nsf) : H(X_{0}) \leq \Bsf \Rbc(\dv).
\label{eq: MBS encoding function def}
\end{align}

\item SBS sidelink communication.
Given $(\dv,Z_h,X_0)$, SBS $h\in [\Hsf]$ broadcasts a message $X_{h}$ of $\Bsf \, \Rsf_{h}(\dv)$ bits to all other SBSs.
In other words, for encoding functions $\psi_h, h\in [\Hsf]$, we have 
\begin{align}
X_{h} = \psi_h(\dv,Z_h,X_0) : H(X_{h}) \leq \Bsf \Rsf_{h}(\dv), \ \forall h\in[\Hsf].
\label{eq: SBS encoding function def}
\end{align}
The SBS sidelink load  for demand vector $\dv$, is denoted as
\begin{align}
\Rdd(\dv):=\sum_{h\in [\Hsf]}\Rsf_{h}(\dv).
\label{eq: d2d load def}
\end{align} 
\end{enumerate}

Decoding of the demanded files is as follows.
For each user $k\in \Uc_h$, SBS $h\in [\Hsf]$ decodes $F_{d_k}$ from $(Z_h,X_0,X_1,\ldots,X_{\Hsf})$, and then forwards $F_{d_k}$ to user $k$.
In other words, for decoding functions $\xi_k, k\in[1+\Kbc:\Kdd+\Kbc]$, we have
\begin{align}
\widehat{F_{d_k}} = \xi_k(\dv,Z_{\hsf_{k}},X_0,X_1,\ldots,X_{\Hsf}), \ \forall k\in \cup_{h\in [\Hsf]} \Uc_h.
\label{eq: SBS decoding functions def}
\end{align}
The users directly connected to the MBS decode their demanded file from $X_0$ only.
In other words, for decoding functions $\xi_k, k\in[\Kbc]$, we have
\begin{align}
\widehat{F_{d_k}} = \xi_k(\dv,X_0), \ \forall k\in \Uc_0.
\label{eq: MBS decoding functions def}
\end{align}

{\it Objective.}
We say that the 
pair $(\Rbc,\Rdd)$ is said to be achievable for the memory constraint $\Msf$, 
if there exist encoding and decoding functions as defined above 
such that all possible demand vectors can be delivered with delivery load pair $(\Rbc,\Rdd)$
such that $\lim_{\Bsf\to\infty}\Pr[\widehat{F_{d_k}} \not= F_{d_k}, \text{for some $\dv$ or some $k$}]=0$. 
We denote the achievable 
region as $\Rc$.
The goal is to characterize the convex closure of the region $\Rc$.

\section{Topology-aware Fog-RAN Systems: Converse Bound}
\label{sec:main converse}
In this section, we   propose a novel converse bound  for topology-aware  Fog-RAN systems with the outline/intuition on the proof. The details of the proof are given in Appendix~\ref{sec:converse proof}.

\subsection{Main Results and Discussion}
\label{sub:main converse}
Recall that, without loss of generality, we have assumed $\Lsf_1\geq \dots \geq \Lsf_{\Hsf} >0$. Recall that $\Lsf_{\Sc}$ in~\eqref{eq:Lsf_Sc def} is the total occupancy number of  the  SBSs in $\Sc$.
\begin{thm}
\label{thm:converse}
\begin{subequations}
For the $(\Kbc,\Kdd,\Hsf,\Nsf,\Msf)$ topology-aware Fog-RAN system, 
an achievable $(\Rbc^{\star},\Rdd^{\star})$ must satisfy
\begin{align}
&  \Rbc^{\star}\geq \min\{\Nsf,\Kbc\}, \quad  \Rdd^{\star}\geq 0,
\label{eq:trivial converse}
\\
&\Rbc^{\star}+ \Rdd^{\star} \geq 
 \Kbc +  
 \min\{\Lsf_{[s]},\Nsf-\Kbc\}\left[1-\frac{s \Msf}{\Nsf-\Kbc} \right]^+, \forall s\in[\Hsf],
\quad \text{if} \quad \Nsf > \Kbc,
\label{eq:converse N>K0} 
\\
&\Rbc^{\star}+ \left(1-\frac{s}{\Hsf}\right) \Rdd^{\star} \geq  \Kbc+ \frac{s}{\Hsf }\Kdd\left[1-\frac{s \Msf}{\Nsf-\Kbc} \right]^+ , \forall s \in [\Hsf], 
\quad \text{if} \quad \Nsf \geq  \Ksf=\Kbc+\Kdd.
\label{eq:converse N>K=K0+K1} 
\\
&\Rbc^{\star} \geq  
 \Kbc +  
 \min\{\Kdd,\Nsf-\Kbc\}\left[1-\frac{\Hsf \Msf}{\Nsf-\Kbc} \right]^+,
\quad \text{if} \quad \Nsf > \Kbc,
\label{eq:converse N>K0 s=H} 
\end{align}
\label{eq:converse}
\end{subequations}
\end{thm}

Our converse bounds can be interpreted as follows.

\paragraph*{Bound~\eqref{eq:trivial converse}}
The bound in~\eqref{eq:trivial converse} trivially says that, when considering only the cache-less users directly connected to the MBS (and neglecting all other users), the MBS must send in the worst case either all the files in the library or one file per user, whichever is smaller. 

%

\paragraph*{Bound~\eqref{eq:converse N>K0}}
Intuitively the bound in~\eqref{eq:converse N>K0}, which holds for $\Nsf > \Kbc$, can be interpreted as follows. The worst-case overall load cannot be smaller than that of a genie-aided system in which:
(i) the users connected directly to the MBS request $\Kbc$ different files, 
(ii) the users connected to an SBS request one file among the remaining $\Nsf-\Kbc$ files;
(iii) the cache contents of the $s$ most loaded SBSs are aggregated (for an overall cache size of $s\Msf$ files; that cache serves a total of $\Lsf_{[s]}$ users), 
(iv) the demands of the users connected to the remaining $\Hsf-s$ least loaded SBSs are dismissed.
Then the genie-aided system can be thought as the following two systems in parallel:
(S1) a single bottleneck-link system with $\Nsf^\text{(S1)} = \Kbc$ files, serving $\Ksf^\text{(S1)} = \Kbc$ users with cache size $\Msf^\text{(S1)} = 0$ and each user making $\Lsf^\text{(S1)} = 1$ requests; this requires the transmission of $\Rsf^\text{(S1)} = \Ksf^\text{(S1)} = \Kbc$ files; and
(S2) a single bottleneck-link system with $\Nsf^\text{(S2)} = \Nsf-\Kbc$ files, serving a $\Ksf^\text{(S2)} = 1$ user with cache size $\Msf^\text{(S2)} = s\Msf$ and the user making $\Lsf^\text{(S2)} =  \min\{\Lsf_{[s]},\Nsf-\Kbc\}$ requests; this requires the transmission of $\Rsf^\text{(S2)}$ files. 
If we straightforwardly use the cut-set idea in~\cite{dvbt2fundamental} to lower bound $\Rsf^\text{(S2)}$, we obtain $\Rsf^\text{(S2)}\geq    \Lsf^\text{(S2)}-\frac{\Msf^\text{(S2)}}{\left\lfloor\Nsf^\text{(S2)}/\Lsf^\text{(S2)}  \right\rfloor }$ as in~\cite[Theorem~2]{dvbt2fundamental}. In this paper, we use    Han's inequlaity~\cite[Theorem 17.6.1]{hanineq} to get the bound $\Rsf^\text{(S2)}\geq \Lsf^\text{(S2)}\left(1-\frac{\Msf^\text{(S2)}}{\Nsf^\text{(S2)}} \right)$. Notice that in our bound the ``floor operator'' from the cut-set bound has been removed. In addition, as it will be shown in Theorem~\ref{thm:optimality improvement}, the converse bound in~\eqref{eq:converse N>K0} can provide exact optimality results for some memory regime, while the cut-set converse bound cannot.


 Han's inequality was used in~\cite{Sengupta2017multirequest} in order to strengthen the cut-set bound for the single bottleneck-link caching problem with multiple requests, when each user demands the same number of files (see the derivation of~\cite[Equation~(45)]{Sengupta2017multirequest} and the conditional entropies in~\cite[Equation~(45)]{Sengupta2017multirequest} should be conditioned on the same quantity, which imposes that the number of files demanded by each user is the same). Hence, the proof in~\cite{Sengupta2017multirequest} cannot be used in our problem when the occupancy number of each SBS is not identical, because of the inherent asymmetry of our problem. Therefore, the use of Han's inequality in this paper is indeed novel.

\paragraph*{Bound~\eqref{eq:converse N>K=K0+K1}}
The bound in~\eqref{eq:converse N>K=K0+K1} (which holds for $\Nsf > \Kbc+\Kdd$) can intuitively be interpreted as the one in~\eqref{eq:converse N>K0}. In~\eqref{eq:converse N>K=K0+K1}, we consider all subsets $\Sc$ with cardinality $s$, for some $s\in[\Hsf]$. For each  $\Sc$,  we also consider a genie aided system where the SBSs in $\Sc$ are aggregated into a single cache that has to satisfy the demands of the users in $\cup_{h\in\Sc} \Uc_h$, while receiving packets from the MBS and each SBS $h^{\prime} \in [\Hsf]\setminus \Sc$. Hence, we can lower bound  $\Rbc(\dv)+\sum_{h\not\in\Sc} \Rsf_{h}(\dv)$ by the desired files in this genie aided system. In addition, by using
 $\binom{\Hsf-1}{s} \Rdd \geq \sum_{\Sc\subseteq[\Hsf] : |\Sc|=s}\sum_{h\notin \Sc} \Rsf_{h}(\dv)$, we can derive~\eqref{eq:converse N>K=K0+K1}.

\paragraph*{Comparison between~\eqref{eq:converse N>K0} and~\eqref{eq:converse N>K=K0+K1}}
In both cases, we consider a genie aided system where the SBSs in $\Sc$ are aggregated into a single cache.
In~\eqref{eq:converse N>K0}, we use a fixed $\Sc = [s]$, for some $s\in[\Hsf]$. The advantage is that we only consider the SBSs connected to largest numbers of users. However, the limitation is that we simply use $\Rdd \geq \sum_{h\in[\Hsf]} \Rsf_{h}(\dv) \geq \sum_{h\not\in[s]} \Rsf_{h}(\dv)$ when $s\in[\Hsf-1]$, and $\sum_{h\not\in[s]} \Rsf_{h}(\dv)\geq 0$ when $s=\Hsf$.
In~\eqref{eq:converse N>K=K0+K1}, we consider all subsets $\Sc$ with cardinality $s$, for some $s\in[\Hsf]$, and use $\binom{\Hsf-1}{s} \Rdd \geq \sum_{\Sc\subseteq[\Hsf] : |\Sc|=s}\sum_{h\notin \Sc} \Rsf_{h}(\dv)$.  However, the limitation is that we need to consider all SBSs. 
By combining these ideas we obtain Theorem~\ref{thm:converse}.

\paragraph*{Bound~\eqref{eq:converse N>K0 s=H}}
If $\Kdd\leq \Nsf-\Kbc$, the bound in~\eqref{eq:converse N>K0 s=H} trivially says that  the worst-case overall load cannot be smaller than that of a genie-aided system in which 
there is  one user equipped with the    cached content  of all SBSs, demanding  the whole library  and receiving packets from the MBS.
If $\Kdd> \Nsf-\Kbc$,~\eqref{eq:converse N>K0 s=H} can be directly obtained by   letting $s=\Hsf$ in~\eqref{eq:two parts}. 

\bigskip
\paragraph*{Strengthening the Cut-set Bounds in Other Problems}
Our strategy, which improves on the cut-set converse bound by using Han's inequality, 
can be directly used to remove the ``floor operator'' in state-of-the-art cut-set converse bounds for other cache-aided networks, such as for example the converse bound for  single bottleneck-link caching system in~\cite{dvbt2fundamental}, the converse bounds for single bottleneck-link caching system with multi-requests  in~\cite{multiJi2014,Sengupta2017multirequest}, the converse bound for single bottleneck-link caching system with heterogeneous cache sizes in~\cite{wang2015hetero}, etc.
An advantage of our bound is that one can gain a factor of up to $2$ in gap results by the removal of the ``floor operator'' in the cut-set converse bounds.

\bigskip
\paragraph*{Other Uses of Our Approach on Bounding D2D Load}
The strategy of considering a ``cut'' of  caches $\Sc$ and bounding the load to these caches as $\sum_{h\notin \Sc} \Rsf_{h}(\dv)$ (instead of using $\Rdd^{\star}(\dv)$, which is equal to $\sum_{h\in[\Hsf]} \Rsf_{h}(\dv)$), was originally proposed in~\cite{Sengupta2017multirequest} for the D2D caching problem with multi-requests. In~\cite{Sengupta2017multirequest}, the authors only consider the case where the load from each cache is the same  (by the symmetry of their setting) such that one can readily obtain $\sum_{h\notin \Sc} \Rsf_{h}(\dv)\leq \frac{\Hsf- |\Sc| }{\Hsf} \Rdd^{\star},$  for any $\Sc\subseteq[\Hsf]$. The novelty in our work is to show a technique to deal with any asymmetric D2D caching system. 

\begin{figure}
\centerline{\includegraphics[scale=0.6]{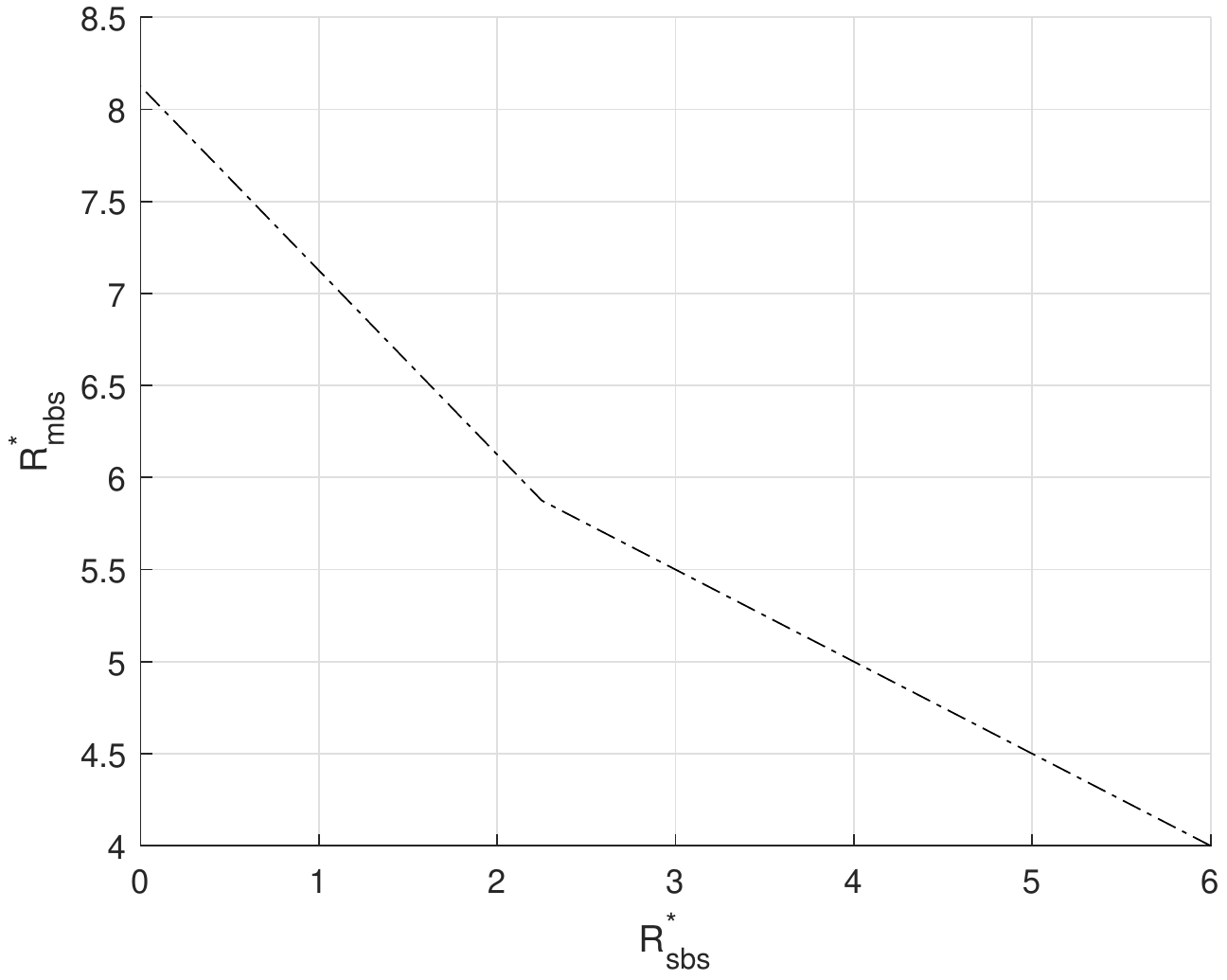}}
\caption{\small  The converse bound region $(\Rdd^{\star},\Rbc^{\star})$ for a topology-aware Fog-RAN system with $\Msf=5$, $\Hsf=4$, $\Ksf=\Nsf=20$, $\Kbc=4$, $\Lsf_1=6$, $\Lsf_2=4$, $\Lsf_3=3$, and $\Lsf_4=3$.}
\label{fig:d2d_bc}
\end{figure}

\bigskip
\paragraph*{On the Corner Points of the Converse Bound}
In Fig.~\ref{fig:d2d_bc}, we plot the outer bound region in Theorem~\ref{thm:converse} for $\Msf=5$, $\Hsf=4$, $\Ksf=\Nsf=20$, $\Kbc=4$, and $\Kdd=16$ (with $\Lsf_1=6$, $\Lsf_2=4$, $\Lsf_3=3$, and $\Lsf_4=3$). The $(\Rdd^{\star},\Rbc^{\star})$ region has three corner points in this case, namely, $(\Rdd^{\star},\Rbc^{\star})\in\{(0,8.125),(2.25,5.875),(6,4)\}$. We identify two class of corner points: 
in the first class, the load from the MBS is lowest possible, here $\Rbc^{\star}=\min\{\Nsf,\Kbc\}=4$, while 
in the second class, the load from the SBSs is lowest possible, here $\Rdd^{\star}=0$.
In the following sections, we shall design achievable schemes to approach those two classes of corner points, which shall suffice to characterize the system performance to within a constant gap. It is part of current work to try to ``match'' the other corner points of the converse region so as to further reduce the gap.


\bigskip
\paragraph*{Some Special Cases}
We now consider some special cases.

{\it Case $\Nsf\leq \Kbc$.}
The proposed converse bound when $\Nsf\leq \Kbc$ gives $\Rbc^{\star}\geq  \Nsf$ and  $\Rdd^{\star}\geq 0$ from~\eqref{eq:trivial converse}. For the worst-case demand, where the users in $\Uc_0$ demand the whole library of $\Nsf$ files, the optimal scheme is to let the MBS transmit all $\Nsf$ files thus achieving $(\Rbc,\Rdd)=(\Nsf,0)$.  In this case there is no need for the SBS sidelink communication (and caches are useless), as stated by the next theorem.
\begin{thm}[Exact Optimality for $\Nsf\leq \Kbc$]
\label{thm:exact optimality N<K0}
For the $(\Kbc,\Kdd,\Hsf,\Nsf,\Msf)$ topology-aware Fog-RAN system, when $\Nsf\leq \Kbc$ the optimal region has one single corner point given by $(\Rbc^{\star},\Rdd^{\star})=(\Nsf,0)$.
\end{thm}

{\it Case $\Hsf=1$ and $\Nsf > \Kbc$.}
In this case there is only one SBS and our Fog-RAN reduces a single bottleneck-link model with two caches of heterogenous sizes, each serving multiple requests.  In Theorem~\ref{thm:exact optimality H=1} we shall show an optimal scheme for this case.

{\it Case $\Msf\geq \Nsf-\Kbc > 0$.} 
In this case, our converse 
region has a single corner point $(\Rbc^{\star}, \Rdd^{\star})=(\Kbc, 0)$.
In Theorem~\ref{thm:optimality N>K0 M>N-K0} we shall show the achievability of this region.
In plain words, if the MBS only broadcasts the files requested by its $\Kbc$ connected users and the cache size of each SBS is such that it can store the ``remaining'' $\Nsf - \Kbc$ files, then there is no need of the SBS sidelink communication (by letting the MBS serve all requests).\footnote{\label{foot:nontrivial} This result is not so trivial as it may appear, since the identity of the $\Kbc$ broadcasted files is unknown in prior at the time of cache placement. Therefore, some coding is needed in order to achieve such optimal performance as if the requests of the $\Kbc$ cache-less users were known in advance.}
 
\section{Topology-aware Fog-RAN Systems: Novel Achievable Scheme with Symmetric File Subpacketization}
\label{sec:main symmetric}

This section considers {\it topology-aware} Fog-RAN systems,
and is organized as follows.
In Section~\ref{sub:main result symm.placement},
we start by introducing an achievable region with {\it symmetric file subpacketization}, 
and (order) optimality results by comparing the resulting achievable region with the converse bound in Section~\ref{sec:main converse}.
The details of the proofs are given in subsequent subsections.
We conclude in Section~\ref{sub:numerical for baseline symm.placement} with some numerical evaluations of the proposed regions.

\subsection{Main Results and Discussion}
\label{sub:main result symm.placement}
As shown in Theorem~\ref{thm:exact optimality N<K0}, it is quite straightforward to achieve the optimal point $(\Rbc^{\star},\Rdd^{\star})=(\Nsf,0)$ when $\Nsf\leq \Kbc$.
 Hence, in the following we only consider the case $\Nsf>\Kbc$.

As anticipated before, we consider achievable schemes targeted to work at the two classes of corner points by letting   $\Rdd=0$ and $\Rbc=\min\{\Nsf,\Kbc\}$, respectively. 
  In passing, we notice that the former minimizes the sum load $\Rdd+\Rbc$ since the caching gain of the single shared broadcast link is larger than that of the SBS sidelink. The following achievable region uses the symmetric file subpacketization as in the MAN scheme.   The proof can be found in Sections~\ref{sub:interfile} to~\ref{sub:delivery D2D point case 2}.

\begin{thm}[Achievable region with symmetric file subpacketization]
\label{thm:symm ach theorem}
 For the $(\Kbc,\Kdd,\Hsf,\Nsf,\Msf)$ topology-aware Fog-RAN system with $\Nsf>\Kbc$, the lower convex envelope of the following points is achievable
\begin{align}
(\Msf, \Rbc,\Rdd) 
  &= \left(\frac{t(\Nsf-\Kbc)}{\Hsf},  \ \Kbc+\min\left\{\frac{\min\{\Nsf-\Kbc,\Kdd\}(\Hsf-t)}{\Hsf},\frac{\sum_{r\in[\Hsf-t]}\Lsf^{\prime}_r \binom{\Hsf-r}{t}}{\binom{\Hsf}{t}} \right\}, \ 0 \right),
\notag\\&\quad   \  t\in[0:\Hsf],
\label{eq:achievable points 1}
\\
(\Msf, \Rbc, \Rdd) 
  &=\left(\frac{t(\Nsf-\Kbc)}{\Hsf}, \ \Kbc, \ \min\left\{\frac{\min\{\Nsf-\Kbc,\Kdd\}(\Hsf-t)}{\Hsf-1}, \right.
\right.\nonumber\\&\left.\left.
  \ \frac{1}{\binom{\Hsf}{t}}\sum_{\Sc\subseteq [\Hsf]:|\Sc|=t+1} \Big(\Lsf^{\prime}_{v_1(\Sc)} +\frac{1}{t} \left[\Lsf^{\prime}_{v_{t+1}(\Sc)} - \Lsf^{\prime}_{v_{1}(\Sc)} + \Lsf^{\prime}_{v_{t}(\Sc)}\right]^+\Big) \right\}   \right), 
   \  t\in[\Hsf],
\label{eq:achievable points 2}
\end{align}
where for a set $\Sc $ the function $v(\cdot)$ sorts the elements in $\Sc$ in   increasing order, i.e., 
$v_1(\Sc)\leq v_2(\Sc) \leq \ldots \leq v_{|\Sc|}(\Sc)$ and $\Lsf^{\prime}_{h}:=\min\{\Lsf_{h},\Nsf-\Kbc\}$ for each $h\in[\Hsf]$. 
\end{thm}

\bigskip
By comparing the converse bound in Theorem~\ref{thm:converse} with the achievable region in Theorem~\ref{thm:symm ach theorem} we have the following (exact or approximate) optimality results.

When there is only one SBS, our Fog-RAN reduces a single bottleneck-link model (as there is no SBS sidelink communication, i.e., $\Rdd^{\star}=0$) with two caches each serving multiple requests. In particular, 
one cache of size $\Msf_0=0$ serves $\Lsf_0 = \Kbc$ requests, and
another cache of size $\Msf_t\in[0,\Nsf]$ served $\Lsf_1 = \Kdd$ requests.
For this model we have exact optimality as stated next.
\begin{thm}[Exact Optimality for $\Hsf=1$ and $\Nsf > \Kbc$]
\label{thm:exact optimality H=1}
For the $(\Kbc,\Kdd,\Hsf=1,\Nsf,\Msf)$ topology-aware Fog-RAN system with a single SBS, 
for $\Nsf > \Kbc$ the optimal load is $\Rbc^{\star}\geq \Kbc+ \min\{\Kdd,\Nsf-\Kbc\} \left[1-\frac{\Msf}{\Nsf-\Kbc} \right]^+$ achieved by Theorem~\ref{thm:symm ach theorem}.
\end{thm}
\begin{IEEEproof}
The proof is rather straightforward. The converse follows from~\eqref{eq:converse N>K0 s=H} (here $\Hsf=1$), 
while the achievability follows by memory-sharing between the points in~\eqref{eq:achievable points 1}  for $t=0$ and $t=1$. 
The result in Theorem~\ref{thm:exact optimality H=1} is a generalization of the line of work on single bottleneck-link systems with two caches of different size, each serving one user~\cite{cao2018twousers,ibrahim2018coded}. The generalization is to the case where each cache receives multiple requests, and one cache has size zero. The general problem of a single bottleneck-link system with heterogenous cache sizes and with multiple requests is open beyond the case $\Lsf_0=\Lsf_1=1$~\cite{cao2018twousers,ibrahim2018coded}.
 \end{IEEEproof}

\bigskip
The next exact optimality result says that no SBS sidelink communication is needed if the cache size at each SBS is enough to store $\Nsf-\Kbc$ files.
\begin{thm}[Optimality for $\Nsf> \Kbc$ and  $\Msf\geq \Nsf-\Kbc$]
\label{thm:optimality N>K0 M>N-K0}
For the $(\Kbc,\Kdd,\Hsf,\Nsf,\Msf)$ topology-aware Fog-RAN system with $\Nsf> \Kbc$ and $\Msf\geq \Nsf-\Kbc$, the optimal region has a single corner point given by $(\Rbc^{\star},\Rdd^{\star})=(\Kbc,0)$, achieved by Theorem~\ref{thm:symm ach theorem}. 
\end{thm}
\begin{IEEEproof}
The proof is rather straightforward. 

Converse. under the conditions of the theorem, the converse region has a single corner point, namely $(\Rbc^{\star},\Rdd^{\star})=(\Kbc,0)$ (because we have $\Rbc^{\star}\geq \min\{\Nsf,\Kbc\}=\Kbc,  \Rdd^{\star}\geq 0$ from~\eqref{eq:trivial converse}, and $\Rbc^{\star}+\Rdd^{\star}\geq \Kbc$ from~\eqref{eq:converse N>K0}). 

Achievability. use $t=1$ in~\eqref{eq:achievable points 1}. 
Intuitively, in this regime, the proposed inter-file coded placement in Theorem~\ref{sub:interfile} is such that each SBS can decode the whole library after receiving the $\Kbc$ files broadcasted by the MBS, and can thus satisfy any requests from its connected users (and in this regime the optimal region does not depend on how many users are connected to each SBS, or on how many SBSs are present, as long as they have enough cache size).
\end{IEEEproof}

\bigskip
At this point we have exact optimality, except for
\begin{align}
\Hsf\geq 2, \  \Nsf> \Kbc, \  \Msf< \Nsf-\Kbc, 
\label{eq:set of params only gap results}
\end{align}
for which we shall develop multiplicative gap results next.

By comparing Theorems~\ref{thm:converse} and~\ref{thm:symm ach theorem}, we have the following order optimality results for $\Nsf\leq \Kbc+\Lsf_1$ (where $\Lsf_1$ is the occupancy number of  the most loaded SBS). The proof can be found in Appendix~\ref{sub:opt proof N<K0+L1}. 
\begin{thm}[Order Optimality for $\Kbc<\Nsf\leq \Kbc+\Lsf_1$ within the parameter set in~\eqref{eq:set of params only gap results}]
\label{thm:order optimality N<K0+L1 M>N-K0}
For the $(\Kbc,\Kdd,\Hsf,\Nsf,\Msf)$ topology-aware Fog-RAN system with $\Kbc<\Nsf\leq \Kbc+\Lsf_1$  and  within the parameter set in~\eqref{eq:set of params only gap results}, 
the caching scheme in Theorem~\ref{thm:symm ach theorem} is order optimal to within a factor no larger than $2$. More precisely, for any $(\Msf,\Rbc^{\star},\Rdd^{\star})$ in the converse region of Theorem~\ref{thm:converse}, the caching scheme in Theorem~\ref{thm:symm ach theorem} achieves $(\Msf, \Rbc^{\star},  \frac{\Hsf}{\Hsf-1}  \Rdd^{\star})$, where $\frac{\Hsf}{\Hsf-1}\leq 2$ for $\Hsf \geq 2$ .
\end{thm}
We note that the gap in Theorem~\ref{thm:order optimality N<K0+L1 M>N-K0} is the largest for $\Hsf=2$ SBSs, and approaches one when the number of SBSs is large.

\bigskip
Finally, for $\Nsf> \Kbc+\Lsf_1$  (where $\Lsf_1$ is the occupancy number of   the most loaded SBS) we have the following order optimality results. The proofs can be found in Appendices~\ref{sub:order opt proof2} and~\ref{sub:order opt proof}, respectively.
\begin{thm}[Order Optimality for $\Nsf > \Kbc+\Lsf_1$ within the parameter set in~\eqref{eq:set of params only gap results}]
\label{thm:order optimality2}
For the $(\Kbc,\Kdd,\Hsf,\Nsf,\Msf)$ topology-aware Fog-RAN system with $\Nsf> \Kbc+\Lsf_1$  and within the parameter set in~\eqref{eq:set of params only gap results},
the caching scheme in Theorem~\ref{thm:symm ach theorem} is order optimal to within a factor of   $2g$, where $g:=\min\left\{\Hsf,\frac{\Nsf-\Kbc}{\Msf}\right\}$.
More precisely, for any $(\Msf,\Rbc^{\star},\Rdd^{\star})$ in the converse region of Theorem~\ref{thm:converse}, the caching scheme in Theorem~\ref{thm:symm ach theorem} can achieve $(\Msf, 2g \ \Rbc^{\star}, 2g \ \Rdd^{\star})$.
\end{thm}

 From Theorem~\ref{thm:order optimality2}, we see that the worst factor approximation is for the ``very low'' cache size $\Msf< \frac{\Nsf-\Kbc}{\Hsf}$ where that gap is bounded by $\Hsf$. The gap is decreasing in $\Msf$ for $\frac{\Nsf-\Kbc}{\Hsf} < \Hsf < \Nsf-\Kbc$; for $\Msf \geq \Nsf-\Kbc$ we have exact optimality result from Theorem~\ref{thm:optimality N>K0 M>N-K0}. We try to partially remedy this, by deriving a constant gap result for the case where
the occupancy number of  each SBS is identical. 
\begin{thm}[Order Optimality for $\Nsf> \Kbc+\Lsf_1$ and symmetric SBS  occupancy numbers) 
within the parameter set in~\eqref{eq:set of params only gap results}]
\label{thm:order optimality}
For the $(\Kbc,\Kdd,\Hsf,\Nsf,\Msf)$ topology-aware Fog-RAN system within the parameter set in~\eqref{eq:set of params only gap results}, 
with in addition  $\Lsf_h = \Lsf$ for all $h$ and $\Nsf> \Kbc+\Lsf$,
the caching scheme in Theorem~\ref{thm:symm ach theorem} is order optimal to within a factor of $22$. More precisely, for any $(\Msf,\Rbc^{\star},\Rdd^{\star})$ in the converse region of Theorem~\ref{thm:converse}, the caching scheme in Theorem~\ref{thm:symm ach theorem} can achieve $(\Msf,22\Rbc^{\star},22\Rdd^{\star})$.
\end{thm}


\subsection{Inter-file Coded Cache Placement and Prototype Delivery}
\label{sub:interfile}

The proposed achievable scheme achieving the region in Theorem~\ref{thm:symm ach theorem} for $\Nsf >\Kbc$ works as follows. 

\paragraph*{Inter-file Coded Cache Placement} 
Let $t\in [0:\Hsf]$. 
Partition each file $F_i, i\in[\Nsf],$ into $\binom{\Hsf}{t}$ equal-length subfiles of size  $\Bsf/\binom{\Hsf}{t}$ bits. Denote the $i$-th file as $F_{i}=\{F_{i,\Wc}:\Wc\subseteq [\Hsf],|\Wc|=t\}$. Define  
\begin{align}
\Fc_{\Wc}:=\{F_{i,\Wc}:i\in[\Nsf]\},\  \forall \Wc\subseteq [\Hsf] : |\Wc|=t,
\label{eq:Fc_Wc def}
\end{align}
as the set that contains the $\Bsf \Nsf/\binom{\Hsf}{t}$ bits, which will be placed coded in the caches of the SBSs indexed by $\Wc$.
Each SBS $h\in[\Hsf]$ caches $|\Fc_{\Wc}|(1-\Kbc/\Nsf)$ random linear combinations
\footnote{\label{footnote:RLC}Instead of random linear combinations, the proposed inter-file coded cache placement could also be achieved by using the parity check matrix of MDS code as in~\cite{cao2018twousers,ibrahim2018coded} or Cauchy matrix as in~\cite{ourisitinnerbound}. These matrices, with dimension $m_1\times m_2$ where $m_1\leq m_2$, have the properties that every $m_1$ columns are linearly independent. In this paper, for the sake of simplicity, we use random linear combinations.
Using random linear combinations comes with a caveat. We can partition each subfile into equal-length symbols over some finite field. Since $\Bsf$ is as large as desired, the above random linear combinations are linearly independent with high probability if operations are on a finite field of sufficiently large size. 
} 
of all bits in $\Fc_{\Wc}$ in~\eqref{eq:Fc_Wc def} if $h\in \Wc$, thus needing a cache of size 
\begin{align}
\Msf = \binom{\Hsf-1}{t-1}(\Nsf-\Kbc)/\binom{\Hsf}{t}=t\frac{\Nsf-\Kbc}{\Hsf}. 
\end{align}

\paragraph*{Delivery Phase}
We first satisfy the demands of the users in $\Uc_0$ while letting each SBS $h\in[\Hsf]$ recover all bits in $\Fc_{\Wc}$ in~\eqref{eq:Fc_Wc def} if $h\in \Wc$ (which requires receiving $|\Fc_{\Wc}| \Kbc/\Nsf$ linear combinations of all the bits in $\Fc_{\Wc}$). If some users are not yet satisfied after this, we then let the MBS broadcast some random linear combinations of all bits of the files desired by these users. More precisely, let 
\begin{align}
\Dc_{\rm{mbs}}(\dv) := \cup_{k\in\Uc_0}  \{d_k\}
\end{align}
be the files demanded by the users directly served by the MBS, and 
\begin{align}
\Dc_{\rm{sbs}}(\dv) := \big(\cup_{h\in[\Hsf], k\in\Uc_h} \{d_k\}\big)  \backslash \Dc_{\rm{mbs}}(\dv)
\end{align}
be the files demanded exclusively by the users served by some SBS. We distinguish two cases. 
\begin{itemize}

\item
{\it Case~1}: If $|\Dc_{\rm{sbs}}(\dv)| + |\Dc_{\rm{mbs}}(\dv)| \leq \Kbc$. 
In this case the delivery has a single step.

We let the MBS broadcast all the files in $\Dc_{\rm{mbs}}(\dv)$ (so that each user in $\Uc_0$ can recover its desired file), and all the files in $\Dc_{\rm{sbs}}(\dv)$ (so that all the remaining users can receive their demanded file by the SBS to which they connect). The delivery is finished. The loads are $\Rbc(\dv)=\Kbc=\min\{\Nsf,\Kbc\}$ and $\Rdd(\dv)=0$.

\item
{\it Case~2}: if $|\Dc_{\rm{sbs}}(\dv)| + |\Dc_{\rm{mbs}}(\dv)| > \Kbc$.
In this case the delivery has two steps.

Case~2.Step~1:
we let the MBS broadcast all the files in $\Dc_{\rm{mbs}}(\dv)$, and $\Kbc - |\Dc_{\rm{mbs}}(\dv)|$ files in $\Dc_{\rm{sbs}}(\dv)$. This enables each user in $\Uc_0$ to recover its desired file,  each SBS  $h\in[\Hsf]$ to recover all the bits in $\Fc_{\Wc}$ in~\eqref{eq:Fc_Wc def} if $h\in \Wc$, and some users connected to an SBS to recover its desired file.  The delivery is however not finished. 
Define $\Dc^{\prime}_{\rm{sbs}}(\dv)$ as the set of demanded files in $\Dc_{\rm{sbs}}(\dv)$, which have not been transmitted in Case~2.Step~1. 

Case~2.Step~2:
There are various ways in which the files in $\Dc^{\prime}_{\rm{sbs}}(\dv)$ can be delivered. We shall propose in the following subsections some delivery methods that collectively prove Theorem~\ref{thm:symm ach theorem}.
\end{itemize}

\subsection{Case~2.Step~2 for~\eqref{eq:achievable points 1}:   No SBS Sidelink Communication  Takes Place}
\label{sub:delivery shared point case 2}
We introduce two approaches, where in the first one the MBS transmits some random linear combinations of the files in $\Dc^{\prime}_{\rm{sbs}}(\dv)$, and in the second one we use the single bottleneck-link caching scheme with shared caches in~\cite{parrinello2018sharedcache} to deliver the files in $\Dc^{\prime}_{\rm{sbs}}(\dv)$.

\paragraph{Case~2.Step~2.Approach~1 for~\eqref{eq:achievable points 1}}
Since each SBS $h\in[\Hsf]$ has recovered $\Fc_{\Wc}$ if $h\in \Wc$ where $|\Wc|=t$, SBS $h$ has recovered $\binom{\Hsf-1}{t-1}\Bsf/ \binom{\Hsf}{t}=t \Bsf/\Hsf$ bits of each file in $\Dc^{\prime}_{\rm{sbs}}(\dv)$. We let the MBS transmit $\binom{\Hsf-1}{t}\Bsf/ \binom{\Hsf}{t}=\Bsf(1-t/\Hsf)$ random linear combinations of each file in  $\Dc^{\prime}_{\rm{sbs}}(\dv)$, such that each SBS can recover each file in  $\Dc^{\prime}_{\rm{sbs}}(\dv)$ and forward it to the demanding users.

The worst case is when  $|\Dc_{\rm{mbs}}(\dv)| = \Kbc$ and $|\Dc_{\rm{sbs}}(\dv)| = \min\{\Nsf-\Kbc,\Kdd\}$, and the loads are 
\begin{align}
& \Rbc(\dv)=\Kbc+ \frac{ \min\{\Nsf-\Kbc,\Kdd\}(\Hsf-t)}{\Hsf}, \  \Rdd(\dv)=0, \ \forall t\in[0:\Hsf].\label{eq:case 2 shared 1}
\end{align}

It will be shown in Appendix~\ref{sub:opt proof N<K0+L1} that, for $\Nsf-\Kbc\leq \Lsf_1$, Approach~1 is optimal.
 

\paragraph{Case~2.Step~2.Approach~2 for~\eqref{eq:achievable points 1}}
For each SBS $h\in[\Hsf]$, if several users connected to SBS $h$ have the same demand, we  consider them as one user. 
We let $\Uc^{\prime}_{h}$  
be the set of users connected to SBS $h\in[\Hsf]$ whose demanded files are distinct and in  $\Dc^{\prime}_{\rm{sbs}}(\dv)$. The delivery contains $\max_{h\in[\Hsf]} |\Uc^{\prime}_{h}| $ rounds. In each round $j$, we pick one unpicked user connected to each SBS $h\in[\Hsf]$ if still any; denote the set of picked users as $\Kc_j$, and the user in $\Kc_j$ connected to SBS $h\in[\Hsf]$ as $k_j(h)$. We let the MBS broadcast the MAN-type multicast messages 
\begin{align}
\underset{h\in\Sc}{\oplus} F_{d_{k_j(h)},\Sc \setminus \{h\}}, \ \forall \Sc \subseteq [\Hsf] \textrm{ where }|\Sc|=t+1,
\label{eq:MAN multicast}
\end{align}
so as each SBS $h$ can recovers $F_{d_{k_j(h)}}$ and then forwards it to user $k_j(h)$.

Recall that $\Lsf^{\prime}_h:=\min\{\Lsf_h,\Nsf-\Kbc\}$ for each $h\in [\Hsf]$.
The worst case is when  $|\Dc_{\rm{mbs}}(\dv)| = \Kbc$ and 
 $|\Uc^{\prime}_{h}|=\Lsf^{\prime}_h$ for each $h\in [\Hsf]$, and the loads are 
\begin{align}
\Rbc(\dv)=\Kbc+\frac{\sum_{r\in[\Hsf-t]}\Lsf^{\prime}_r \binom{\Hsf-r}{t}}{\binom{\Hsf}{t}}, \  \Rdd(\dv)=0, \ \forall t\in[0:\Hsf] \label{eq:case 2 shared 2}
\end{align}
as in~\eqref{eq:achievable points 1}, where the computation of $\Rbc(\dv)$ in~\eqref{eq:case 2 shared 2} is the same as in~\cite[Equation (29)]{parrinello2018sharedcache}.

By combining the achieved loads in~\eqref{eq:case 2 shared 1} and~\eqref{eq:case 2 shared 2}, we prove~\eqref{eq:achievable points 1}.


\begin{rem}[Improvement for $\Lsf_1<\Nsf-\Kbc<\Kdd$]
\label{rem:improvement N<K}
When $\Kbc+\Lsf_1<\Nsf<\Kbc+\Kdd=\Ksf$, we can improve the scheme leading to~\eqref{eq:case 2 shared 2} by further leveraging intrinsic multicast opportunities. In each round, instead of using the MAN multicast messages, we can use the caching scheme~\cite{exactrateuncoded}, similarly to the delivery schemes in~\cite{multireqWei2017,multishared2018Karat,yang2018centralizedcorrected}. More precisely,  we choose one leader user among the picked users for each demanded files in this round, and each multicast message in~\eqref{eq:MAN multicast} is transmitted if in there is at least one  SBS  in $\Sc$ connected to some leader.

The achieved load depends on how we divide the users for the various rounds. There are many ways to partition users into groups, such as for example the greedy search method based on the best division as proposed in~\cite{multireqWei2017} (which is extremely complex), or the random division as proposed in~\cite{multishared2018Karat,yang2018centralizedcorrected}. It is generally very hard to obtain a closed form expression for the achieved loads in these cases. Thus, in this paper, we stick to the MAN multicast messages scheme for each round, possibly forging some saving in number of transmissions, but instead obtaining a closed form expression for the loads that we shall use next for our optimality results.
\end{rem}

\subsection{Case~2.Step~2 for~\eqref{eq:achievable points 2}:   the SBS Sidelink Communication  Takes Place}
\label{sub:delivery D2D point case 2}
We introduce two approaches, where in the first one the SBSs transmit some random linear combinations of the files in $\Dc^{\prime}_{\rm{sbs}}(\dv)$, and in the second one we propose a novel D2D delivery scheme to deliver the files in $\Dc^{\prime}_{\rm{sbs}}(\dv)$ through the SBS sidelink.
The latter can be applied to the D2D caching scenario with shared caches.

\paragraph{Case~2.Step~2.Approach~1 for~\eqref{eq:achievable points 2}}
For each subfile $F_{i,\Wc}$ where $i\in \Dc^{\prime}_{\rm{sbs}}(\dv)$, $\Wc\subseteq [\Hsf]$, and $|\Wc|=t$, we partition $F_{i,\Wc}$ into $t$ equal-length sub-pieces, $F_{i,\Wc}=\{F_{i,\Wc,h_1}:h_1\in \Wc\}$, where each sub-piece has $\frac{\Bsf}{\binom{\Hsf}{t}t}$ bits.
For each file $F_i$ where $i\in \Dc^{\prime}_{\rm{sbs}}(\dv)$,
we let each SBS  $h\in[\Hsf]$ transmit $\frac{\binom{\Hsf-2}{t-1}}{\binom{\Hsf}{t}t}\Bsf$  random linear combinations of all   $\frac{\binom{\Hsf-1}{t-1}}{\binom{\Hsf}{t}t}\Bsf$ bits in $\{F_{i,\Wc,h}:\Wc\subseteq [\Hsf], |\Wc|=t,h\in \Wc\}$.

Each SBS $h_2\in [\Hsf]$ knows $\frac{\binom{\Hsf-2}{t-2}}{\binom{\Hsf}{t}t}\Bsf$ bits in $\{F_{i,\Wc,h}:\Wc\subseteq [\Hsf], |\Wc|=t, \{h,h_2\}\subseteq \Wc\}$ for any $h\in [\Hsf]\setminus\{h_2\}$. In addition, SBS $h_2$ receives  $\frac{\binom{\Hsf-2}{t-1}}{\binom{\Hsf}{t}t}\Bsf$  random linear combinations of all   $\frac{\binom{\Hsf-1}{t-1}}{\binom{\Hsf}{t}t}\Bsf$ bits in $\{F_{i,\Wc,h}:\Wc\subseteq [\Hsf], |\Wc|=t,h\in \Wc\}$. Hence, SBS $h_2$ can recover $\{F_{i,\Wc,h}:\Wc\subseteq [\Hsf], |\Wc|=t,h\in \Wc\}$. 
Considering all $h\in [\Hsf]\setminus\{h_2\}$, it can be seen that SBS $h_2$ can recover $F_i$ and the forward it to the demanding user.

The worst case is when  $|\Dc_{\rm{mbs}}(\dv)| = \Kbc$ and $|\Dc_{\rm{sbs}}(\dv)| = \min\{\Nsf-\Kbc,\Kdd\}$, and the loads are 
\begin{align}
 \Rbc(\dv)=\Kbc, \ \Rdd(\dv)=\min\{\Nsf-\Kbc,\Kdd\} \Hsf\frac{\binom{\Hsf-2}{t-1}}{\binom{\Hsf}{t}t}, \ \forall t\in [\Hsf].
\label{eq:case 2 d2d}
\end{align}
Note that the sum of the loads in~\eqref{eq:case 2 d2d} is a factor $\frac{\Hsf}{\Hsf-1}$ larger than in~\eqref{eq:case 2 shared 1}.

\paragraph{Case~2.Step~2.Approach~2 for~\eqref{eq:achievable points 2}}
We here propose a novel D2D delivery scheme to deliver the files in $\Dc^{\prime}_{\rm{sbs}}(\dv)$. Our key idea is illustrated by the following example.

\begin{example}
\label{ex:ex1}
\rm
We focus on the network illustrated in Fig.~\ref{fig: system_model} with $\Hsf=3$, $\Kdd=4$, $\Kbc=2$, $\Nsf=6$ and $\Msf=8/3$. Hence, we have $t=\Hsf \frac{\Msf}{\Nsf-\Kbc}=2$.

In the cache placement phase, we partition each file $F_i$ into $\binom{\Hsf}{t}=3$ equal-length subfiles, $F_{i}=\{F_{i,\{1,2\}},F_{i,\{1,3\}},F_{i,\{2,3\}}\}$ and each subfile has $\Bsf/3$ bits.  We let each SBS $h\in[\Hsf]$ cache $\Bsf(\Nsf-\Kbc)/\binom{\Hsf}{t}=4\Bsf/3$ random linear combinations of all bits in $\Fc_{\{h,h_1\}}$, where $h_1\in ([\Hsf]\setminus \{h\})$. It can be seen that each SBS caches $8\Bsf/3$ bits, satisfying the cache size constraint. 

Assume $\dv=(1,2,3,4,5,6)$.
In the first step of the delivery phase, we let the MBS broadcast $F_5$ and $F_6$ so that the users in $\Uc_0=\{5,6\}$ can recover their desired files. Simultaneously, each SBS $h\in[\Hsf]$ can recover $\Fc_{\{h,h_1\}}$, where $h_1\in ([\Hsf]\setminus \{h\})$. 
 
In the second step, if we use the second approach in Section~\ref{sub:delivery shared point case 2} (equivalent to the single bottleneck-link caching scheme with shared caches in~\cite{parrinello2018sharedcache}) we have two rounds of transmissions:
in the first round, the MBS serves the group of users $\{2,3,4\}$ by broadcasting $F_{2,\{2,3\}}\oplus F_{3,\{1,3\}}\oplus F_{4,\{1,2\}}$, and in the second round the group of users $\{1\}$ by sending $F_{1,\{2,3\}}$;
hence, the load in the second step is $2\times 1/3=4/6$.
We can extend this scheme to the SBS sidelink communication by splitting each subfile used in the first round into two non-overlapping and equal-length pieces, i.e., 
$F_{i,\{h_1,h_2\}}=\{F_{i,\{h_1,h_2\},h_1}, \ F_{i,\{h_1,h_2\},h_2}\}$.
We then let 
SBS~$1$ transmit $F_{3,\{1,3\},1} \oplus F_{4,\{1,2\},1}$, 
SBS~$2$ transmit $F_{2,\{2,3\},2} \oplus F_{4,\{1,2\},2}$, and
SBS~$3$ transmit $F_{2,\{2,3\},3} \oplus F_{3,\{1,3\},3}$; finally we 
let SBS~$2$ transmit $F_{1,\{2,3\}}$;
hence, the SBS sidelink load in the second step is $3\times 1/6+1/3=5/6$.

The question is whether one can achieve the ``single bottleneck-link'' load of $4/6$ in a ``D2D'' fashion.
The answer is  positive and the key is to change the grouping of the users. 
Consider groups $\{2,3\}$ and $\{1,4\}$.
To serve the users in $\{2,3\}$, we let SBS~$3$ transmit $F_{2,\{2,3\}}\oplus F_{3,\{1,3\}}$. 
To serve the users in $\{1,4\}$, we let SBS~$2$ transmit $F_{1,\{2,3\}}\oplus F_{4,\{1,2\}}$. 
Hence, the load in the second step achieved by the new grouping method is $1/3+1/3=4/6$.

The achievable triple with the new grouping method is $(\Msf,\Rbc,\Rdd)=(8/3,2,2/3)$.  Theorem~\ref{thm:converse} shows the optimality of this triple, e.g., there does not exist a caching scheme achieving  $(8/3,\Rsf^{\prime}_{\rm{mbs}},\Rsf^{\prime}_{\rm{sbs}})$, where  $\Rsf^{\prime}_{\rm{mbs}}< 2$ and $\Rsf^{\prime}_{\rm{sbs}}= 2/3$, or $\Rsf^{\prime}_{\rm{mbs}}= 2$ and $\Rsf^{\prime}_{\rm{sbs}}< 2/3$.
 \hfill$\square$
\end{example}

We are now ready to generalize Example~\ref{ex:ex1}. 
The main idea is to partition the users into groups  to achieve the largest possible ``multicasting gain'' when transmitting each subfile from an SBS. To avoid heavy notation, assume $|\Uc^{\prime}_{i}|\geq |\Uc^{\prime}_{j}|$ for any $1\leq i<j\leq \Hsf$   (else consider a suitable permutation of the SBS indices).

We focus on each set $\Sc\subseteq [\Hsf]$ of cardinality $|\Sc|=t+1$, and we aim to let each user $k \in \Uc^{\prime}_h$  to recover $F_{d_k,\Sc\setminus\{h\}}$, for each $h\in\Sc$. 
We partition the users connected to the SBSs in $\Sc$ into $|\Uc^{\prime}_{v_1(\Sc)}|$  groups  as in Section~\ref{sub:delivery shared point case 2}, where 
in each group we pick one unpicked user connected to each SBS $h\in\Sc$ if still any.
Recall for a set $\Sc $ the function $v(\cdot)$ sorts the elements in $\Sc$ in   increasing order, i.e., $v_1(\Sc)\leq v_2(\Sc) \leq \ldots v_{|\Sc|}(\Sc)$.

Since each SBS in $\Sc$ is connected to at least  $|\Uc^{\prime}_{v_{t+1}(\Sc)}|$  users whose demands are in $\Dc^{\prime}_{\rm{sbs}}(\dv)$,
it can be seen that each group with index in $[|\Uc^{\prime}_{v_{t+1}(\Sc)}|]$  contains $t+1$ users.
Similarly, each group in $[|\Uc^{\prime}_{v_{t+1}(\Sc)}|+1:|\Uc^{\prime}_{v_{t}(\Sc)}|]$ contains exactly $t$ users and
each group in $[|\Uc^{\prime}_{v_{t}(\Sc)}|+1:|\Uc^{\prime}_{v_1(\Sc)}| ]$ contains  strictly  less than $t$ users.
 We start with the partition of the users into groups as in Section~\ref{sub:delivery shared point case 2}; if there are groups with different sizes we proceed as follows. 
If $|\Uc^{\prime}_{v_{t+1}(\Sc)}| \leq |\Uc^{\prime}_{v_1(\Sc)}|-|\Uc^{\prime}_{v_{t}(\Sc)}|$, from each group $g \in [|\Uc^{\prime}_{v_{t+1}(\Sc)}|]$, we move the user connected to SBS $v_{t+1}(\Sc)$ to group  $|\Uc^{\prime}_{v_{t}(\Sc)}|+g$; otherwise, from each group $g \in [|\Uc^{\prime}_{v_1(\Sc)}|-|\Uc^{\prime}_{v_{t}(\Sc)}|]$, 
we move the user connected to SBS $v_{t+1}(\Sc)$ to group  $|\Uc^{\prime}_{v_{t}(\Sc)}|+g$. 
\footnote{ 
Let us see how this grouping algorithm gives what we used in Example~\ref{ex:ex1}.
In Example~\ref{ex:ex1}, we have $v_1(\Sc)=1$, $v_t(\Sc)=2$, and $v_{t+1}(\Sc)=3$. Originally, in group 
$1$ we have users $\{2,3,4\}$, and in group $2$ we have user $\{1\}$. Since $|\Uc^{\prime}_{v_{t+1}(\Sc)}|=1 \leq (|\Uc^{\prime}_{v_{1}(\Sc)}|-|\Uc^{\prime}_{v_{t}(\Sc)}|)=1$, we move user $4$ connected to SBS $3$ to group $2$, as we did in the example.
}

Recall that the SBS to which user $k$ is connected  is denoted by $\hsf_k$. 
Next, we generate coded packets for each group as follows. 
\begin{itemize}

\item {\it Case~A}:
If group $g$ of set $\Sc$ contains $t+1$ users, for each user $k$ in group $g$, we partition each subfile $F_{d_{k},\Sc\setminus \{\hsf_k\}}$ into $t$ equal-length pieces, 
$
F_{d_{k},\Sc\setminus \{\hsf_k\}}=\left\{F_{d_{k},\Sc\setminus \{\hsf_k\},h}:h \in (\Sc\setminus \{\hsf_k\})\right\}.
$
Each SBS $h\in \Sc$ transmits
\begin{align}
\underset{k \textrm{ is in group }g:\hsf_k\neq h}{\oplus} F_{d_{k},\Sc\setminus \{\hsf_k\},h}.
\end{align}
Hence, the transmitted load for   group $g$ is $\frac{t+1}{\binom{\Hsf}{t} t}$. Notice that there are $\left[|\Uc^{\prime}_{v_{t+1}(\Sc)}| - |\Uc^{\prime}_{v_1(\Sc)}| + |\Uc^{\prime}_{v_{t}(\Sc)}| \right]^+$ such groups for set $\Sc$.

\item {\it Case~B}:
If group $g$ of set $\Sc$ contains strictly less than $t+1$ users, we can choose one SBS $h\in \Sc$   not serving any users in group $g$,  and let it transmit 
\begin{align}
\underset{k \textrm{ is in group }g}{\oplus} F_{d_{k},\Sc\setminus \{\hsf_k\}}.
\end{align}
Hence, the transmitted load for   group $g$ is $\frac{1}{\binom{\Hsf}{t}}$. The total transmitted load  for set $\Sc$ is 
\begin{align}
\frac{|\Uc^{\prime}_{v_1(\Sc)}|+\left[ |\Uc^{\prime}_{v_{t+1}(\Sc)}| - |\Uc^{\prime}_{v_1(\Sc)}| + |\Uc^{\prime}_{v_{t}(\Sc)}|\right]^+/t}{\binom{\Hsf}{t}}.
\label{eq:d2d load appr 2}
\end{align}
\end{itemize}

After considering all the subsets $\Sc\subseteq[\Hsf]$ with cardinality $|\Sc|=t+1$, each SBS can decode the files which its connected users demand and then forward the demanded files to its connected users. The worst case is when  $|\Dc_{\rm{mbs}}(\dv)| = \Kbc$ and 
 $|\Uc^{\prime}_{h}|=\Lsf^{\prime}_h$ for each $h\in [\Hsf]$.
 
Combining the two approaches, we prove~\eqref{eq:achievable points 2}.


\subsection{Application to the Single Bottleneck-link Caching Problem with Heterogeneous Cache Sizes}
\label{rem:extension to heterogeneous}
We can extend the inter-file coded cache placement and the above delivery scheme to the single bottleneck-link caching problem with heterogeneous cache sizes, where there exist some users without cache. In this setting, assume the library has $\Nsf$ files and there are in total $\Ksf$ users, $\Kbc$ of which do not have cache.

For any caching scheme with uncoded cache placement which is symmetric across the $\Nsf$ files (e.g., the caching schemes in~\cite{wang2015hetero,bidokhti2018noisycaching,ibrahim2017hetero,asadi2018unequal,daniel2017hetero}), we 
partition each file $F_i$ into subfiles as $F_{i}=\{F_{i,\Wc}:\Wc\subseteq [\Ksf]\}$, where $F_{i,\Wc}$ represents the bits exclusively cached by users in $\Wc$, where by ``symmetry'' we mean that $|F_{i,\Wc}|=|F_{j,\Wc}|$ for any $i,j \in [\Nsf]$. To meet the cache size constraints we must have $\sum_{i\in[\Nsf]}\sum_{\Wc\subseteq [\Ksf]:k\in \Wc} |F_{i,\Wc}| \leq \Msf_k\Bsf$, for each user $k\in[\Ksf]$.

In this problem, there are $\Kbc$ users without cache and thus the server must transmit the whole files demanded by them. So we can use the proposed inter-file coded cache placement to leverage those transmissions. We thus propose to let user $k$ store $\Msf_k\Bsf(\Nsf-\Kbc)/\Nsf$ random linear combinations of all bits in $\{F_{i,\Wc}:i\in[\Nsf], \Wc\subseteq [\Ksf], k\in \Wc\}$. 
In the delivery phase, let $\Dc_{0}(\dv) $ be the files demanded by the $\Kbc$ users without cache, and $\Dc_{1}(\dv)$ be the files demanded exclusively by the users equipped with cache. 
If $|\Dc_{0}(\dv)| + |\Dc_{1}(\dv)| \leq \Kbc$, we let the server broadcast all the files in $\Dc_{0}(\dv)\cup\Dc_{1}(\dv)$ and the delivery is finished; otherwise,  we let the server broadcast all the files in $\Dc_{0}(\dv)$, and $\Kbc - |\Dc_{0}(\dv)|$ files in $\Dc_{1}(\dv)$. Hence, each user $k$ can recover all bits in $\{F_{i,\Wc}:i\in[\Nsf], \Wc\subseteq [\Ksf]\}$ if $k\in \Wc$. We then use the corresponding delivery phase of the original caching scheme to let the remaining users recover their desired files. 

In conclusion, with our inter-file coded cache placement, we can achieve the same worst-cast load but with a lower memory size $\Msf^{\prime}_k= \Msf_k (\Nsf-\Kbc)/\Nsf$ compared to the needed memory size $\Msf_k$ of the original caching scheme with uncoded cache placement, for any user $k\in [\Ksf]$. Instead, if an uncoded cache placement is used (e.g.,~\cite{cless2018Em,bidokhti2017cacheassignment}), the files transmitted to the $\Kbc$ users without cache cannot help the SBSs ``increase''  their cache size (i.e., decode coded contents in their caches).

Notice that compared to the MDS inter-file coded cache placement proposed in~\cite{cao2018twousers,ibrahim2018coded}, which works with their proposed delivery schemes for single bottleneck-link caching systems with two and three users, respectively, our inter-file coded cache placement can be used in any caching scheme with symmetric cache placement for any single bottleneck-link caching system where there exist some users without cache.

\subsection{Numerical Evaluations}
\label{sub:numerical for baseline symm.placement}
\begin{figure}
    \centering
    \begin{subfigure}[t]{0.5\textwidth}
        \centering
        \includegraphics[scale=0.6]{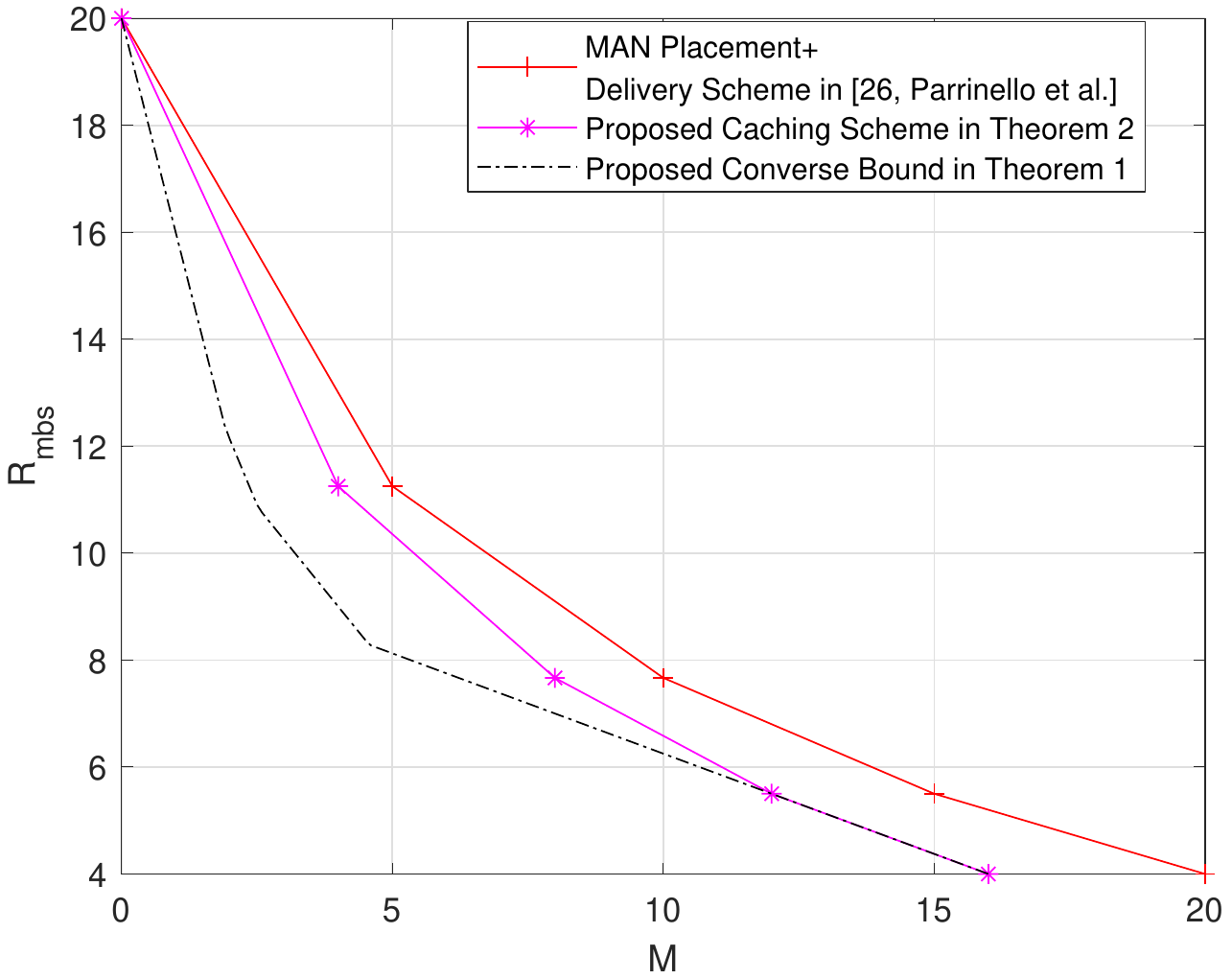}
        \caption{\small $(\Msf,\Rbc)$ tradeoff for $\Rdd=0$.}
        \label{fig:numerical 1a}
    \end{subfigure}%
    ~ 
    \begin{subfigure}[t]{0.5\textwidth}
        \centering
        \includegraphics[scale=0.6]{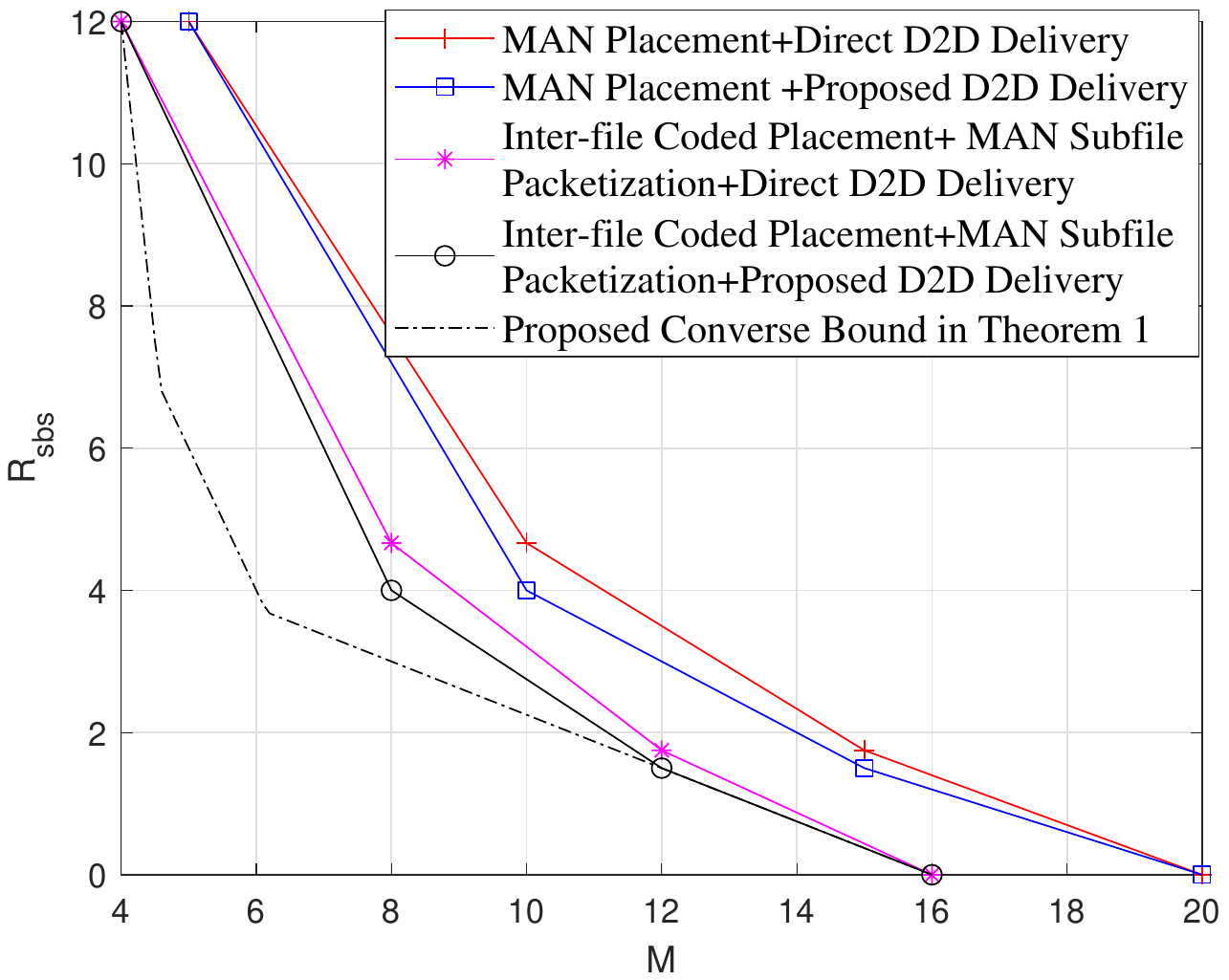}
        \caption{\small $(\Msf,\Rdd)$ tradeoff for $\Rbc=4$. }
        \label{fig:numerical 1b}
    \end{subfigure}
    \caption{\small Performance of a topology-aware Fog-RAN system  with $\Hsf=4$, $\Kdd=16$, $\Kbc=4$, and $\Nsf=20$.}
    \label{fig:numerical 1}
\end{figure}

We provide numerical evaluations of the proposed caching scheme with symmetric subpacketization. In Fig.~\ref{fig:numerical 1}, we consider the network with $\Hsf=4$, $\Kdd=16$, $\Kbc=4$, $\Nsf=20$, $\Lsf_1=6$, $\Lsf_0=4$, $\Lsf_3=3$ and $\Lsf_4=3$. We first fix $\Rdd=0$ and plot the memory-load tradeoff $(\Msf,\Rbc)$.  Fig.~\ref{fig:numerical 1a} shows the proposed scheme in~\eqref{eq:achievable points 1}, outperforms the caching scheme based on the ``MAN Placement" without inter-file coded cache placement (as the cache placemen phase in~\cite{cless2018Em}) and the delivery scheme in~\cite{parrinello2018sharedcache}. When $\Msf\geq 12$, the proposed single bottleneck-link scheme in~\eqref{eq:achievable points 1} is exactly optimal.

We then fix $\Rbc=\Kbc=4$  and plot the memory-load tradeoff $(\Msf,\Rdd)$ in Fig.~\ref{fig:numerical 1b}. 
From~\eqref{eq:trivial converse} and~\eqref{eq:converse N>K0 s=H}, we see that in order to have $\Rbc^{\star}=\min\{\Nsf,\Kbc\}=\Kbc$, we need $\Hsf \Msf \geq \Nsf-\Kbc$.  Hence, we must have  $\Msf\geq  \frac{\Nsf-\Kbc}{\Hsf}=4$. 
To compare the proposed caching scheme, we also plot other caching schemes with ``MAN Placement" (without inter-file coded cache placement) or with ``Direct D2D Delivery" (the D2D delivery scheme with shared caches by directly extending the single bottleneck-link delivery scheme with shared caches in~\cite{parrinello2018sharedcache}).
It can be seen that the novel D2D caching scheme in~\eqref{eq:achievable points 2} outperforms the other schemes.  
When $\Msf\geq 12$, the proposed D2D caching scheme is also exactly optimal.
In other words, if $\Msf\geq 12$, both of the proposed single bottleneck-link caching scheme in~\eqref{eq:achievable points 1}  and the D2D caching scheme in~\eqref{eq:achievable points 2}  are optimal.

\section{Topology-aware Fog-RAN Systems: Improved Caching Scheme with Asymmetric File Subpacketization }
\label{sec:improvement}

The proposed inter-file coded cache placement in Theorem~\ref{thm:symm ach theorem} uses the MAN symmetric file subpacketization of~\cite{dvbt2fundamental}. 
The proposed placement with such file subpacketization is topology-partially-agnostic (i.e., it only uses the value of $\Kdd$ which is the total number of users connected to some SBS). 
In the following example, we show that we can improve on Theorem~\ref{thm:symm ach theorem} by a file subpacketization  that is topology-aware.

\begin{example}
\label{ex:ex2}
\rm
We  focus on the network  illustrated in Fig.~\ref{fig: system_model} with $\Hsf=3$, $\Kdd=4$, $\Kbc=2$, $\Nsf=6$ and $\Msf=2$. 
With $\Rdd=0$,  the  scheme in Theorem~\ref{thm:symm ach theorem} yields the 
triple  $(\Msf,\Rbc,\Rdd)=(2,19/6,0)$.  We will improve on it as follows.

\paragraph*{Placement Phase}
We partition each file $F_{i}, i\in [\Nsf],$ into two equal-length subfiles as $F_{i}=\{F_{i,\{1\}},F_{i,\{2,3\}}\}$. Each subfile has $\Bsf/2$ bits. We let SBS $1$ cache $2\Bsf$ random linear combinations of all bits in $\{F_{i,\{1\}}:i\in [\Nsf]\}$. We also let each of SBS $2$ and SBS $3$ cache $ 2\Bsf$ random linear combinations of all bits in $\{F_{i,\{2,3\}}:i\in [\Nsf]\}$.

\paragraph*{Delivery Phase}
Assume  $\dv=(1, \ldots,6)$. 
We first let the MBS broadcast $F_5$ and $F_6$ to satisfy the demands of users $5$ and $6$  directly connected to the MBS. Simultaneously, SBS $1$ can recover all bits in $\{F_{i,\{1\}}:i\in [\Nsf]\}$, and SBS $2$ and SBS $3$ can recover all bits in $\{F_{i,\{2,3\}}:i\in [\Nsf]\}$.
We then let the MBS broadcast $F_{1,\{2,3\}}\oplus F_{3,\{1\}}$ to satisfy the demands of users $1$ and $3$, and broadcast $F_{2,\{2,3\}}\oplus F_{4,\{1\}}$ to satisfy the demands of users $2$, $4$. So we  achieve the memory-loads triple $(2,3,0)$. By the converse bound in Theorem~\ref{thm:converse} with $s=1$, this memory-loads point is optimal.
\hfill$\square$
\end{example}

This section is organized as follows.
In Section~\ref{sub:main result Asymm.placement} we state our results with asymmetric file subpacketization.
The details of the proofs are given in subsequent subsections.
We conclude in Section~\ref{sub:numerical for baseline Asymm.placement} with some numerical evaluations of the proposed regions.

\subsection{Main Results and Discussion}
\label{sub:main result Asymm.placement}
In this section we consider the case where $\Nsf\geq \Kbc$ only, with the aim to improve on the achievable bounds and the optimality results in Section~\ref{sub:main result symm.placement}. The main idea is come up with schemes that employ an asymmetric file subpacketization that is topology-aware. We start with a definition.

\begin{defn}[Topology-aware $G$-way partition]
\label{def:cent G-way partition}
For an integer $G \in [\Hsf]$, we say that $\Phi = \{ \Gc^\Phi_1, \ldots, \Gc^\Phi_G\}$ is a {\it $G$-way partition of $[\Hsf]$} if it partitions $[\Hsf]$ into $G$ subsets/groups  
such that $\sum_{h \in \Gc^\Phi_i} \Lsf_h \geq \sum_{h \in \Gc^\Phi_j} \Lsf_h$ for $i \leq j$. 
The set of all the $G$-way partitions is denoted by $\mathbf{Q}_G$, for each integer $G\in[\Hsf]$. 
\end{defn}

The main idea is to group the SBSs by a $G$-way partition and to let each SBS in the same group have the {\it same cache content}.
Notice that for any network, there always exists a $\Hsf$-way partition, where each group contains a single SBS and $\Lsf_1\geq \dots \geq \Lsf_{\Hsf}$.

\begin{defn}[Aggregate occupancy number]
\label{def:Aggregate Occupancy number}
For each $G$-way partition $\Phi$ and 
  each subset of groups $\Sc\subseteq [G]$, we sort  by the total occupancy number of the SBSs in each group in a descending order,
as $(q_1(\Phi,\Sc),\ldots,q_{|\Sc|}(\Phi,\Sc))$  where $q_1(\Phi,\Sc):=\max_{g\in\Sc}\sum_{j\in \Gc^{\Phi}_{g}} \Lsf_j$ and $q_{|\Sc|}(\Phi,\Sc):=\min_{g\in\Sc}\sum_{j\in \Gc^{\Phi}_{g}} \Lsf_j$. We also define for each $j\in |\Sc|$,
 $$
 q^{\prime}_j(\Phi,\Sc):=\min\{q_j(\Phi,\Sc),\Nsf-\Kbc\}.
 $$ 
\end{defn}
\bigskip
\begin{thm}[Achievable region with asymmetric file subpacketization] 
\label{thm:improvement}
For the $(\Kbc,\Kdd,\Hsf,\Nsf,\Msf)$ topology-aware Fog-RAN system with $\Nsf\geq \Kbc$, the lower convex envelop of the following corner points is achievable
\begin{align}
 & (\Msf,\Rbc,\Rdd)= \left(\frac{t(\Nsf-\Kbc)}{G}, \ \Kbc+\min_{\Phi \in \mathbf{Q}_G}\frac{\sum_{r=1}^{G-t} q^{\prime}_r(\Phi,[G]) \binom{G-r}{t}}{\binom{G}{t}},0  \right), 
\label{eq:achievable improvement}\\
&(\Msf, \Rbc,\Rdd)=\Big(\frac{t(\Nsf-\Kbc)}{G}, \ \Kbc, \ \min_{\Phi \in \mathbf{Q}_G} \negmedspace\negmedspace  \sum_{\Sc\subseteq [G]:|\Sc|=t+1}  \negmedspace\negmedspace\negmedspace\negmedspace    \frac{q^{\prime}_1(\Phi,\Sc)+\frac{1}{t} \left[q^{\prime}_{t+1}(\Phi,\Sc)-q^{\prime}_{1}(\Phi,\Sc)+q^{\prime}_{t}(\Phi,\Sc)\right]^+ }{\binom{G}{t}}  \Big),
\label{eq:D2D achievable improvement}\\
&(\Msf, \Rbc,\Rdd)=\Big(\frac{ \Nsf-\Kbc }{\Hsf}, \ \Kbc, \ \min\{\Nsf-\Kbc,\Kdd\} \Big),
\label{eq:D2D achievable improvement app1}
\end{align}
for all $t\in[G]$ and all $G\in[\Hsf]$.
\end{thm}
The proofs of~\eqref{eq:achievable improvement} and~\eqref{eq:D2D achievable improvement}  can be found in Section~\ref{sub:proof of eq:achievable improvement}, while~\eqref{eq:D2D achievable improvement app1} is achieved by the first approach in Section~\ref{sub:delivery D2D point case 2} by letting $t=1$.

\begin{cor}
\label{cor:cover}
The achievable region in Theorem~\ref{thm:improvement} contains the one in Theorem~\ref{thm:symm ach theorem}.
\end{cor}
\begin{IEEEproof}
It can be seen that if $G=1$, the scheme in~\eqref{eq:achievable improvement} achieves the same region as
the first approach  in Section~\ref{sub:delivery shared point case 2}. By memory-sharing between~\eqref{eq:D2D achievable improvement app1} and $(\Msf, \Rbc,\Rdd)=( \Nsf-\Kbc, \Kbc,0 )$, we achieve the   the same region as the first approach
   in Section~\ref{sub:delivery D2D point case 2}. 
In addition, if $G=\Hsf$, the scheme in Theorem~\ref{thm:improvement} achieves the same region as
the second approaches    in Sections~\ref{sub:delivery shared point case 2} and~\ref{sub:delivery D2D point case 2}. 
\end{IEEEproof}

From Theorem~\ref{thm:improvement}, with the knowledge of the network topology and the memory size, we can compute the achieved loads of all possible partitions, and thus we can choose the best  $G$-way partition  leading to the minimal load. 
As a result, asymmetric subpacketizations improve the symmetric MAN subpacketization. This is due to the network asymmetry; if we use MAN symmetric subpacketization and the multi-rounds MAN delivery, in some rounds some multicast messages of the form~\eqref{eq:MAN multicast} are not ``full'' (some subfiles in the binary sum do not exist) and this reduces the coded caching gain. However, the grouping method proposed in this paper ``balances'' the number of users accessing each cache and reduce the load. 
The complexity on evaluation Theorem~\ref{thm:improvement} depends on the number of possible partitions, while for each partition the evaluation is very  simple. 


\bigskip
From the achievable scheme in Theorem~\ref{thm:improvement}, we can derive the following optimality result, whose proof can be found in Appendix~\ref{sub:proof of optimality improvement}. 
\begin{thm}[Exact optimality for~\eqref{eq:achievable improvement}]
\label{thm:optimality improvement}
For the $(\Kbc,\Kdd,\Hsf,\Nsf,\Msf)$ topology-aware Fog-RAN system with $\Nsf\geq \Kbc+\Lsf_1$, if there exists a $G$-way partition $\Phi$ such that $\Gc^{\Phi}_1=\{1\}$, the achievable triplet $(\Msf,\Rbc,\Rdd)$ is exactly optimal with $\Msf\geq \left(1-\frac{1}{G}\right)(\Nsf-\Kbc)$, thus giving $\Rbc^\star =\Kbc+\Lsf_1\big(1-\frac{\Msf}{\Nsf-\Kbc} \big)$ and $\Rdd^\star=0$.
\end{thm}
Note that, since a $\Hsf$-way partition always exists, the achieved load in~\eqref{eq:achievable improvement} is exactly optimal when $\Msf\geq (1-1/\Hsf)(\Nsf-\Kbc)$ for any topology with $\Nsf\geq \Kbc+\Lsf_1$.

\bigskip
Similarly, by comparing the achievable scheme in Theorem~\ref{thm:improvement} and the converse bound in Theorem~\ref{thm:converse}, we have the following order optimality result, whose proof can be found in Appendix~\ref{sub:proof of D2D optimality improvement}.
\begin{thm}[Order optimality for Theorem~\ref{thm:improvement}]
\label{thm:D2D optimality improvement}
For the $(\Kbc,\Kdd,\Hsf,\Nsf,\Msf)$ topology-aware Fog-RAN system with $\Nsf\geq \Kbc+\Lsf_1$, if there exists one $G$-way partition $\Phi$ such that $\Gc^{\Phi}_1=\{1\}$, for any  $(\Msf,\Rbc^{\star},\Rdd^{\star})$ in the converse region with $\Msf\geq \left(1-\frac{1}{G}\right)(\Nsf-\Kbc)$, the caching scheme in Theorem~\ref{thm:improvement} can achieve the triplet $ \left(\Msf,\Rbc^{\star},\frac{G}{G-1}\Rdd^{\star}\right)$.
\end{thm}

\begin{rem}[On the sub-optimlaity of MAN subpacketization for the single bottleneck-link caching problem with shared caches in~\cite{parrinello2018sharedcache}]
\label{rem:shared-cache}
Theorem~\ref{thm:improvement} for $\Kbc=0$ gives an achievable region for the single bottleneck-link problem with shared caches 
with uncoded cache placement as studied in~\cite{parrinello2018sharedcache}.
By Theorem~\ref{thm:optimality improvement} for $\Kbc=0$, 
our scheme is exactly optimal in large memory size regime.
This implies that the scheme in~\cite{parrinello2018sharedcache}, which was proved to be optimal under the constraint of uncoded cache placement and of placement that is independent of the network topology (i.e., the placement cannot depend on the occupancy number of each cache), is strictly suboptimal. 
For example, 
in Example~\ref{ex:ex2} with $\Kbc=0$, our scheme achieves the optimal load equal to $2$ when $\Msf=2$, while the scheme in~\cite{parrinello2018sharedcache} only achieves $19/6$.
In general, if the knowledge of the occupancy number of each cache is known a  priori and can be used in the design of the caching placement, improvements over the ``topology-agnostic'' scheme in~\cite{parrinello2018sharedcache} are possible.

\end{rem}

\begin{rem}[On reduce the subpacketization level while achieving optimal load]
\label{rem:subpacketization}
Dividing users into groups and letting the users in the same group have the same cache content was originally proposed in~\cite{finiteanalysis} for the single bottleneck-link caching problem in order to reduce the subpacketization level at the expense of a higher load compared to the original MAN scheme. In this paper, we showed that for ``asymmetric'' cache-aided networks, one can come up with a grouping strategy that {\it reduces the subpacketization level 
while it still achieves the optimal load}; this cannot be achieved by the MAN symmetric placement (because the converse bound on the load under the constraint of the MAN placement was given in~\cite{parrinello2018sharedcache}).\footnote{Similar phenomenon using efficient grouping strategy to reduce the subpacketization level while maintaining the optimal load achieved by the MAN placement, has also been found in D2D caching networks~\cite{zhang2019D2D}. Different from the work in~\cite{zhang2019D2D}, we also show the grouping strategy can achieve a strictly lower load than the MAN placement.}

We can compare the subpacketization levels of the proposed placement and the   MAN placement used in~\cite{parrinello2018sharedcache}.
 Consider a $G$-way partition, memory size $\Msf$, and integers $t_1=\Hsf\Msf/\Nsf$ and $t_2=G \Msf/\Nsf$. The subpacketization level of the MAN placement used in~\cite{parrinello2018sharedcache} is $\binom{\Hsf}{\Hsf\Msf/\Nsf}$, which is strictly  higher than the subpacketization level of the proposed asymmetric placement which is $\binom{G}{G\Msf/\Nsf}$ if $G<\Hsf$.
Assume $\Hsf \gg \Msf/\Nsf$ and $G\gg \Msf/\Nsf$, it can be seen that $\binom{\Hsf}{\Hsf\Msf/\Nsf}/\binom{G}{G\Msf/\Nsf}\approx (\Nsf e/\Msf-0.5e)^{(\Hsf-G)\Msf/\Nsf}$ by the  Stirling's formula ($e$ is the Euler's number), which is a large multiplicative gap.
\end{rem}

\subsection{Proof of Theorem~\ref{thm:improvement}}
\label{sub:proof of eq:achievable improvement}

\paragraph*{Placement Phase}
Consider a $G$-way partition $\Phi$, for some $G\in[\Hsf]$. 
For $\Msf=\frac{t(\Nsf-\Kbc)}{G}$ where $t\in [0:G]$, we partition each file $F_i, i\in[\Nsf],$ into $\binom{G}{t}$ equal-length subfiles $F_{i}=\{F_{i,\Wc}:\Wc\subseteq [G],|\Wc|=t\}$.  Thus each subfile has $\Bsf/\binom{G}{t}$ bits. Recall that  
$
\Fc_{\Wc}:=\{F_{i,\Wc}:i\in[\Nsf]\},
$
for each  $\Wc\subseteq [G]$ where $|\Wc|=t$. It can be seen that $\Fc_{\Wc}$ contains $\Bsf \Nsf/\binom{G}{t}$ bits. 
For each $g\in [G]$, we then let each SBS $h\in \Gc^{\Phi}_{g}$ cache the same  $|\Fc_{\Wc}|(\Nsf-\Kbc)/\Nsf$ random linear combinations of all bits in $\Fc_{\Wc}$ for each $\Wc\subseteq [G]$ where $|\Wc|=t$ and $g\in \Wc$. Hence, it can be seen that each SBS   caches $\binom{G-1}{t-1} \Bsf(\Nsf-\Kbc)/\binom{G}{t}=\Bsf t(\Nsf-\Kbc)/G$ bits, satisfying the cache size constraint. 

\paragraph*{Delivery Phase}
Since the SBSs in the same group have the same cache content, the system is equivalent to one where all users connected to the SBSs in the same group are connected to a single virtual SBS. 
Hence, we can use the delivery phase described in Section~\ref{sub:delivery shared point case 2} to achieve~\eqref{eq:achievable improvement}, and use the delivery phase in Section~\ref{sub:delivery D2D point case 2} to achieve~\eqref{eq:D2D achievable improvement}.

\subsection{Numerical Evaluations}
\label{sub:numerical for baseline Asymm.placement}

We conclude with some numerical results. 

\begin{figure}
    \centering
    \begin{subfigure}[t]{0.5\textwidth}
        \centering
        \includegraphics[scale=0.6]{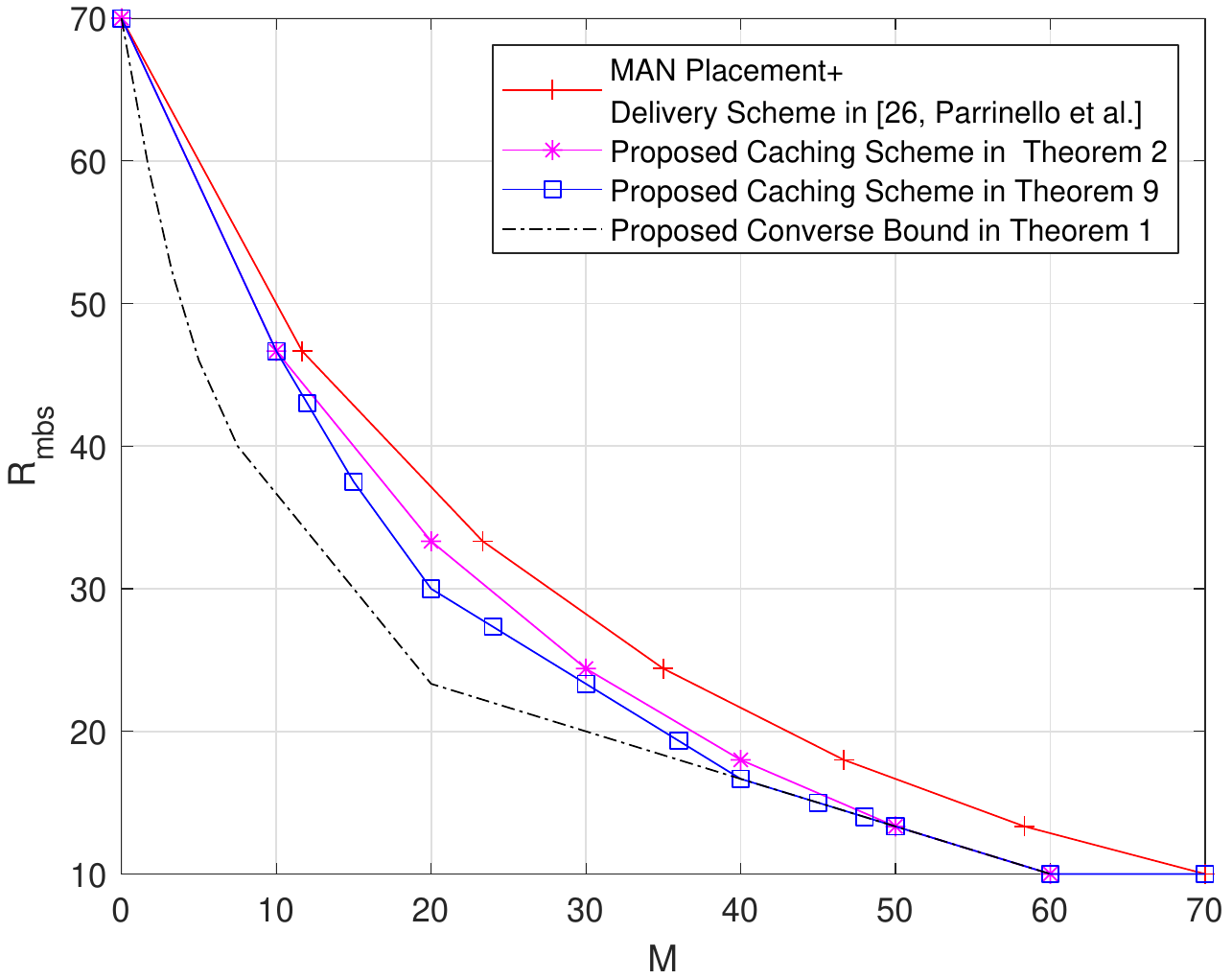}
        \caption{\small $(\Msf,\Rbc)$ tradeoff for $\Rdd=0$.}
        \label{fig:numerical 2a}
    \end{subfigure}%
    ~ 
    \begin{subfigure}[t]{0.5\textwidth}
        \centering
        \includegraphics[scale=0.6]{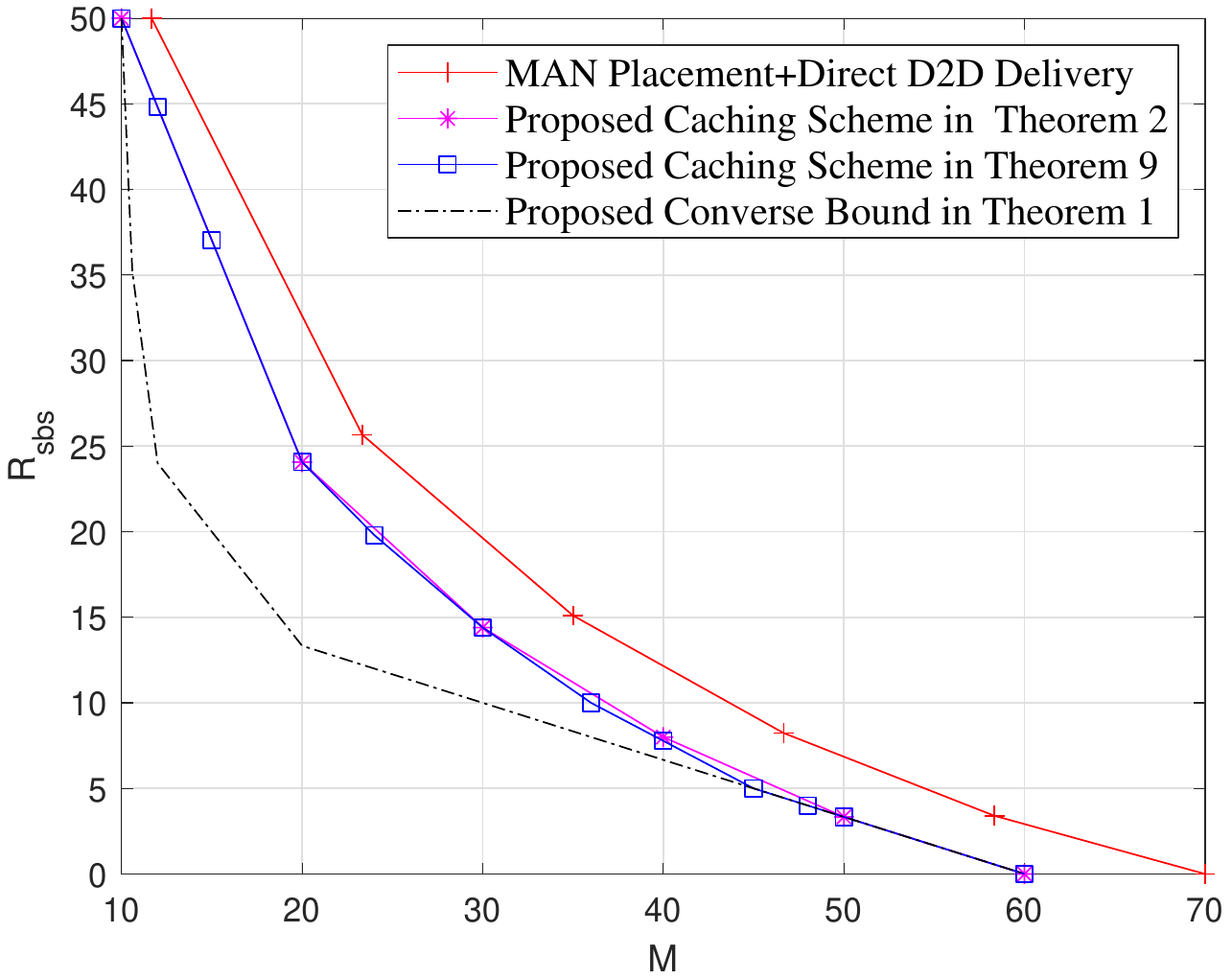}
        \caption{\small $(\Msf,\Rdd)$ tradeoff for $\Rbc=10$. }
        \label{fig:numerical 2b}
    \end{subfigure}
    \caption{\small Performance of a topology-aware Fog-RAN system with $\Hsf=6$, $\Nsf=70$, $\Kbc=10$, $\Kdd=60$ ($\Lsf_1=20$, $\Lsf_2=20$, $\Lsf_3=8$, $\Lsf_4=6$, $\Lsf_5=4$ and $\Lsf_6=2$).}
    \label{fig:numerical 2}
\end{figure}

In Fig.~\ref{fig:numerical 2}, we consider the network with $\Hsf=6$, $\Nsf=70$, $\Kbc=10$, $\Kdd=60$ (in particular, $\Lsf_1=20$, $\Lsf_2=20$, $\Lsf_3=8$, $\Lsf_4=6$, $\Lsf_5=4$ and $\Lsf_6=2$). 
We first fix $\Rdd=0$ and plot the memory-load tradeoff $(\Msf,\Rbc)$ in Fig~\ref{fig:numerical 2a}. We can see that the proposed inter-file coded cache placement with the novel   file subpacketization  in~\eqref{eq:achievable improvement} outperforms the other schemes. In addition, it can also be  seen that when $\Msf\geq 40$, the proposed single bottleneck-link scheme in~\eqref{eq:achievable improvement} is optimal.
We then fix $\Rbc=\Kbc=10$ and plot the memory-load tradeoff $(\Msf,\Rdd)$ in Fig~\ref{fig:numerical 2b}. We can see that the proposed inter-file coded cache placement with the novel file subpacketization in~\eqref{eq:D2D achievable improvement} outperforms the other schemes. In addition, it can also be  seen that when $\Msf\geq 45$, the 
D2D scheme in~\eqref{eq:D2D achievable improvement} is optimal. 
In other words, if $\Msf\geq 45$, both of the proposed single bottleneck-link caching scheme in Theorem~\ref{thm:improvement}  and the  D2D caching scheme in Theorem~\ref{thm:improvement}  are optimal. 

\begin{figure}
\centerline{\includegraphics[scale=0.6]{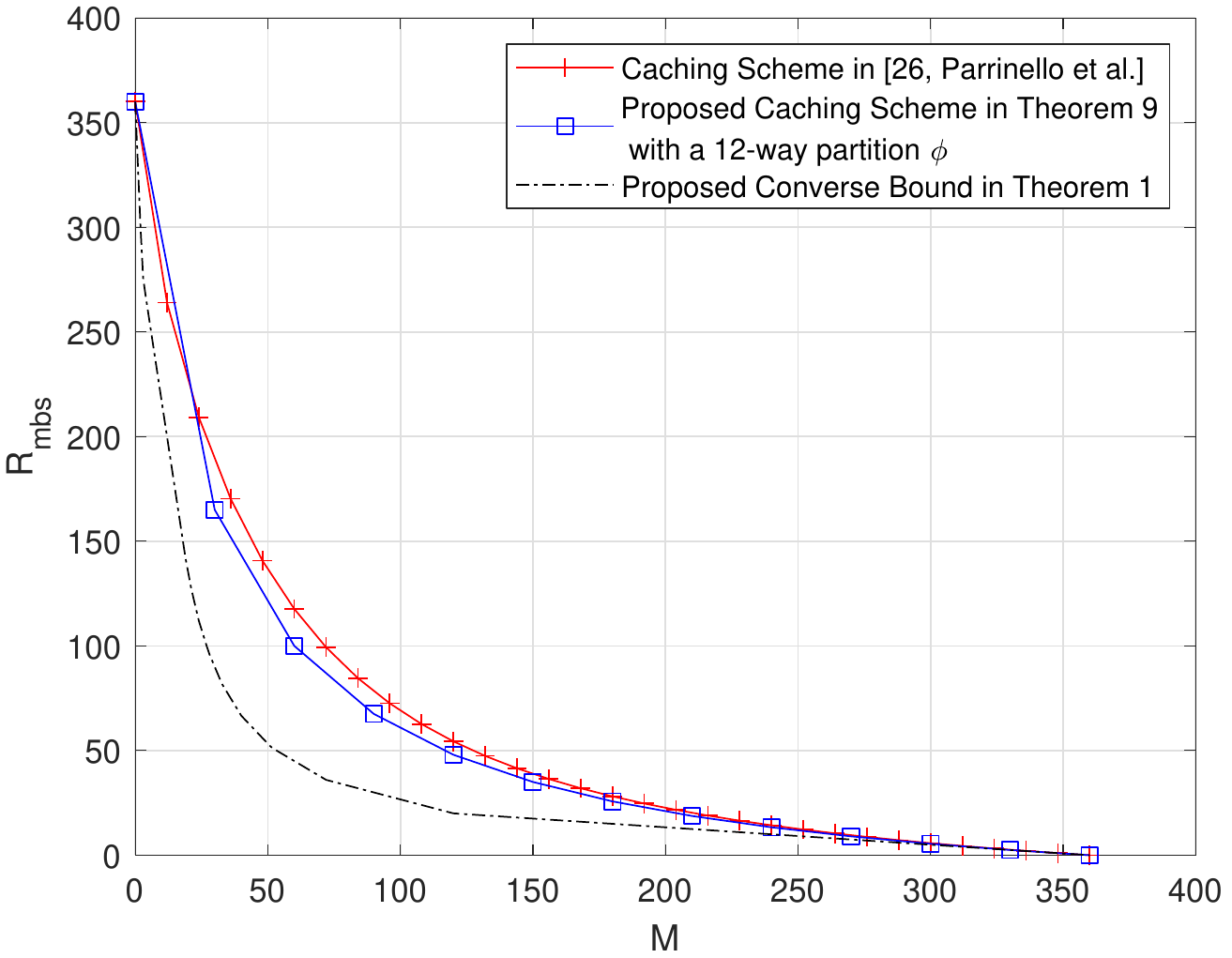}}
\caption{\small Performance of a topology-aware Fog-RAN system with $\Hsf=30$, $\Nsf=360$, $\Kbc=0$, $\Kdd=360$ (with $\Lsf_h=30$ for $h\in[10]$, and $\Lsf_h=3$ for $h\in[11:30]$). We used a $12$-way partition $\Phi$ where $\Gc^{\Phi}_i=\{i\}$ for $i\in [10]$, $\Gc^{\Phi}_{11}=[11:20]$, and $\Gc^{\Phi}_{12}=[21:30]$. Here $\Rdd=0$.}
\label{fig:numerical 3}
\end{figure}

In Fig.~\ref{fig:numerical 3} we  consider  the network  with $\Hsf=30$, $\Nsf=360$, $\Kbc=0$, $\Kdd=360$ (with $\Lsf_h=30$ for $h\in[10]$, and $\Lsf_h=3$ for $h\in[11:30]$).  We choose a $12$-way partition $\Phi$ where $\Gc^{\Phi}_i=\{i\}$ for $i\in [10]$, $\Gc^{\Phi}_{11}=[11:20]$, and $\Gc^{\Phi}_{12}=[21:30]$. With this partition, we can see $\Gc^{\Phi}_1$ only contains one SBS and 
each group contains the same number of users; we do not claim however that this partition is optimal.
By fixing $\Rdd=0$, the problem is the same as the single bottleneck-link caching problem with shared caches as considered in~\cite{parrinello2018sharedcache}.
We compare the performance of the scheme in~\cite{parrinello2018sharedcache} with our scheme that uses an asymmetric file subpacketization based on this $12$-way partition. It can be seen that our proposed scheme with this $12$-way partition achieves a lower load than the scheme in~\cite{parrinello2018sharedcache} when $\Msf\geq 24$. 
Moreover, the subpacketization level with our scheme is in general lower that that of~\cite{parrinello2018sharedcache}; for example, when $\Msf=\Nsf/2=180$, the subpacketization levels of the scheme in~\cite{parrinello2018sharedcache} and of the proposed scheme with the  $12$-way partition are $\binom{30}{15}\approx 1.55 \times 10^8$ and $\binom{12}{6}=924$, respectively,  that is, a five order of magnitude difference.
 
Furthermore, we can see that compared to the MBS broadcast load in Fig.~\ref{fig:numerical 2a}, the improvement from the proposed asymmetric file subpacketization on the SBS sidelink load in Fig.~\ref{fig:numerical 2b} is smaller. As we explained in Section~\ref{sub:main result Asymm.placement}, the  proposed asymmetric file subpacketization improves on the symmetric one because  the latter may  generate redundant multicast messages  (i.e., some subfiles in the multicast message are empty). By the asymmetric file subpacketization, we reduce this redundancy in a way that most multicast messages are ``full''. 
The proposed  D2D delivery scheme  in Section~\ref{sub:delivery D2D point case 2} leverages the redundancy   in the   multicast messages. Hence, if we combine the asymmetric file subpacketization and the proposed D2D delivery scheme, the improvement on the load is not large compared to the caching scheme with MAN  file subpacketization and the proposed D2D delivery scheme. However, with the asymmetric file subpacketization, we have a much lower subpacketization level.

\section{Topology-agnostic Systems for Random Topology}
\label{sub:decentralized}

The novel inter-file coded cache placement and asymmetric file subpacketization in the previous sections applies 
to {\it topology-aware} systems, since the placement exploits the knowledge of the network topology. 
Topology-aware  systems are not suited for the case of nomadic/mobile users (who move around the network), 
because in the framework of coded caching the caches are updated on a much longer time scale than that of user mobility, 
and thus the network topology during placement may in general be different from that during delivery. 
In the context of our Fog-RAN network, in this section we consider the case of random topology (i.e., $\Kbc, \Lsf_1, \ldots, \Lsf_{\Hsf}$ are 
 jointly distributed random variables), and use the previously developed topology-aware strategies 
for a ``typical'' network configuration. In fact, the identity and mobility of the users is irrelevant to the system performance, 
as long as the number of users served by the MBS downlink and the SBSs is given. Hence, we can design the system for a typical realization of the network, 
considering that for large mobile networks such numbers will fluctuate around typical values which can be learned from network statistics accumulated over time. 
In this case our novel inter-file coded cache placement and asymmetric file subpacketization can still be used to leverage 
``statistical knowledge'' about the SBS occupancy numbers.

In the following, $(\Kbc, \Lsf_1, \ldots, \Lsf_\Hsf)$ is a random vector with known distribution.
We denote a realization of $(\Kbc, \Lsf_1, \ldots, \Lsf_\Hsf)$ by $(\ksf_0,\lsf_1,\ldots,\lsf_\Hsf)$.
For topology-agnostic systems, the objective function is the   average worst-case  loads, 
where the average is taken over all the realizations of $(\Kbc, \Lsf_1, \ldots, \Lsf_\Hsf)$ and the worst case is take over all the possible demands for one realization $(\ksf_0,\lsf_1,\ldots,\lsf_\Hsf)$.
For   sake of simplicity, we only consider the realizations that $\Nsf>\Kbc$ because if $\Nsf\leq \Kbc$, it is trivial that 
delivering all the $\Nsf$ files from the MBS is optimal.

\subsection{Main Results and Discussion}
 
 A     converse bound  for topology-agnostic  systems, can be obtained by taking the average of the   converse bound  for topology-aware systems in Theorem~\ref{thm:converse}    over all realization of users. However, since the number of $\Kbc$ is not known a priori for the placement phase, 
 this converse bound is loose in general. Hence, this will not be pursued in the following.
 
We start by extending the notion of $G$-way partition in Definition~\ref{def:cent G-way partition} to the topology-agnostic systems.
\begin{defn}[Topology-agnostic $G$-way partition]
\label{def:dec G-way partition}
A topology-agnostic $G$-way partition is simply a partition of the $\Hsf$ SBSs into $G$ groups, for some $G\in [\Hsf]$. 
Compared to the topology-aware partition in Definition~\ref{def:cent G-way partition}, a topology-agnostic $G$-way partition $\varphi$ is done without knowledge of  the exact number of users to be served during delivery. Hence, one cannot ensure that $\sum_{h\in \Gc^\varphi_{g_1}}\Lsf_h \geq \sum_{h\in \Gc^\varphi_{g_2}}\Lsf_h $ with probability $1$ for $1\leq g_1\leq g_2 \leq G$ as in Definition~\ref{def:cent G-way partition}.
The set of all the topology-agnostic $G$-way partitions is denoted by $\mathbf{A}_G$, for each $G\in[\Hsf]$.
\end{defn}

As Definition~\ref{def:Aggregate Occupancy number},
for each $G$-way partition $\varphi$ and each subset of groups $\Sc\subseteq [G]$, we sort  by the total occupancy number of the SBSs in each group
  in a descending order as $(q_1(\varphi,\Sc),\ldots,q_{|\Sc|}(\varphi,\Sc))$  where $q_1(\varphi,\Sc):=\max_{g\in\Sc}\sum_{j\in \Gc^{\varphi}_{g}} \Lsf_j$ and $q_{|\Sc|}(\varphi,\Sc):=\min_{g\in\Sc}\sum_{j\in \Gc^{\varphi}_{g}} \Lsf_j$. In addition, $q^{\prime}_j(\varphi,\Sc):=\min\{q_j(\varphi,\Sc),\Nsf-\Kbc\}$ for each $j\in |\Sc|$.

We first consider the case where the SBS sidelink load is $\Rdd=0$ regardless of the network realization. 

%
The following region is achievable, whose proof can be found in Section~\ref{sub:proof of thm:dec shared link}.
\begin{thm}
\label{thm:dec shared link}
For the topology-agnostic Fog-RAN system,  the following memory-load  points are achievable,  
\begin{align}
 & (\Msf,\Rbc,\Rdd)=\bigg(
 \frac{t(\Nsf-n)}{G}, 
 \nonumber\\& 
 \mathbb{E}_{(\Kbc, \Lsf_1, \ldots, \Lsf_\Hsf)}\left[ \max\{\Kbc,n\}+\frac{\sum_{h\in [\Hsf]}\Lsf_h-[n-\Kbc]^+}{\binom{G}{t}\sum_{h\in [\Hsf]}\Lsf_h}\sum^{G-t}_{r=1}q^{\prime}_r(\varphi,[G]) \binom{G-r}{t}\right],
 0 \bigg),
\label{eq:achieveable dec}
\end{align}
for $n\in [0:\Nsf-1]$, $t\in[0:G]$, $\varphi \in \mathbf{A}_G$, and $G\in[\Hsf]$.
\end{thm}
The parameter $n$ in~\eqref{eq:achieveable dec} corresponds to the inter-file coded placement. More precisely, for each $\Wc\subseteq [G]$ where $|\Wc|=t$, we let each SBS in group $\Gc^\varphi_{g}\in[G]$ caches  $|\Fc_{\Wc}|(1-n/\Nsf)$ random linear combinations of all bits in $\Fc_{\Wc}$. In a topology-aware system, the exact value of $\Kbc$ is known in the placement and   we let $n=\Kbc$, such that during the delivery phase, after the MBS transmits $\Kbc$ files to satisfy the $\Kbc$ users without cache, each SBS can recover $\Fc_{\Wc}$. In a topology-agnostic system, the exact value of $\Kbc$ is unknown at the time of placement, thus for a 
given memory size and joint distribution of the users, we choose the value of $n\in [0:\Nsf-1]$ that leads to the minimal average worst-case loads.

\bigskip
Then we consider a second class corner points, where the single bottleneck-link load is minimal  for any network realization.  
The following region is achievable, whose proof can be found in Section~\ref{sub:proof of thm:D2D dec}.
\begin{thm}
\label{thm:D2D dec}
For the topology-agnostic Fog-RAN system, the following memory-load 
points are achievable  
 \begin{align}
&(\Msf, \Rbc,\Rdd)=\bigg(
\frac{t(\Nsf-n)}{G}, 
\mathbb{E}[\Kbc], \ \mathbb{E}_{(\Kbc, \Lsf_1, \ldots, \Lsf_\Hsf)}\left[ \frac{G}{G-1}[n-\Kbc]^+   +  \right.
\nonumber\\&
 \left.   \min\left\{\frac{\min\{\Nsf-\Kbc,\Kdd\}(G-t)}{G-1} , \ \sum_{\Sc\subseteq [G]:|\Sc|=t+1}  \frac{q^{\prime}_1(\varphi,\Sc) + [q^{\prime}_{t+1}(\varphi,\Sc)-q^{\prime}_{1}(\varphi,\Sc)+q_{t}(\varphi,\Sc)]^+ /t }{\binom{G}{t}} \right\}\right]
\bigg) 
\label{eq:D2D achieveable dec}
\end{align}
for $n\in [0:\Nsf-1]$, $t\in[G]$, $\varphi \in \mathbf{A}_G$, $G\in[2:\Hsf]$, where $\frac{\Nsf-n}{\Nsf}t \geq 1$.
\end{thm}
Note that the constraint $\frac{\Nsf-n}{\Nsf}t \geq 1$ in~\eqref{eq:D2D achieveable dec} guarantees that, if the instantaneous realization of $\Kbc$ is zero, the whole library is stored across all the SBSs and thus successful delivery is possible without the transmission from the MBS.

\subsection{Proof of Theorems~\ref{thm:dec shared link} and~\ref{thm:D2D dec}}
\label{sub:proof of thm:dec shared link}
\label{sub:proof of thm:D2D dec}

Fix an $n\in [0:\Nsf-1]$, a $t\in[0:G]$, a $G\in[\Hsf]$ and a topology-agnostic $G$-way partition $\varphi$.

\paragraph*{Placement Phase}
We partition each file $F_i, i\in[\Nsf],$ into $\binom{G}{t}$ equal-length subfiles as $F_{i}=\{F_{i,\Wc}:\Wc\subseteq [G],|\Wc|=t\}$.  Thus each subfile has $\Bsf/\binom{G}{t}$ bits. Recall that  
$
\Fc_{\Wc}:=\{F_{i,\Wc}:i\in[\Nsf]\},
$
for each  $\Wc\subseteq [G]$ where $|\Wc|=t$. It can be seen that $\Fc_{\Wc}$ contains $\Bsf \Nsf/\binom{G}{t}$ bits. 
For each $g\in [G]$, we  let each SBS $h\in \Gc^{\varphi}_{g}$ cache $|\Fc_{\Wc}|(\Nsf-n)/\Nsf$ random linear combinations of all bits in $\Fc_{\Wc}$ for each $\Wc\subseteq [G]$ where $|\Wc|=t$ and $g\in \Wc$. Hence, each SBS caches $\binom{G-1}{t-1} \Bsf(\Nsf-n)/\binom{G}{t}=\Bsf t(\Nsf-n)/G$ bits, and thus the memory size $\Msf=\frac{t(\Nsf-n)}{G}$. 

\paragraph*{Delivery Phase}
To achieve~\eqref{eq:achieveable dec}, no SBS sidelink communication takes place in the delivery phase.
At this point we assume that all nodes in the system are aware of the realization of $(\Kbc, \Lsf_1, \ldots, \Lsf_\Hsf)$.  
We still define the users without connection to any SBS as $\Uc_0$ and the users connected to SBS $h\in[\Hsf]$ as $\Uc_h$.
As in the proposed caching scheme in Section~\ref{sub:interfile} for topology-aware systems, we first satisfy the demands of the users in $\Uc_0$ while letting each SBS $h\in[\Hsf]$ recover all bits in $\Fc_{\Wc}$  if $h\in \Wc$.
 
Recall that $\Dc_{\rm{mbs}}(\dv) := \cup_{k\in\Uc_0}  \{d_k\}$ and $\Dc_{\rm{sbs}}(\dv) := \big(\cup_{h\in[\Hsf], k\in\Uc_h} \{d_k\}\big)  \backslash \Dc_{\rm{mbs}}(\dv)$. In a topology-agnostic system, since the realization of $|\Dc_{\rm{mbs}}(\dv)|$ is unknown at the time of placement, the chosen $n$ may be strictly less than $\Dc_{\rm{mbs}}(\dv)$  (a case that was not present in topology-aware systems). Hence, we consider the following two cases:
\begin{itemize}

\item
{\it Case~1}: 
$|\Dc_{\rm{mbs}}(\dv)|\geq n$. 
We let the MBS broadcast all the files in $\Dc_{\rm{mbs}}(\dv)$ (so that each user in $\Uc_0$ can recover its desired file). In addition, each SBS $h$ can recover all the non-cached bits in $\Fc_{\Wc}$ if $h\in \Gc^{\varphi}_{g}$ and  $g\in \Wc$.

\medskip
\paragraph{For Theorem~\ref{thm:dec shared link}}
we then use the single bottleneck-link caching scheme with shared caches in~\cite{parrinello2018sharedcache} to satisfy the requests from the users whose demanded file is in  $\Dc_{\rm{sbs}}(\dv)$.

\medskip
\paragraph{For Theorem~\ref{thm:D2D dec}}
we then use the proposed D2D caching scheme with shared caches in Section~\ref{sub:delivery D2D point case 2} to satisfy the requests from the users   whose demanded file is in  $\Dc_{\rm{sbs}}(\dv)$.

\item
{\it Case~2}:
$|\Dc_{\rm{mbs}}(\dv)|< n$.  
We first let the MBS broadcast $F_i$ where $i\in \Dc_{\rm{mbs}}(\dv)$ such that each user in $\Uc_0$ can recover its desired file. 

\medskip
\paragraph{For Theorem~\ref{thm:dec shared link}}
we then partition each subfile $F_{j,\Wc}$ where $j\in \Dc_{\rm{sbs}}(\dv)$, $\Wc\subseteq [G]$, and $|\Wc|=t$, into two parts as $F_{j,\Wc}=\{F^1_{j,\Wc},F^2_{j,\Wc}\}$ with lengths $|F^1_{j,\Wc}|=\frac{ n-|\Dc_{\rm{mbs}}(\dv)|}{|\Dc_{\rm{sbs}}(\dv)|}|F_{j,\Wc}|$ and $|F^2_{j,\Wc}|=\frac{|\Dc_{\rm{sbs}}(\dv)|-(n-|\Dc_{\rm{mbs}}(\dv)|)}{|\Dc_{\rm{sbs}}(\dv)|}|F_{j,\Wc}|$. It can be seen that if $|\Dc_{\rm{mbs}}(\dv)|\geq n$, $|F^1_{j,\Wc}|=0$. We then let the MBS broadcast  $F^1_{j,\Wc}$ for each $j\in \Dc_{\rm{sbs}}(\dv)$.
Hence, For each $\Wc\subseteq [G]$ where $|\Wc|=t$, each SBS $h\in \Gc^{\varphi}_{g}$ with $g\in \Wc$ receives
\begin{align}
 \frac{|\Dc_{\rm{mbs}}(\dv)| }{\Nsf} |\Fc_{\Wc}|+\frac{ n-|\Dc_{\rm{mbs}}(\dv)| }{\Nsf}|F_{\Wc}|= \frac{n}{\Nsf}|\Fc_{\Wc}|
\end{align}
bits of $\Fc_{\Wc}$. Thus, each SBS $h$ can recover $\Fc_{\Wc}$ from its cached content and the MBS transmission. 

We then use the single bottleneck-link caching scheme with shared caches in~\cite{parrinello2018sharedcache} to     let each 
SBS  $h\in[\Hsf]$ recover $F^2_{d_k,\Wc}$ where $\Wc\subseteq [G]$, $|\Wc|=t$,   $d_k\in \Dc_{\rm{sbs}}(\dv)$, and  $k\in \Uc_{h}$. So SBS $h$   recovers all the demanded files by its connected users. 
 
\medskip
\paragraph{For Theorem~\ref{thm:D2D dec}}
we let the SBSs transmit packets such that each SBS in $\Gc^{\varphi}_{g}$ where $g\in [G]$ can recover $\Fc_{\Wc}$ where $\Wc\subseteq [G]$, $|\Wc|=t$ and $g\in \Wc$. More precisely, for each $\Wc \subseteq [G]$ where $|\Wc|=t$ and each $g\in \Wc$, we choose one SBS $h_1 \in \Gc^{\varphi}_{g}$ to transmit $\frac{(n-|\Dc_{\rm{mbs}}(\dv)|)|\Fc_{\Wc}|}{\Nsf(G-1)}$  random linear combinations of $\Fc_{\Wc}$. It can be seen that for each $\Wc \subseteq [G]$ where $|\Wc|=t$ and each $g\in \Wc$, each SBS $h \in  \Gc^{\varphi}_{g}$ receives  $\frac{(n-|\Dc_{\rm{mbs}}(\dv)|)|\Fc_{\Wc}|}{\Nsf}$  random linear combinations of $\Fc_{\Wc}$. Furthermore, SBS $h$ obtains $|\Dc_{\rm{mbs}}(\dv)| |\Fc_{\Wc}|/\Nsf$ bits of  $\Fc_{\Wc}$ from the MBS, and $\frac{(\Nsf-n)|\Fc_{\Wc}|}{\Nsf}$   random linear combinations of  $\Fc_{\Wc}$ from its cache, such that it can recover $\Fc_{\Wc}$. This step can be done if for each $\Wc \subseteq [G]$ where $|\Wc|=t$, the union cached bits of $\Fc_{\Wc}$  by all the SBSs  is $\Fc_{\Wc}$, and thus we have the constraint $\frac{\Nsf-n}{\Nsf}t \geq 1$.

Then, we use the proposed D2D caching scheme with shared caches in Section~\ref{sub:delivery D2D point case 2} to satisfy
 the requests from the users whose demanded file is in $\Dc_{\rm{sbs}}(\dv)$.

\end{itemize}

 \paragraph*{Performance}
For a give realization of the number of users,  the worst case is when  $|\Dc_{\rm{mbs}}(\dv)| = \Kbc$ and  
$$
|\Dc_{\rm{sbs}}(\dv)\cap (\cup_{h\in \Gc^{\varphi}_{g}}\cup_{k\in \Uc_h} \{d_k\})|=q^{\prime}_g(\varphi,[G])
$$ 
 for each $g\in [G]$.
The achieved loads are as in~\eqref{eq:achieveable dec}  for Theorem~\ref{thm:dec shared link}, 
and as in~\eqref{eq:D2D achieveable dec}  for Theorem~\ref{thm:D2D dec}.

\subsection{Numerical Evaluations}
\begin{figure}
    \centering
    \begin{subfigure}[t]{0.5\textwidth}
        \centering
        \includegraphics[scale=0.6]{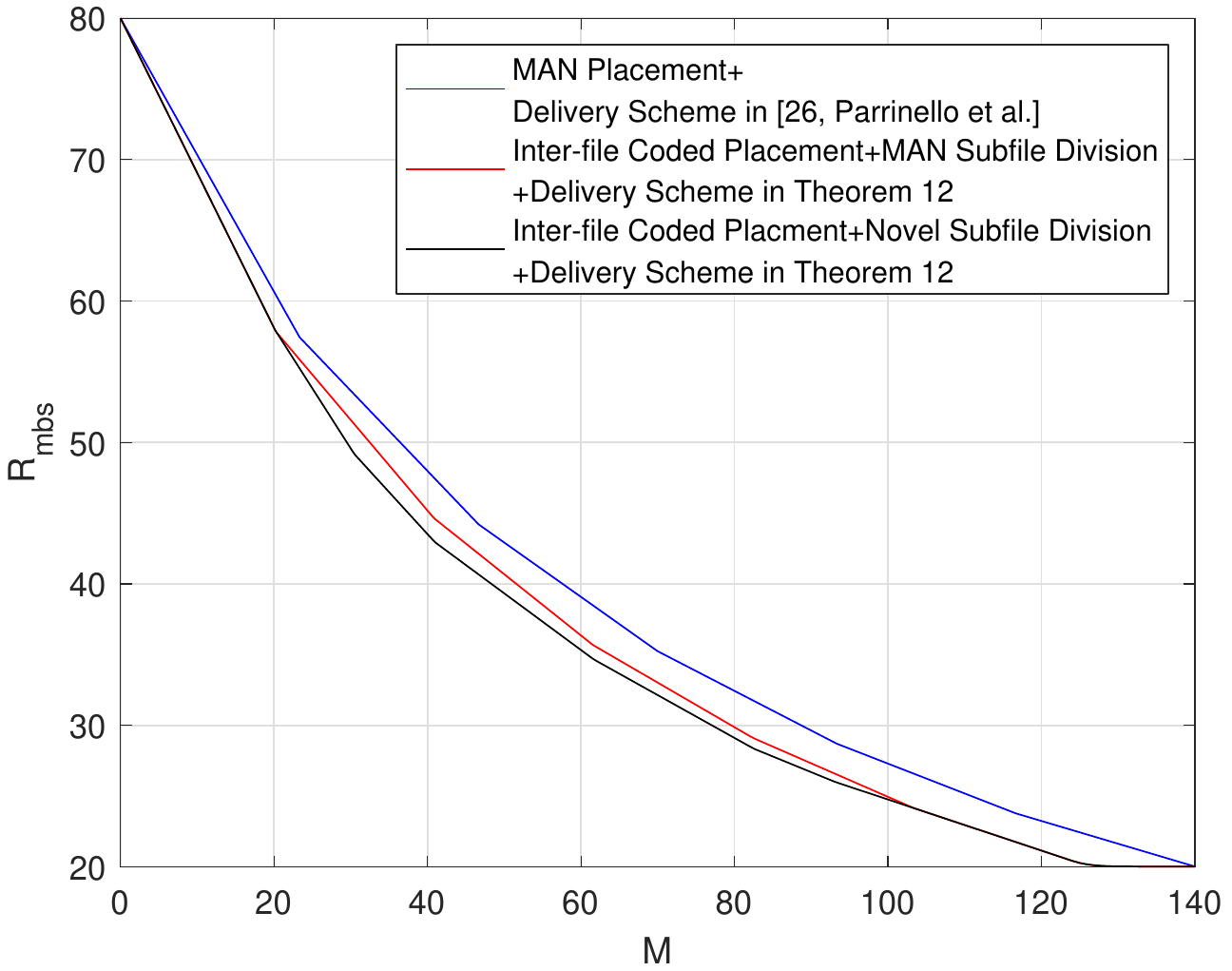}
        \caption{\small $(\Msf,\Rbc)$ tradeoff with $\Rdd=0$.}
        \label{fig:numerical 4a}
    \end{subfigure}%
    ~ 
    \begin{subfigure}[t]{0.5\textwidth}
        \centering
        \includegraphics[scale=0.6]{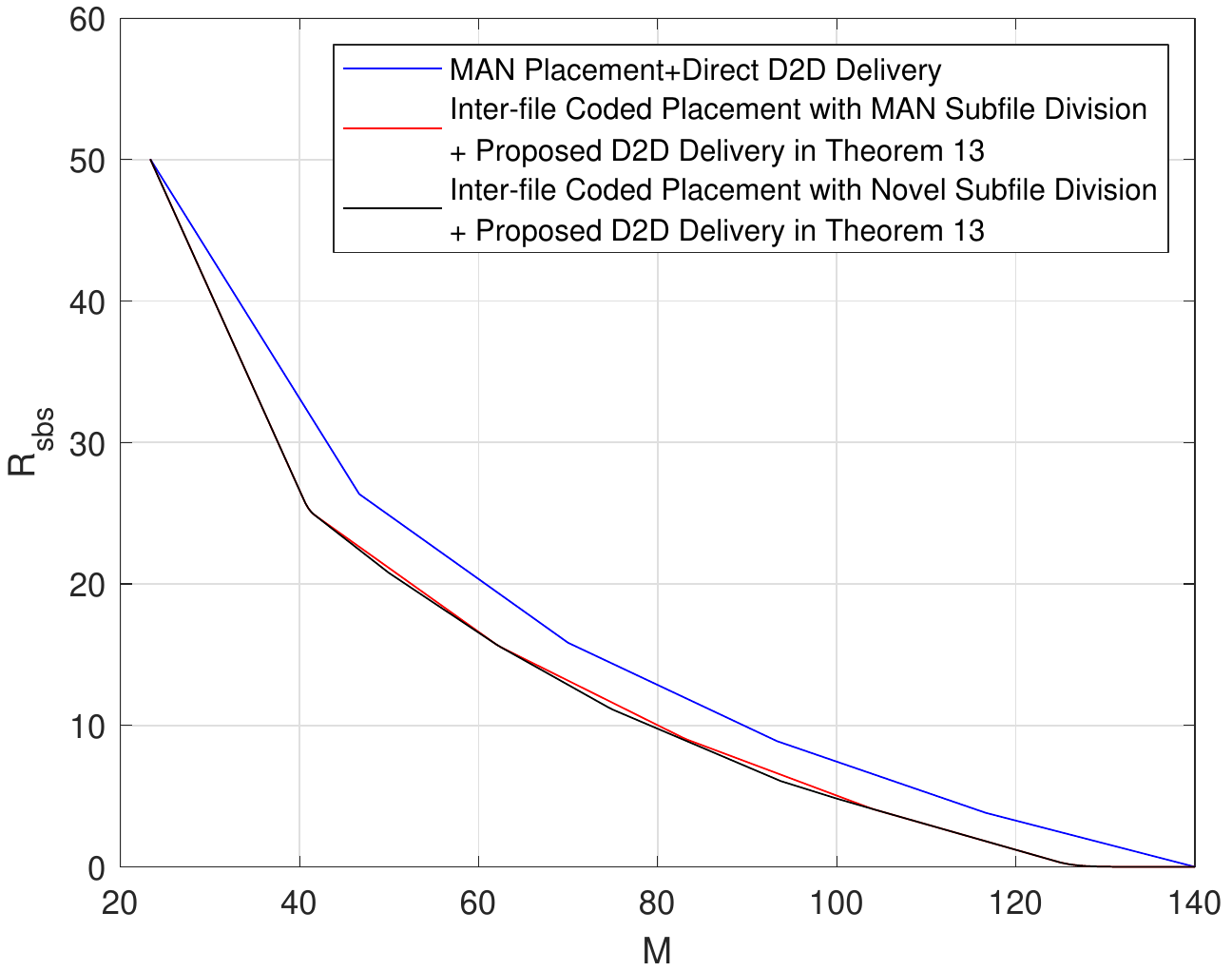}
        \caption{\small $(\Msf,\Rdd)$ tradeoff where the single bottleneck-link load is minimal.}
        \label{fig:numerical 4b}
    \end{subfigure}
    \caption{\small Consider a topology-agnostic Fog-RAN system  with  $\Hsf=6$, and $\Nsf=140$.}
    \label{fig:numerical 4}
\end{figure}

We present some numerical results on the topology-agnostic systems.   In Fig.~\ref{fig:numerical 4}, we consider  the network  with 
$\Hsf=6$, $\Nsf=140$. We use the notation  $Y\sim \textrm{Pois}(\lambda)$ to denote that the distribution of the random variable $Y$ is   Poisson distribution with factor $\lambda$, i.e., $\Pr(Y=k)=\frac {\lambda ^{k}e^{-\lambda }}{k!}$. 
We consider that the number of connected users to each SBS is according to an
 independent Poisson distribution~\cite{andrews2016poisson}, $\Lsf_{1}\sim \textrm{Pois}(20)$, $\Lsf_{2}\sim \textrm{Pois}(20)$, $\Lsf_{3}\sim \textrm{Pois}(8)$, $\Lsf_{4}\sim \textrm{Pois}(6)$, $\Lsf_{5}\sim \textrm{Pois}(4)$, $\Lsf_{6}\sim \textrm{Pois}(2)$, and $\Kbc\sim \textrm{Pois}(20)$. From the Chernoff bound we have $\Pr\{\sum_{h\in [\Hsf]} \Lsf_h+\Kbc >\Nsf \}<1.07\times 10^{-8}$, in other words, we have $\Nsf\geq \Ksf$.
 
In Theorems~\ref{thm:dec shared link} and~\ref{thm:D2D dec} for topology-agnostic Fog-RAN systems,  the average over all realizations is not easy to compute. 
To simplify the computation, we give an alternative way to approximately compute~\eqref{eq:achieveable dec} and~\eqref{eq:D2D achieveable dec} but with a much lower complexity.
For each $G\in [2:\Hsf]$, we choose the topology-agnostic $G$-way  partition with minimal variance of total number of users in each group, given by
\begin{align}
 \arg \min_{\varphi \in \mathbf{A}_G} \mathbb{E}_{(  \Lsf_1, \ldots, \Lsf_\Hsf)}\left[ \sum_{i\in [G]} \left(\sum_{j\in \Gc^{\varphi}_{i}} \Lsf_j -\frac{60}{G}\right)^2\right],\label{eq:approximation}
\end{align}
where $60/G$ is the average number of users in each group, because if the variance  increases, there may be more multicast messages which are not ``full'' (some subfiles in the binary sum do not exist) and this reduces the coded caching gain. 

We first fix $\Rdd=0$ and plot the memory-load tradeoff $(\Msf,\Rbc)$ in Fig.~\ref{fig:numerical 4a}. 
We then minimize the single bottleneck-link load and plot the memory-load tradeoff $(\Msf,\Rdd)$ in Fig.~\ref{fig:numerical 4b}. We can see that in the topology-agnostic systems, both of the proposed inter-file coded cache placement and asymmetric file subpacketization can reduce the load achieved by MAN placement.

\section{Conclusions}
\label{sec:conclusion}

In this paper, we introduced a novel cache-aided Fog-RAN network model with downlink and sidelink communication.
For topology-aware systems, we proposed a converse bound, 
a novel inter-file coded cache placement and a D2D caching scheme with shared caches. 
Some exact optimality results were derived for either small library size or larger memory size regime, while
order optimality within constant factor could also be characterized when the occupancy number of each SBS is identical or when the memory size is not too small.
We also proposed a novel asymmetric cache placement based on a novel file subpacketization  dependent additionally on the  SBS occupancy numbers, which is exactly optimal in some memory size regimes.
We then showed that the proposed inter-file coded cache placement and novel file subpacketization are also extensible to topology-agnostic systems with statistical knowledge of the network topology. 
Mathematical analysis and numerical results showed that the proposed schemes improve on the performance of  straightforward extensions of  existing schemes in  the considered systems.

\appendices

\section{Proofs of the Proposed Converse Bounds in Section~\ref{sec:main converse}}
 \label{sec:converse proof}
 \subsection{Proof of~\eqref{eq:trivial converse}}
We disregard the requests from the users served by the SBSs and only focus on the requests by the users connected directly to the MBS.
In this case, the system is equivalent to single bottleneck-link problem where the users have no caches, for which $\Rbc^{\star}\geq \min\{\Nsf,\Kbc\}$ in~\eqref{eq:trivial converse} is trivially optimal.

\subsection{Proof of~\eqref{eq:converse N>K0}}
\label{sec:proof of converse bound}

We focus next on the case $\Nsf> \Kbc$.
In this converse proof, we only consider those demand vectors that have the largest possible number of distinct entries. 
For future use, let 
\begin{align}
   \Fc_{\Uc_h} := \{F_{d_i}:i\in \Uc_h\}, \ \forall h\in[0:\Hsf],
\label{eq: demanded files in Uh def}
\end{align}
denote the set of files demanded by the users in $\Uc_h$, that is, $\Fc_{\Uc_0}$ are the files demanded by the users directly served by the MBS, $\Fc_{\Uc_1}$ those demanded by the users served by the first SBS, etc; note $|\Fc_{\Uc_h}| \leq \min\{\Nsf,|\Uc_h|\} \leq |\Uc_h|\leq \Lsf_h$ for all $h\in[0:\Hsf]$, with $\Lsf_0=\Kbc$.
Also define 
\begin{align}
X_{0}^{\Hsf} := \{X_{j}:j\in [0:\Hsf]\}
\label{eq: allX def}
\end{align}
as the collection of all transmitted signals. 
Note that we do not indicate the dependance on $\dv$ in~\eqref{eq: demanded files in Uh def} and in~\eqref{eq: allX def} so as not to clutter the notation.

For any feasible memory-load tuplet, we must have
\begin{subequations}
\begin{align}
\Bsf(\Rbc^{\star}+\Rdd^{\star}) 
  &\geq \Bsf(\Rbc^{\star}(\dv)+\Rdd^{\star}(\dv))
   \geq \sum_{h\in [0:\Hsf]} H(X_h)
   \geq H( X_{0}^{\Hsf} )
\label{eq:two parts by def}
\\&= H\big( X_{0}^{\Hsf} \big|  \Fc_{\Uc_0} \big) 
   + H\big( \Fc_{\Uc_0} \big)
   - H\big( \Fc_{\Uc_0} \big|  X_{0}^{\Hsf} \big),
\label{eq:two parts TBC}
\end{align}
\end{subequations}
where~\eqref{eq:two parts by def} follows by the definition of the loads in~\eqref{eq: MBS encoding function def}-\eqref{eq: d2d load def} and by the ``independence bound on entropy.''
We now focus on bounding the term $H\big( X_{0}^{\Hsf} \big| \Fc_{\Uc_0} \big)$ in~\eqref{eq:two parts TBC}. 
From the non-negativity of entropy and for every $s\in[\Hsf]$, we have 
\begin{subequations}
\begin{align}
  &H\big( X_{0}^{\Hsf} \big| \Fc_{\Uc_0} \big)
\label{eq:bound of R starting point}  
\\&\geq H\big( X_{0}^{\Hsf} \big| \big\{Z_h : h\in[s]\},\Fc_{\Uc_0} \big)
\\&\geq I\big( \{\Fc_{\Uc_h} : h\in[s]\};  X_{0}^{\Hsf} \big| \{Z_h : h\in[s]\}, \Fc_{\Uc_0} \big)
\\&= H\big( \{\Fc_{\Uc_h} : h\in[s]\} \big| \{Z_h : h\in[s]\}, \Fc_{\Uc_0} \big) 
   - H\big( \{\Fc_{\Uc_h} : h\in[s]\} \big| \{Z_h : h\in[s]\}, \Fc_{\Uc_0}, \  X_{0}^{\Hsf} \big)
\label{eq:bound of R first step h}  
\\&= H\big( \{\Fc_{\Uc_h} : h\in[s]\} \big| \Fc_{\Uc_0} \big) 
- I\big( \{\Fc_{\Uc_h} : h\in[s]\} ; \{Z_h : h\in[s]\} |\Fc_{\Uc_0} \big)
\notag\\& \qquad
- H\big( \{\Fc_{\Uc_h} : h\in[s]\} \big| \{Z_h : h\in[s]\}, \  X_{0}^{\Hsf} \big).
\label{eq:before summing}
\end{align}
\end{subequations}

The term $H\big( \Fc_{\Uc_0} \big|  X_{0}^{\Hsf} \big)$ in~\eqref{eq:two parts TBC} and
the term $H\big( \{\Fc_{\Uc_h} : h\in[s]\} \big| \{Z_h : h\in[s]\}, \  X_{0}^{\Hsf} \big)$ in~\eqref{eq:before summing}
can be bounded by  Fano's inequality (with $\lim_{\Bsf\to\infty}\varepsilon_{\Bsf}=0$),
from~\eqref{eq: MBS decoding functions def} and~\eqref{eq: SBS decoding functions def},
as 
\begin{subequations}
\begin{align}
H( \Fc_{\Uc_0}|  X_{0}^{\Hsf}) 
&\leq H( \Fc_{\Uc_0} | X_0)
\leq |\Fc_{\Uc_0}| \Bsf \varepsilon_{\Bsf},
\\
H\big( \{\Fc_{\Uc_h} : h\in[s]\} \big| \{Z_h : h\in[s]\}, \  X_{0}^{\Hsf} \big)
&\leq \sum_{h\in[s]} H( \Fc_{\Uc_h}|Z_h, \  X_{0}^{\Hsf})
\leq \sum_{h\in[s]} |\Fc_{\Uc_h}| \Bsf \varepsilon_{\Bsf}.
\end{align}
\label{eq:FanosIneq}
\end{subequations}
The sum of
the term $H( \Fc_{\Uc_0})$ in~\eqref{eq:two parts TBC} and
the term $H(\{\Fc_{\Uc_h} : h\in[s]\} | \Fc_{\Uc_0})$ in~\eqref{eq:before summing}
equals
\begin{align}
H(\{\Fc_{\Uc_h} : h\in[s]\} , \Fc_{\Uc_0}) = \left| \cup_{h\in[0:s]} \Fc_{\Uc_h} \right| \Bsf.
\label{eq:BitsEq}
\end{align}
We thus write~\eqref{eq:two parts TBC}, by plugging in~\eqref{eq:before summing},~\eqref{eq:FanosIneq}, and~\eqref{eq:BitsEq}, as
\begin{align}
\Rbc^{\star}+\Rdd^{\star}
\geq 
\left| \cup_{h\in[0:s]} \Fc_{\Uc_h} \right| (1- \varepsilon_{\Bsf})
- \frac{1}{\Bsf}I\big( \{\Fc_{\Uc_h} : h\in[s]\} ;\{Z_h : h\in[s]\} \big| \Fc_{\Uc_0} \big),
\label{eq:two parts}
\end{align}
where the term $\left| \cup_{h\in[0:s]} \Fc_{\Uc_h} \right|$ is the number of distinct demands by the users connect either to the MBS or the $s$ most loaded SBSs.

Now we bound the term $I\big( \{\Fc_{\Uc_h} : h\in[s]\} ;\{Z_h : h\in[s]\} \big| \Fc_{\Uc_0} \big)$ in~\eqref{eq:two parts}, which is the novel key contribution in the converse result. Trivially we have
\begin{align}
  &I\big( \{\Fc_{\Uc_h} : h\in[s]\} ;\{Z_h : h\in[s]\} \big| \Fc_{\Uc_0} \big) \leq 
  \sum_{h\in[s]} H(Z_h) \leq s \Msf \Bsf,
\label{eq:BoundITrivial}
%
\end{align}
which we shall use for the case  $\Nsf \leq \Kbc + \Lsf_{[s]}$ (as it will become clear in the following).
%
For the case $\Nsf > \Kbc + \Lsf_{[s]}$ we can provide a better bound than~\eqref{eq:BoundITrivial} as follows.
We only consider those demand vectors where the users in $\Uc_0$ demand distinct files in $[\Nsf-\Kbc+1,\Nsf]$, and the users in $\cup_{h\in [s]}\Uc_h$ demand distinct  files in $[\Nsf-\Kbc]$. 
Consider next the $\binom{\Nsf-\Kbc}{\Lsf_{[s]}}$ demand vectors, where the users in $[\Lsf_{[s]}]$ have distinct requests, and the union of the demanded files by the users in $[\Lsf_{[s]}]$ for each demand vector is different from the one of other considered $\binom{\Nsf-\Kbc}{\Lsf_{[s]}}-1$ demand vectors. Hence, among these $\binom{\Nsf-\Kbc}{\Lsf_{[s]}}$ demand vectors, each file in $[\Nsf-\Kbc]$ is demanded by the users in $[\Lsf_{[s]}]$ for $\binom{\Nsf-\Kbc-1}{\Lsf_{[s]}-1}$ times. For each such a demand vector, we have an inequality as in~\eqref{eq:two parts}. We consider all of such demand vectors and sum all of these inequalities to obtain 
\begin{subequations}
\begin{align}
  &\sum_{\Vc \in [\Nsf-\Kbc]:|\Vc|=\Lsf_{[s]}} I\big( \cup_{i\in \Vc} \{F_i\} ; \{Z_h : h\in[s]\} \big| \Fc_{\Uc_0} \big)
\\&=  
\sum_{\Vc \in [\Nsf-\Kbc]:|\Vc|=\Lsf_{[s]}}H\big( \cup_{i\in \Vc} F_i \big| \Fc_{\Uc_0} \big)
-\sum_{\Vc \in [\Nsf-\Kbc]:|\Vc|=\Lsf_{[s]}}  H\big( \cup_{i\in \Vc} \{F_i\}   \big| \{Z_h : h\in[s]\}, \Fc_{\Uc_0} \big)
\label{eq:before submodularity}
\\&=
 \binom{\Nsf-\Kbc-1}{\Lsf_{[s]}-1}   H\big( \{F_i:i\in [\Nsf-\Kbc]\} \big)
-\sum_{\Vc \in [\Nsf-\Kbc]:|\Vc|=\Lsf_{[s]}}  H\big( \cup_{i\in \Vc} \{F_i\}   \big| \{Z_h : h\in[s]\}, \Fc_{\Uc_0} \big)
\\&\leq 
 \binom{\Nsf-\Kbc-1}{\Lsf_{[s]}-1}  \left( H\big( \{F_i:i\in [\Nsf-\Kbc]\} \big)
-  H\big( \{F_i:i\in [\Nsf-\Kbc]\} \big| \{Z_h : h\in[s]\}, \Fc_{\Uc_0} \big) \right)
\label{eq:KEY STEP} 
\\&= 
\binom{\Nsf-\Kbc-1}{\Lsf_{[s]}-1}  I\big( \{F_i:i\in [\Nsf-\Kbc]\};  \{Z_h : h\in[s]\}\big| \Fc_{\Uc_0} \big)
\label{eq:def biomial coefficient} 
\\&\leq 
 \binom{\Nsf-\Kbc-1}{\Lsf_{[s]}-1}  s \Msf\Bsf  
\label{eq:cache size constraint} 
\\&= 
 \binom{\Nsf-\Kbc}{\Lsf_{[s]}} \frac{\Lsf_{[s]}}{\Nsf-\Kbc} s \Msf\Bsf  .
\label{eq:final step}
\end{align} 
\label{eq:long bound for I}
\end{subequations}
where~\eqref{eq:KEY STEP} comes from Han's inequlaity~\cite[Theorem 17.6.1]{hanineq} which leads 
$$
\frac{\sum_{\Vc \in [\Nsf-\Kbc]:|\Vc|=\Lsf_{[s]}}  H\big( \cup_{i\in \Vc} \{F_i\}  \big| \{Z_h : h\in[s]\}, \Fc_{\Uc_0} \big) }{\binom{\Nsf-\Kbc-1}{\Lsf_{[s]}-1}  } \geq  H\big( \{F_i:i\in [\Nsf-\Kbc]\} \big| \{Z_h : h\in[s]\}, \Fc_{\Uc_0} \big),
$$
 and where~\eqref{eq:cache size constraint} is from the cache size constraint in~\eqref{eq: placement functions def}.

Finally, from~\eqref{eq:two parts} (or by summing many bounds as in~\eqref{eq:two parts}) with 
vanishing $\varepsilon_{\Bsf}$, 
$\left| \cup_{h\in[0:s]} \Fc_{\Uc_h} \right| =  \Kbc  + \min\{\Lsf_{[s]},\Nsf-\Kbc\}$, and 
the best between~\eqref{eq:BoundITrivial} and~\eqref{eq:long bound for I},
we get
\begin{subequations}
\begin{align}
\Rbc^{\star}+\Rdd^{\star} 
  &\geq \Kbc  + \min\{\Lsf_{[s]},\Nsf-\Kbc\} - \min\left\{1, \frac{\Lsf_{[s]}}{\Nsf-\Kbc} \right\} s \Msf
\\&= \Kbc 
+ \min\left\{\Lsf_{[s]}, \Nsf-\Kbc \right\} \left(1- \frac{s \Msf}{\Nsf-\Kbc} \right),
\ \forall s\in[\Hsf],
\end{align}
\label{eq:converse N>K0 FINALLY}
\end{subequations}
which proves~\eqref{eq:converse N>K0}. 

\begin{rem}
\label{rem:cutset}
We could have used the cut-set strategy from~\cite{dvbt2fundamental} to bound $H\big( X_{0}^{\Hsf} \big| \Fc_{\Uc_0} \big)$ in~\eqref{eq:two parts TBC} for the case  $\Nsf > \Kbc + \Lsf_{[s]}$.  We show here that, had we done so, we would have obtained a looser bound than the one in Theorem~\ref{thm:converse}.
More precisely, fix an integer $k$ and consider the demand vector with $d_{i}=i + k \Lsf_{[s]}$ for each $i\in [\Lsf_{[s]}]$.
For example, for $k=0$, the users connected to the SBSs indexed by $[s]$ demand the first $\Lsf_{[s]}$ files; for this demand vector, by using~\eqref{eq:bound of R first step h}, we obtain
\begin{align}
H\big( X_{0}^{\Hsf} \big| \Fc_{\Uc_0} \big)
\geq H\big( \{F_{i}:i\in [\Lsf_{[s]}]\} \big| \{Z_h : h\in[s]\},\Fc_{\Uc_0} \big) - \Lsf_{[s]} \Bsf \varepsilon_{\Bsf}.
\label{eq:F1}
\end{align}
We can obtain bounds as in~\eqref{eq:F1} as long as $\Lsf_{[s]}(1+k) \leq \Nsf-\Kbc$ (as we have assumed demand vectors where the users connected to the SBSs request a file not demanded by the $\Kbc$ users connected directly to the MBS) and obtain in total $\left\lfloor \frac{\Nsf-\Kbc}{\Lsf_{[s]}} \right\rfloor$ inequalities. 
By summing all together such inequalities we obtain
\begin{subequations}
\begin{align}
  &\left\lfloor \frac{\Nsf-\Kbc}{\Lsf_{[s]}} \right\rfloor 
H\big( X_{0}^{\Hsf} \big| \Fc_{\Uc_0} \big)
\\&\geq H\Big(\big\{F_i:i\in \left[\Lsf_{[s]}\left\lfloor \frac{\Nsf-\Kbc}{\Lsf_{[s]}} \right\rfloor \right]\big\}|\{Z_h : h\in[s]\},\Fc_{\Uc_0}\Big)
 - \Lsf_{[s]} \left\lfloor \frac{\Nsf-\Kbc}{\Lsf_{[s]}} \right\rfloor \Bsf \varepsilon_{\Bsf}
\label{eq:F3}
\\&\geq H\Big(\big\{F_i:i\in \left[\Lsf_{[s]} \left\lfloor \frac{\Nsf-\Kbc}{\Lsf_{[s]}} \right\rfloor \right]\big\}|\Fc_{\Uc_0}\Big) 
 - H(\{Z_h : h\in[s]\}|\Fc_{\Uc_0})
 - \Lsf_{[s]} \left\lfloor \frac{\Nsf-\Kbc}{\Lsf_{[s]}} \right\rfloor \Bsf \varepsilon_{\Bsf}
\label{eq:F4}
\\&\geq \Lsf_{[s]} \left\lfloor \frac{\Nsf-\Kbc}{\Lsf_{[s]}} \right\rfloor
 - s\Msf
 - \Lsf_{[s]} \left\lfloor \frac{\Nsf-\Kbc}{\Lsf_{[s]}} \right\rfloor \Bsf \varepsilon_{\Bsf},
 \ \forall s\in[\Hsf].
\label{eq:F5} 
\end{align}
\end{subequations}
By using~\eqref{eq:F5} into~\eqref{eq:two parts} with $|\Fc_{\Uc_0}|= \Kbc$ (because we consider only those demand vectors with distinct entries), we have 
\begin{align}
\Rbc^{\star}+\Rdd^{\star} \geq \Kbc+\left[\Lsf_{[s]}-\frac{s \Msf}{  \left\lfloor \frac{\Nsf-\Kbc}{\Lsf_{[s]} } \right\rfloor }\right]^+, \ \forall s\in[\Hsf],\label{eq:cut-set converse}
\end{align}
which is clearly looser than the proposed converse bound in~\eqref{eq:converse N>K0}.
\end{rem}


\subsection{Proof of~\eqref{eq:converse N>K=K0+K1}}
\label{sub:proof of converse N>K}
The LHS of~\eqref{eq:two parts} can actually be taken to be $\Rbc^{\star}(\dv)+\sum_{h\not\in[s]} \Rsf_{h}^{\star}(\dv)$ by starting in~\eqref{eq:two parts by def} with the following reasoning: fix an integer $s, \ s\in [\Hsf]$; consider the SBSs in $\Sc=[s]$ as a single ``aggregate cache'' $\{Z_h : h\in\Sc\}$ with $|\cup_{h\in \Sc}\Uc_h|$ demands; to satisfy these multi-demands (by receiving packets from the MBS, and from the SBSs not in $\Sc$) and the demands of the users directly connected to the MBS
\begin{align}
\Rbc(\dv)+ \sum_{h\notin \Sc} \Rsf_{h}(\dv)
&\geq \left| \cup_{h\in\{0\}\cup \Sc} \Fc_{\Uc_h} \right| (1-\varepsilon_{\Bsf})
- \frac{1}{\Bsf}I\big( \{\Fc_{\Uc_h}: h\in \Sc\};\{Z_h : h\in\Sc\} \big|\Fc_{\Uc_0} \big).
\label{eq:convese n>K equi}
\end{align}
This reasoning is valid for any subset $\Sc$ of SBSs, and not just for those subsets of the form $\Sc=[s]$.
The idea when we continued the bounding of~\eqref{eq:two parts} was as follows. 
We summed inequalities as in~\eqref{eq:two parts} for all possible demand vectors $\dv$ in some collection $\Dc$. 
By doing so, we only wanted to deal with the RHS of those inequalities (i.e., bound the term $I(\{\Fc_{\Uc_h}: h\in \Sc\};\{Z_h : h\in[s]\} |\Fc_{\Uc_0})$ that depends on the demand vector $\dv\in\Dc$ being considered) and not having to also bound terms like $\sum_{h\notin \Sc} \Rsf_{h}(\dv)$ (because it is in general difficult to relate $\sum_{\dv\in\Dc} \sum_{h\notin \Sc} \Rsf_{h}(\dv)$ to the worst-case SBS sum-rate $\Rdd^{\star}$ for a general set $\Sc$).
Here, we propose a way to ``symmetrize'' our bound in~\eqref{eq:convese n>K equi} by working on $\sum_{\Sc \subseteq [\Hsf] : |\Sc| = s} \sum_{\dv\in\Dc}   \sum_{h\notin \Sc} \Rsf_{h}(\dv) =  \binom{\Hsf}{s}\left(1-\frac{s}{\Hsf}\right) \sum_{\dv\in\Dc} \Rdd^{\star}(\dv)$ for some appropriately chosen collection of demand vectors $\Dc$, which requires the assumption $\Nsf\geq \Ksf$ as we shall soon see.

Consider the case $\Nsf\geq \Ksf=\Kbc+\Kdd$. 
Recall that we denote the number of $k$-permutations of $n$ as $P(n,k):=n\cdot(n-1)\cdots(n-k+1) = \frac{n!}{(n-k)!}$.
We consider all $P(\Nsf-\Kbc,\Kdd)$ demand vectors where each user $k\in [\Kbc]$ demands $F_{\Nsf-k+1}$, and each user in $[\Kbc+1:\Ksf]$ demands a distinct file in $[\Nsf-\Kbc]$. We denote the collection of considered demand vectors as $\Dc$. 
Fix one integer $s\in[\Hsf]$, and focus on each subset of $s$ SBSs, which we denoted by $\Sc$ (such that $|\Sc|=s$).
For each of the $P(\Nsf-\Kbc,\Kdd)$ possible $\dv : \dv\in\Dc$, we have a bound as in~\eqref{eq:convese n>K equi}
with $\left| \cup_{h\in\{0\}\cup \Sc} \Fc_{\Uc_h} \right| = \Kbc + \Lsf_{\Sc}$.
By summing all these $P(\Nsf-\Kbc,\Kdd)$ inequalities, we have
\begin{subequations}
\begin{align}
&\frac{1}{P(\Nsf-\Kbc,\Kdd)} \sum_{\dv\in \Dc}
\left(  \Rbc(\dv)+  \sum_{h\notin \Sc} \Rsf_{h}(\dv) \right) 
\\
&\geq  (\Kbc +\Lsf_{\Sc})(1-\varepsilon_{\Bsf}) 
 - \frac{1}{\Bsf\binom{\Nsf-\Kbc}{\Lsf_{\Sc}}}\sum_{\Vc \in [\Nsf-\Kbc]:|\Vc|= \Lsf_{\Sc}} I(\cup_{i\in \Vc} F_i; \{Z_h:h\in \Sc\}| \Fc_{\Uc_0})
\label{eq:eq:convese n>K 1}
\\
&\geq  
  (\Kbc +\Lsf_{\Sc}) (1-\varepsilon_{\Bsf}) 
- \frac{\Lsf_{\Sc}}{\Nsf-\Kbc} |\Sc| \Msf, 
\label{eq:eq:convese n>K 3}
\end{align}
\end{subequations}
where~\eqref{eq:eq:convese n>K 3} directly comes from~\eqref{eq:final step}.

Next, we consider all the inequality as in~\eqref{eq:eq:convese n>K 3} for each subset of SBSs of size $|\Sc|=s$ so as to obtain $\binom{\Hsf}{s}$ inequalities, which we then sum together. For vanishing $\varepsilon_{\Bsf}$, We have
\begin{subequations}
\begin{align}
&\frac{1}{P(\Nsf-\Kbc,\Kdd)} \sum_{\dv\in \Dc}\left( \binom{\Hsf}{s}\Rbc(\dv)+    \sum_{\Sc\subseteq [\Hsf]:|\Sc|=s} \sum_{h\notin \Sc} \Rsf_{h}(\dv) \right)
\\ 
&=  
\frac{1}{P(\Nsf-\Kbc,\Kdd)} \sum_{\dv\in \Dc}\left( \binom{\Hsf}{s}\Rbc(\dv) + \binom{\Hsf-1}{s}  \sum_{h\in [\Hsf]} \Rsf_{h}(\dv) \right)
\label{eq:eq:convese n>K 4before 1}
\\
&=  
\frac{1}{P(\Nsf-\Kbc,\Kdd)} \binom{\Hsf}{s} \sum_{\dv\in \Dc}\left( \Rbc(\dv)+ \frac{\Hsf-s}{\Hsf}  \sum_{h\in [\Hsf]} \Rsf_{h}(\dv) \right)
\label{eq:eq:convese n>K 4before 2}
\\ 
&\geq  \binom{\Hsf}{s}\Kbc 
+ \sum_{\Sc\subseteq [\Hsf]:|\Sc|=s}\Lsf_{\Sc} \left( 1-\frac{s \Msf }{\Nsf-\Kbc} \right)
\label{eq:eq:convese n>K 4}
\\
&= \binom{\Hsf}{s}\Kbc +\binom{\Hsf-1}{s-1}\Kdd \left( 1-\frac{s \Msf }{\Nsf-\Kbc} \right), 
\label{eq:eq:convese n>K 5}
\\
&= \binom{\Hsf}{s}\left(\Kbc + \frac{s}{\Hsf} \Kdd \left( 1-\frac{s \Msf }{\Nsf-\Kbc} \right)\right), 
\label{eq:eq:convese n>K 5bis}
\end{align}
\end{subequations}
where~\eqref{eq:eq:convese n>K 4before 1} follows from $\sum_{\Sc\subseteq [\Hsf]:|\Sc|=s} \sum_{h\notin \Sc} \Rsf_{h}(\dv)=\binom{\Hsf-1}{s}  \sum_{h\in [\Hsf]} \Rsf_{h}(\dv)$, and
where~\eqref{eq:eq:convese n>K 5} follows from $\sum_{\Sc\subseteq [\Hsf]:|\Sc|=s} \Lsf_{\Sc}=\binom{\Hsf-1}{s-1} \Kdd$. 
In addition, by the definition of the SBS sidelink   load, we have
$\Rdd^{\star} \geq \sum_{h\in [\Hsf]} \Rsf_{h}(\dv), \ \forall \dv\in \Dc$, thus with~\eqref{eq:eq:convese n>K 4before 1} and~\eqref{eq:eq:convese n>K 5bis} we obtain~\eqref{eq:converse N>K=K0+K1}.

\section{Proof of Theorem~\ref{thm:order optimality N<K0+L1 M>N-K0}}
\label{sub:opt proof N<K0+L1}
Here we consider the parameter regime 
$\Msf < \Nsf-\Kbc\leq\Lsf_1$.

\paragraph{Converse}
By using~\eqref{eq:converse N>K0} with $s=1$ (and the condition $\Nsf-\Kbc \leq \Lsf_1$)
and~\eqref{eq:converse N>K0 s=H} (and the condition $\Nsf-\Kbc \leq \Lsf_1$ together with $\Lsf_1 \leq \Kdd$),
we have
\begin{subequations}
\begin{align}
&\Rbc^{\star}+\Rdd^{\star} \geq  \Nsf-\Msf, 
\label{eq:converse N>K0 s=1 again}
\\
&\Rbc^{\star} \geq \Nsf-\Hsf\Msf, 
\label{eq:converse N>K0 s=H again} 
\\
&\Rbc^{\star}\geq \Kbc, \quad  \Rdd^{\star}\geq 0.
\label{eq:trivial converse again}
\end{align}
Therefore, in this regime, the above converse region has two corner points
\begin{align}
 &(\Msf,\Rbc^{\star},\Rdd^{\star})=\big(\Msf,\Nsf-\Msf,0\big), \ \
  \text{and } 
\label{eq:proof N<K0+L1:p1}
\\
 &(\Msf,\Rbc^{\star},\Rdd^{\star})=\big(\Msf,\Nsf-\Hsf\Msf, (\Hsf-1)\Msf \big) \ \ 
  \text{if } \Msf\in\left[0, \ \frac{\Nsf-\Kbc}{\Hsf}\right),
\label{eq:proof N<K0+L1:p2 small M}
\\
 &(\Msf,\Rbc^{\star},\Rdd^{\star})=\big(\Msf,\Kbc, \Nsf-\Kbc-\Msf\big) \ \
  \text{if } \Msf\in\left[\frac{\Nsf-\Kbc}{\Hsf}, \ \Nsf-\Kbc\right).
\label{eq:proof N<K0+L1:p2 large M}
\end{align}
\end{subequations}

\paragraph{Achievability of~\eqref{eq:proof N<K0+L1:p1}}
From 
Theorem~\ref{thm:symm ach theorem} (we use the first term in the min expression in~\eqref{eq:achievable points 1} and use the condition $\Msf < \Nsf-\Kbc\leq\Lsf_1$), we can achieve 
\begin{align}
(\Msf,\Rbc,\Rdd) 
 = \Bigg(\underbrace{t\frac{\Nsf-\Kbc}{\Hsf}}_{:=\Msf_t}, \ \underbrace{\Kbc+(\Nsf-\Kbc)\frac{\Hsf-t}{\Hsf}}_{=\Nsf-\Msf_t}, \ 0 \Bigg),
\ t\in[0:\Hsf].
\label{eq:achievable points 1 th5}
\end{align}
Since~\eqref{eq:achievable points 1 th5} is linear in $\Msf_t$, the expression in~\eqref{eq:achievable points 1 th5} already accounts for memory-sharing among the corner points for $t\in[0:\Hsf]$. 
The achievable point in~\eqref{eq:achievable points 1 th5} exactly matches the converse point in~\eqref{eq:proof N<K0+L1:p1} for all $\Msf \leq \Nsf-\Kbc$.

\paragraph{Achievability of~\eqref{eq:proof N<K0+L1:p2 small M}}
Here $\Msf\in\left[0, \ \frac{\Nsf-\Kbc}{\Hsf}\right)$.
By memory-sharing between the point $(\Msf_1,\Rbc, \Rdd )= \left(0,  \Nsf ,  0 \right)$ (i.e., the MBS sends the whole library)
and the point in Theorem~\ref{thm:symm ach theorem} (we use the first term in the min expression in~\eqref{eq:achievable points 2}) given by 
\begin{align}
(\Msf_2,\Rbc,\Rdd) 
  =\Bigg(\underbrace{t\frac{\Nsf-\Kbc}{\Hsf}}_{:=\Msf_t}, \ \Kbc, \ \underbrace{(\Nsf-\Kbc)\frac{\Hsf-t}{\Hsf-1}}_{=(\Nsf-\Kbc-\Msf_t) \frac{\Hsf}{\Hsf-1}} \Bigg),
   \  t\in[\Hsf],
\label{eq:achievable points 2 th5}
\end{align}
for $t=1$ we obtain the point 
\begin{align}
(\Msf,\Rbc,\Rdd)  
  =\left(\Msf, \ \Nsf-\Hsf\Msf, \ \Hsf\Msf\right), \ \forall \Msf\in\left[0, \ \frac{\Nsf-\Kbc}{\Hsf}\right],
\label{eq:p2 small N<K0+L1}
\end{align}
which is to within a factor $\frac{\Hsf}{\Hsf-1}$ of the converse point~\eqref{eq:proof N<K0+L1:p2 small M}.

\paragraph{Achievability of~\eqref{eq:proof N<K0+L1:p2 large M}}
Here $\Msf\in\left[\frac{\Nsf-\Kbc}{\Hsf}, \ \Nsf-\Kbc\right)$.
Since~\eqref{eq:achievable points 2 th5} is linear in $\Msf_t$, the expression in~\eqref{eq:achievable points 2 th5} already accounts for memory-sharing among the corner points for $t\in[\Hsf]$. 
Thus~\eqref{eq:achievable points 2 th5} achieves the converse point~\eqref{eq:proof N<K0+L1:p2 large M} to within a factor $\frac{\Hsf}{\Hsf-1}$ for all $\frac{\Nsf-\Kbc}{\Hsf} \leq \Msf\leq \Nsf-\Kbc$.

This concludes the proof of  Theorem~\ref{thm:order optimality N<K0+L1 M>N-K0}.

\section{Proof of Theorem~\ref{thm:order optimality2}}
\label{sub:order opt proof2}
We consider here the case $\max\{\Msf,\Lsf_1\} < \Nsf-\Kbc$.
 
By using~\eqref{eq:converse N>K0} with $s=1$ (and the condition $\Nsf-\Kbc > \Lsf_1$)
and~\eqref{eq:converse N>K0 s=H},
we have
\begin{subequations}
\begin{align}
&\Rbc^{\star}+\Rdd^{\star} \geq  \Kbc +  \Lsf_{1} \left(1-\frac{\Msf}{\Nsf-\Kbc} \right),
\label{eq:converse N>K0 s=1 again again}
\\
&\Rbc^{\star} \geq \Kbc+\min\{\Nsf-\Kbc,\Kdd\}\left[1-\frac{\Hsf\Msf}{\Nsf-\Kbc} \right]^+,
\label{eq:converse N>K0 s=H again again} 
\\
&\Rbc^{\star}\geq \Kbc, \quad  \Rdd^{\star}\geq 0.
\label{eq:trivial converse again again}
\end{align}
Therefore, in this regime, the above converse region has two corner points
\begin{align}
 &(\Msf,\Rbc^{\star},\Rdd^{\star})=\left(\Msf, \ \Kbc+ \Lsf_{1} \left(1-\frac{\Msf}{\Nsf-\Kbc}\right),0\right), \ \
  \text{and } 
\label{eq:proof N>K0+L1:p1}
\\
 &(\Msf,\Rbc^{\star},\Rdd^{\star})=\left(\Msf, \ \Kbc, \ \Lsf_{1} \left(1-\frac{\Msf}{\Nsf-\Kbc}\right)\right) \ \
  \text{if } \Msf\in\left[\frac{\Nsf-\Kbc}{\Hsf}, \ \Nsf-\Kbc\right).
\label{eq:proof N>K0+L1:p2 large M}
\\
 &(\Msf,\Rbc^{\star},\Rdd^{\star})=\left(\Msf, \ \Kbc+\theta_1, \ \max\{\theta_2-\theta_1,0\}\right) \ \ 
  \text{if } \Msf\in\left[0, \ \frac{\Nsf-\Kbc}{\Hsf}\right),
\label{eq:proof N>K0+L1:p2 small M}
\\
 & \quad
   \theta_1 := \min\{\Nsf-\Kbc,\Kdd\}\left(1-\frac{\Hsf\Msf}{\Nsf-\Kbc} \right), \ 
   \theta_2 := \Lsf_{1} \left(1-\frac{\Msf}{\Nsf-\Kbc} \right).
\end{align}
\end{subequations}
\paragraph{Achievability of~\eqref{eq:proof N>K0+L1:p1}}
From~\eqref{eq:achievable points 1} (we use the second term in the min expression, and let $\Msf_t := t \frac{\Nsf-\Kbc}{\Hsf}$ for $t\in[0:\Hsf]$) we have 
\begin{subequations}
\begin{align}
\Rdd=0, \ 
\left.\Rbc\right|_{\Msf = \Msf_t}
&\leq \Kbc+\frac{\sum_{r\in[\Hsf-t]}\Lsf^{\prime}_r \binom{\Hsf-r}{t}}{\binom{\Hsf}{t}}
\\
&\leq \Kbc+\Lsf_1 \frac{\sum_{r\in[\Hsf-t]}\binom{\Hsf-r}{t}}{\binom{\Hsf}{t}}
\label{eq:L1def}
\\
&=\Kbc+\Lsf_1 \frac{\binom{\Hsf}{t+1}}{\binom{\Hsf}{t}}
\label{eq:pascal}
\\
&=\Kbc+\Lsf_1 \frac{\Hsf-t}{t+1}
\\
&=\Kbc+\Lsf_1 \Hsf\frac{1-\frac{\Msf_t}{\Nsf-\Kbc}}{1+\Hsf\frac{\Msf_t}{\Nsf-\Kbc}} \label{eq:new function}
\\&\leq \frac{\Hsf}{1+\Hsf\frac{\Msf_t}{\Nsf-\Kbc}} \left(\Kbc+ \Lsf_{1} \left(1-\frac{\Msf_t}{\Nsf-\Kbc}\right)\right)
\label{eq:final M prime}
\end{align}
\end{subequations}
where~\eqref{eq:L1def} is from $\Lsf_{1}:=\max_{h\in[\Hsf]}\Lsf_{h}$, and
where~\eqref{eq:pascal} is from the Pascal's triangle identity.
%
We can see that the corner point in~\eqref{eq:proof N>K0+L1:p1} is thus achieved to within a factor no larger than $\frac{\Hsf}{1+\Hsf\frac{\Msf}{\Nsf-\Kbc}} \leq \min\left\{\Hsf, \frac{\Nsf-\Kbc}{\Msf} \right\} \leq \Hsf,$ for all $\Msf \leq \Nsf-\Kbc$.

 We next focus   $\frac{\Nsf-\Kbc}{\Hsf} \leq \Msf^{\prime} \leq \Nsf-\Kbc$. Let $t\in [\Hsf]$ be the largest integer in $[\Hsf]$ where $(\Nsf-\Kbc) t/ \Hsf \leq \Msf^{\prime}$. Hence, we have 
\begin{align}
\frac{\Msf^{\prime}}{\Nsf-\Kbc} \leq \frac{\Msf_t}{\Nsf-\Kbc}+\frac{1}{\Hsf}.\label{eq:Mt N-Kbc} 
\end{align} 
From~\eqref{eq:new function}, we define 
\begin{align}
 f(\Msf=\Msf_t):= \Rbc|_{\Msf = \Msf_t} -\Kbc \leq  \Lsf_1 \Hsf\frac{1-\frac{\Msf_t}{\Nsf-\Kbc}}{1+\Hsf\frac{\Msf_t}{\Nsf-\Kbc}}\label{eq:function f} 
\end{align}
By the convexity of the function $f(\Msf)$ in terms of $\Msf$ and $f(\Nsf-\Kbc)= 0$, we have
\begin{align}
\frac{f(\Msf^{\prime}) }{1-\frac{\Msf^{\prime}}{\Nsf-\Kbc}} \leq  \frac{f(\Msf_t)  }{1-\frac{\Msf_t}{\Nsf-\Kbc}}.\label{eq:convexity}
\end{align}
Hence, we have 
\begin{subequations}
\begin{align}
 &\Rbc|_{\Msf =\Msf^{\prime}}-\Kbc \leq    \left(1-\frac{\Msf^{\prime}}{\Nsf-\Kbc} \right) \frac{\Rbc|_{\Msf =\Msf_t}-\Kbc }{1-\frac{\Msf_t}{\Nsf-\Kbc}}\\
 &\leq \left(1-\frac{\Msf^{\prime}}{\Nsf-\Kbc} \right) \frac{ \Lsf_1   }{1/\Hsf+\frac{\Msf_t}{\Nsf-\Kbc}}\\
 &\leq \left(1-\frac{\Msf^{\prime}}{\Nsf-\Kbc} \right) \frac{ \Lsf_1   }{ \frac{\Msf^{\prime}}{\Nsf-\Kbc}},\label{eq:final convexity}
\end{align}
\end{subequations}
where~\eqref{eq:final convexity} comes from~\eqref{eq:Mt N-Kbc}. Hence, we proved when $\Msf\geq \frac{\Nsf-\Kbc}{\Hsf} $, the corner point in~\eqref{eq:proof N>K0+L1:p1} is thus achieved to within a factor no larger than $(\Nsf-\Kbc)/\Msf$.

 In conclusion, we proved for any $\Msf$, the corner point in~\eqref{eq:proof N>K0+L1:p1} is thus achieved to within a factor no larger than $g:=\min\left\{\Hsf,\frac{\Nsf-\Kbc}{\Msf}\right\}$.

\paragraph{Achievability of~\eqref{eq:proof N>K0+L1:p2 large M}}
Here $\Msf\in\left[\frac{\Nsf-\Kbc}{\Hsf}, \ \Nsf-\Kbc\right)$.
We consider the second class corner points in~\eqref{eq:achievable points 2}   (we use the  second  term in the min expression which is less than or equal to $\frac{t+1}{t}\frac{\sum_{r\in[\Hsf-t]}\Lsf^{\prime}_r \binom{\Hsf-r}{t}}{\binom{\Hsf}{t}}$ because we transmit each MAN multicast message including $\frac{\Bsf}{\binom{\Hsf}{t}}$   bits  at most by $\frac{(t+1)\Bsf}{t \binom{\Hsf}{t}}$   bits in the SBS sidelink communication), for which $\Rbc=\Kbc$. 
By a similar argument to the one above, we can achieve the corner point in~\eqref{eq:proof N>K0+L1:p2 large M} to within
$2 \min\left\{\Hsf,\frac{\Nsf-\Kbc}{\Msf}\right\}$, where the factor $2$ is from bounding the term $\frac{t+1}{t}, t\in[\Hsf]$.

\paragraph{Achievability of~\eqref{eq:proof N>K0+L1:p2 small M}}
Here $\Msf\in\left[0, \ \frac{\Nsf-\Kbc}{\Hsf}\right)$.

 When $ \theta_1 \geq  \theta_2$, the corner point in the converse bound is $(\Msf,\Rbc^{\star},\Rdd^{\star})=\left(\Msf, \ \Kbc+\theta_1, 0\right)$. 
By   memory-sharing between the corner points in~\eqref{eq:achievable points 1}  (here we use the first term in the min expression for $t=0$ and $t=1$) 
\begin{align}
&(\Msf_1,\Rbc,\Rdd)=(0,\min\{\Nsf,\Kdd+\Kbc\} ,0),\\
& 
(\Msf_2,\Rbc, \Rdd) 
  =\left(\frac{\Nsf-\Kbc}{\Hsf}, \Kbc+ \min\{\Nsf-\Kbc,\Kdd\}\frac{\Hsf-1}{\Hsf}, 0 \right),
\end{align}
we can achieve 
\begin{align}
(\Msf,\Rbc,\Rdd)=\left(
\Msf, 
\Kbc + \min\{\Nsf-\Kbc,\Kdd\}\left(1-\frac{ \Msf}{\Nsf-\Kbc}\right),
0
\right),
\label{eq:p2 small N>K0+L1 theta1>}
\end{align}
for $\Msf\leq\frac{\Nsf-\Kbc}{\Hsf}$. 
For the achievable point in~\eqref{eq:p2 small N>K0+L1 theta1>}, as compared to the converse point $(\Msf,\Rbc^{\star},\Rdd^{\star})=\left(\Msf, \ \Kbc+\theta_1, 0\right)$ in~\eqref{eq:proof N>K0+L1:p2 small M}, we have 
\begin{subequations}
\begin{align}
&\Rbc=\Kbc + \min\{\Nsf-\Kbc,\Kdd\}\left(1-\frac{ \Msf}{\Nsf-\Kbc}\right)\\
&\leq \Hsf\left(\Kbc +\Lsf_1 \left(1-\frac{ \Msf}{\Nsf-\Kbc}\right)  \right)\\
&=\Hsf (\Kbc +\theta_2)\leq \Hsf (\Kbc +\theta_1).
\end{align}
\end{subequations}
This concludes the proof for  $ \theta_1 \geq  \theta_2$.

When $\theta_1 <  \theta_2$, the corner point in the converse bound is $(\Msf,\Rbc^{\star},\Rdd^{\star})=\left(\Msf, \ \Kbc+\theta_1, \theta_2-\theta_1\right)$. 
By   memory-sharing among the corner points in~\eqref{eq:achievable points 1}  (here we use the first term in the min expression for $t=0$ and $t=1$) 
\begin{align}
&(\Msf_1,\Rbc,\Rdd)=(0,\min\{\Nsf,\Kdd+\Kbc\} ,0),\\
 &
(\Msf_2,\Rbc, \Rdd) 
  =\left(\frac{\Nsf-\Kbc}{\Hsf}, \Kbc+ \min\{\Nsf-\Kbc,\Kdd\}\frac{\Hsf-1}{\Hsf}, 0 \right),
\end{align}
and the corner point in~\eqref{eq:achievable points 2}  (here we use the first term in the min expression for $t=1$) 
\begin{align}
& 
(\Msf_3,\Rbc, \Rdd) 
  =\left(\frac{\Nsf-\Kbc}{\Hsf}, \Kbc, \min\{\Nsf-\Kbc,\Kdd\}  \right),
\end{align}
we can achieve $(\Msf,\Rbc,\Rdd)$ where,
\begin{align}
\Msf&=\Msf_1 \left(1-\frac{\Hsf\Msf}{\Nsf-\Ksf_0}\right) +\Msf_2 \left(\frac{ \theta_1}{ \theta_2}-\big( 1-\frac{\Hsf\Msf}{\Nsf-\Ksf_0}\big)\right)+\Msf_3 \left(1-\frac{ \theta_1}{ \theta_2}\right).\label{eq:MS of M} 
\end{align}
Notice that $\Nsf-\Kbc>\Lsf_1$ and thus $\frac{ \theta_1}{ \theta_2}-\big( 1-\frac{\Hsf\Msf}{\Nsf-\Ksf_0}\big)\geq 0$.

For $\Rbc$, as compared to the corner point in the converse bound in~\eqref{eq:proof N>K0+L1:p2 small M} which shows $\Rbc^{\star}\geq  \Kbc+\theta_1$,  we can achieve
\begin{subequations}
\begin{align}
&\Rbc-\Kbc= \left(1-\frac{\Hsf\Msf}{\Nsf-\Ksf_0}\right) \min\{\Nsf-\Kbc,\Kdd\}+  \left(\frac{ \theta_1}{ \theta_2}-\big( 1-\frac{\Hsf\Msf}{\Nsf-\Ksf_0}\big)\right) \min\{\Nsf-\Kbc,\Kdd\}\frac{\Hsf-1}{\Hsf} \\
&= \left(1-\frac{\Hsf\Msf}{\Nsf-\Ksf_0}\right) \min\{\Nsf-\Kbc,\Kdd\} \frac{1}{\Hsf}+\frac{ \theta_1}{ \theta_2}  \min\{\Nsf-\Kbc,\Kdd\}\frac{\Hsf-1}{\Hsf} \\
&\leq \frac{ \theta_1}{\Hsf}+\frac{ \theta_1(\Hsf-1) }{1-\frac{\Msf}{\Nsf-\Kbc}} \label{eq:MS of Rbc1}\\
&\leq \frac{ \theta_1}{\Hsf}+ \theta_1 \Hsf \label{eq:MS of Rbc2}\\
&\leq  2\Hsf \theta_1 ,
\end{align}
\end{subequations}
where~\eqref{eq:MS of Rbc1} comes from $\Lsf_1 \Hsf\geq \min\{\Nsf-\Kbc,\Kdd\}$ and~\eqref{eq:MS of Rbc2} comes from $\Msf\leq (\Nsf-\Kbc)/\Hsf$. 

Simiarly, for $\Rdd$, as compared to the corner point in the converse bound in~\eqref{eq:proof N>K0+L1:p2 small M} which shows $\Rdd^{\star}\geq  \theta_2-\theta_1$,  we can achieve
\begin{subequations}
\begin{align}
&\Rdd=\left(1-\frac{ \theta_1}{ \theta_2}\right)  \min\{\Nsf-\Kbc,\Kdd\} \\
&\leq \frac{\Hsf(\theta_2-\theta_1)}{1-\frac{\Msf}{\Nsf-\Kbc}}\\
&\leq \frac{\Hsf^2 (\theta_2-\theta_1)}{\Hsf-1}\\
&\leq 2\Hsf(\theta_2-\theta_1).
\end{align}
\end{subequations}
This concludes the proof for  $ \theta_1 <  \theta_2$.

\section{Proof of Theorem~\ref{thm:order optimality}}
\label{sub:order opt proof}
Here consider the case $\Nsf > \Kbc+\Lsf_1$ and $\Lsf=\Lsf_1=\dots=\Lsf_\Hsf$.

\paragraph{Converse}
We consider those demand vectors where the users in $\Uc_0$ demand distinct files in $[\Nsf-\Kbc+1,\Nsf]$, and the users in $[\Kbc+1:\Ksf]$ demand    files in $[\Nsf-\Kbc]$. From~\eqref{eq:two parts TBC}, we have
\begin{align}
\Rbc^{\star}+\Rdd^{\star}\geq \frac{1}{\Bsf}H\big( X_{0}^{\Hsf} \big| \Fc_{\Uc_0} \big) + \Kbc(1 -\varepsilon_{\Bsf}).\label{eq:direct from two parts}
\end{align}
It was proved in~\cite[Appendix B]{Sengupta2017multirequest} that for single bottleneck-link caching problem including $\Nsf-\Kbc$ files and $\Hsf$ users,  where each  user  demands $\Lsf$ files, the number of broadcasted bits is lower bounded by  $\frac{\Nsf-\Kbc-\Msf}{11} $ for $\Msf\leq 1.275 \max\{\frac{\Nsf-\Kbc}{\Hsf}, \Lsf\}$, and lower bounded by the lower convex envelop of $\frac{\Lsf (\Hsf-t)}{11(t+1)}$ where $t=\Hsf\Msf/(\Nsf-\Kbc) \in [\Hsf]$  for $\Msf \geq 1.275 \max\{\frac{\Nsf-\Kbc}{\Hsf}, \Lsf\}$. 
In addition, when $\Msf\leq \frac{\Nsf-\Kbc}{\Hsf}$, we have $  \Rbc^{\star}\geq \Kbc+\min\{\Nsf-\Kbc,\Kdd\}\left(1-\frac{\Hsf\Msf}{\Nsf-\Kbc} \right)$ from~\eqref{eq:converse N>K0 s=H again}.

\paragraph{Achievability}
 $\Msf \geq 1.275 \max\{\frac{\Nsf-\Kbc}{\Hsf}, \Lsf\}$. We can achieve $$(\Msf,\Rbc,\Rdd)=\left(t \frac{\Nsf-\Kbc}{\Hsf} , \Kbc+\frac{\Lsf (\Hsf-t)}{ t+1 }, 0  \right)$$ and $(\Msf,\Rbc,\Rdd)=\left(t \frac{\Nsf-\Kbc}{\Hsf} , \Kbc, \frac{\Lsf (\Hsf-t)}{ t  }   \right)$ for each $t \in [\Hsf]$. By memory-sharing between this two points, we can prove the order optimality for $\Msf \geq 1.275 \max\{\frac{\Nsf-\Kbc}{\Hsf}, \Lsf\}$.

 $ (\Nsf-\Kbc)/\Hsf \leq \Msf\leq 1.275 \max\{\frac{\Nsf-\Kbc}{\Hsf}, \Lsf\}$. We can achieve $(\Msf,\Rbc,\Rdd)=\left(\Msf , \Nsf -\Msf, 0  \right)$ and $(\Msf,\Rbc,\Rdd)=\left(\Msf  , \Kbc, (\Nsf-\Kbc -\Msf)\frac{\Hsf}{\Hsf-1} \right)$. By memory-sharing between this two points, we can prove the order optimality for  $ (\Nsf-\Kbc)/\Hsf \leq \Msf\leq 1.275 \max\{\frac{\Nsf-\Kbc}{\Hsf}, \Lsf\}$.

$\Msf \leq (\Nsf-\Kbc)/\Hsf$. 
By a similar reasoning as in Appendix~\ref{sub:order opt proof2}, we can prove the order optimality for this case.
 Notice that in   Appendix~\ref{sub:order opt proof2}, when $\Msf \leq (\Nsf-\Kbc)/\Hsf$, we use the converse bound $\Rbc^{\star}+\Rdd^{\star}\geq \Kbc+\Lsf_1\left(1-\frac{\Msf}{\Nsf-\Kbc} \right)$ to obtain the order optimality to $\frac{2\min\{\Nsf-\Kbc,\Kdd\}}{\Lsf_1}\leq 2\Hsf$. In this proof, when $\Msf \leq (\Nsf-\Kbc)/\Hsf$,  we use the converse bound $\Rbc^{\star}+\Rdd^{\star}\geq \Kbc+  \frac{\Nsf-\Kbc}{11}\left(1-\frac{\Msf}{\Nsf-\Kbc} \right) $ to obtain the order optimality to $\frac{2\min\{\Nsf-\Kbc,\Kdd\}}{ (\Nsf-\Kbc)/11}\leq 22$, by a similar reasoning.

\section{Proof of Theorem~\ref{thm:optimality improvement}}
\label{sub:proof of optimality improvement}

\paragraph{Converse}
From~\eqref{eq:converse N>K0} with $s=1$ and $\Msf\geq \frac{(G-1)(\Nsf-\Kbc)}{G}$, we have 
\begin{align}
\Rbc^{\star}+\Rdd^{\star} \geq  
  \Kbc +  \Lsf_{1} \left(1-\frac{\Msf}{\Nsf-\Kbc} \right).
\label{eq:s=1 converse (2)}
\end{align}

\paragraph{Achievability}
Since $\Gc^{\Phi}_1=\{1\}$  and $\Nsf\geq \Kbc+\Lsf_1$, we have $q^{\prime}_1(\Phi,[G])=\Lsf_1$.
Hence, when  $t=G-1$ (i.e., $\Msf=\left(1-\frac{1}{G}\right)(\Nsf-\Kbc)$), 
from~\eqref{eq:achievable improvement}, the following triplet is achievable 
\begin{align}
  &\left(\frac{(G-1)(\Nsf-\Kbc)}{G}, \Kbc+\min_{\Phi \in \mathbf{Q}_G}\frac{\sum_{r=1}^{G-t} q^{\prime}_r(\Phi,[G]) \binom{G-r}{t}}{\binom{G}{t}},0  \right)
\notag
\\&=\left(\left(1-\frac{1}{G}\right)(\Nsf-\Kbc), \Kbc+\frac{\Lsf_1}{G},0  \right)
\notag
\\&=\left(\left(1-\frac{1}{G}\right)(\Nsf-\Kbc), \Kbc+\Lsf_1 \big(1-\frac{\Msf}{\Nsf-\Kbc} \big) ,0 \right).
\label{eq:ach for thm:optimality improvement}
\end{align}
When $\Msf= \Nsf-\Kbc$, we have  the achieved load $\Rbc= \Kbc$ with $t=G$ from~\eqref{eq:achievable improvement}.
By memory-sharing between $\Msf=\left(1-\frac{1}{G}\right)(\Nsf-\Kbc)$ and $\Msf=\Nsf-\Kbc$, we can prove the claim of Theorem~\ref{thm:optimality improvement}.

\section{Proof of Theorem~\ref{thm:D2D optimality improvement}}
\label{sub:proof of D2D optimality improvement}

\paragraph{Converse}
The converse is as in~\eqref{eq:s=1 converse (2)}.

\paragraph{Achievability}
First class corner points.
When $\Msf= \left(1-\frac{1}{G}\right)(\Nsf-\Kbc)$, 
we can see the achieved loads are $\Rdd= 0, \Rbc= \Kbc +  \Lsf_{1} \left(1-\frac{\Msf}{\Nsf-\Kbc} \right)$ in~\eqref{eq:ach for thm:optimality improvement}.
When $\Msf= \Nsf-\Kbc$, the achieved loads are $\Rdd= 0, \Rbc= \Kbc$.

\begin{subequations}
Second class corner points. 
Form~\eqref{eq:D2D achievable improvement} with $t=G-1$ (that is, $\Msf=\left(1-\frac{1}{G}\right)(\Nsf-\Kbc)$), we have $\Rbc=\Kbc$ and
\begin{align}
\Rdd&\leq \sum_{\Sc\subseteq [G]:|\Sc|=t+1}     \frac{q^{\prime}_1(\Phi ,\Sc)+\frac{1}{t}\max \{q^{\prime}_{t+1}(\Phi ,\Sc)-q^{\prime}_{1}(\Phi, \Sc)+q^{\prime}_{t}(\Phi, \Sc),0 \} }{\binom{G}{t}} \\
&\leq    \frac{q^{\prime}_1(\Phi, [G])+\frac{1}{G-1} q^{\prime}_{1}(\Phi,[G] )  }{G}\label{eq:use q1}\\
&= \Lsf_1 \big(1-\frac{\Msf}{\Nsf-\Kbc} \big)\frac{G}{G-1},\label{eq:use T}
\end{align}
where~\eqref{eq:use q1} is because $q^{\prime}_{1}(\Sc)\geq q^{\prime}_{t}(\Sc)\geq q^{\prime}_{t+1}(\Sc)$, and~\eqref{eq:use T} because $\Msf=\frac{(G-1)(\Nsf-\Kbc)}{G}$ and $q^{\prime}_1(\Phi,[G])=\Lsf_1$ (recall that  $\Nsf\geq \Kbc+\Lsf_1$).
When $\Msf= \Nsf-\Kbc$, we have  the achieved loads are $\Rbc= \Kbc, \Rdd=0$.
\end{subequations}

By memory-sharing between the above two classes of  corner points, between $\Msf=\frac{(G-1)(\Nsf-\Kbc)}{G}$ and $\Msf=\Nsf-\Kbc$, we can prove Theorem~\ref{thm:D2D optimality improvement}.
 
\bibliographystyle{IEEEtran}
\bibliography{IEEEabrv,IEEEexample}


\end{document}